\documentclass[]{aa}  

\usepackage{graphicx}
\usepackage{color}
\usepackage{txfonts}
\begin{document}

   \title{Non-equilibrium ionization by a periodic electron beam}
   \subtitle{I. Synthetic coronal spectra and implications for interpretation of observations}

   \author{E. Dzif\v{c}\'{a}kov\'{a}\inst{1} 
          \and J. Dud\'ik\inst{1}\thanks{RS Newton International Alumnus}
          \and \v{S}. Mackovjak\inst{2}
          }

   \institute{Astronomical Institute, Czech Academy of Sciences, Fri\v{c}ova 298, 25165 Ond\v{r}ejov, Czech Republic
              \email{elena@asu.cas.cz}
         \and {Department of Space Physics, Institute of Experimental Physics, SAS, Watsonova 47, 040 01 Košice, Slovak Republic}
             }
   \date{Received ; accepted }

  \abstract
   {Coronal heating is currently thought to proceed via the mechanism of nanoflares, small-scale and possibly recurring heating events that release magnetic energy.}
  % context heading (optional)
   {We investigate the effects of a periodic high-energy electron beam on the synthetic spectra of coronal Fe ions.}
  % aims heading (mandatory)
   {Initially, the coronal plasma is assumed to be Maxwellian with a temperature of 1 MK. The high-energy beam, described by a $\kappa$-distribution, is then switched on every period $P$ for the duration of $P/2$. The periods are on the order of several tens of seconds, similar to exposure times or cadences of space-borne spectrometers. Ionization, recombination, and excitation rates for the respective distributions are used to calculate the resulting non-equilibrium ionization state of Fe and the instantaneous and period-averaged synthetic spectra.}
  % methods heading (mandatory)
   {Under the presence of the periodic electron beam, the plasma is out of ionization equilibrium at all times. The resulting spectra averaged over one period are almost always multithermal if interpreted in terms of ionization equilibrium for either a Maxwellian or a $\kappa$-distribution. Exceptions occur, however;  the EM-loci curves appear to have a nearly isothermal crossing-point for some values of $\kappa_\mathrm{s}$. The instantaneous spectra show fast changes in intensities of some lines, especially those formed outside of the peak of the respective EM$(T)$ distributions if the ionization equilibrium is assumed.}
  % results heading (mandatory)
  {}
%   {The possible multithermal interpretation of the radiation from non-equilibrium plasmas shows that the non-equilibrium effects are potentially important when interpreting the coronal spectra, and deserve further study.}
  % conclusions heading (optional), leave it empty if necessary 

   \titlerunning{}
   \authorrunning{Dzif\v{c}\'akov\'a et al.}
   \maketitle
%
%________________________________________________________________
\section{Introduction}
\label{Sect:1}

The solar corona, the upper atmosphere of the Sun, has temperatures of up to several million Kelvin. Its radiation is characterized by a large number of emission lines originating from multiply ionized metal ions, such as Fe, Ca, and Si \citep[e.g.,][]{Landi02,Curdt04,Landi09}. This emission typically comes from coronal loops and unresolved background located within active regions \citep[see, e.g.,][for recent observations]{DelZanna03,Cirtain05,Young09,Young12,Tripathi09,Schmelz09,Schmelz11,Warren12,Ugarte12,Teriaca12,DelZanna13,Subramanian14,Gupta15}.

It is currently thought that the active region corona is heated by nanoflares, i.e., impulsive and possibly recurring processes of unknown origin releasing small quantities of magnetic energy \citep[e.g.,][]{Parker88,Cargill94,Cargill04,Cargill14,Klimchuk06,Klimchuk10,Tripathi10,Warren11,Viall11,Viall12,Viall15,Bradshaw12,Reep13,Winebarger13,Testa13,Klimchuk15}. It has been recognized that such impulsive energy releases could occur on timescales shorter than the ionization equilibration timescales \citep[e.g.,][]{Bradshaw03b,Bradshaw06,Reale08,Smith10,Bradshaw11}, leading to significant departures of the ion composition of the plasma from its equilibrium value for the given local electron temperature. This non-equilibrium ionization must then be taken into account when modeling the solar corona and the arising spectra \citep[see, e.g.,][]{Bradshaw03a,Bradshaw09,Olluri13a,Olluri13b,Olluri15}.

The  presence of the non-equilibrium ionization, however, may  not be the only difficulty connected to the impulsive energy release in nanoflares. Accelerated particles may be present as well, especially if the nanoflares happen via magnetic reconnection \citep[e.g.,][]{Fletcher11,Cargill12,Gontikakis13,Gordovskyy13,Gordovskyy14,Burge14}, wave-particle interactions \citep{Vocks08,Che14}, or turbulence with a diffusion coefficient inversely proportional to velocity \citep{Hasegawa85,Laming07,Bian14}. The presence of energetic particles have been indirectly detected using analyses of emission lines originating in active regions \citep{Dzifcakova11,Testa14,Dudik15}.% In particular, \citet{Dudik15} detected the presence of the non-Maxwellian $\kappa$-distributions in a transient coronal loop using the ratio-ratio method involving line ratios sensitive to energetic particles.

Nevertheless, the current analyses of coronal spectra are subject to various difficulties and uncertainties involved and therefore the departures from equilibrium are still largely ignored. This paper presents an exploratory study of the combined effects of energetic particles and the non-equilibrium ionization on the coronal spectra. A simple model of a periodic electron beam is assumed in Sect. \ref{Sect:2}. The resulting behavior of the non-equilibrium plasma is described in Sect. \ref{Sect:3}, while the consequences for interpretation of observed spectra are given in Sect. \ref{Sect:4}. A summary is presented in Sect. \ref{Sect:5}.

%
%---------------------------------------------------- FIGURE 1
\begin{figure*}[ht]
\sidecaption
        \centering
        \includegraphics[width=6.415cm,clip,bb= 25  0 450 440]{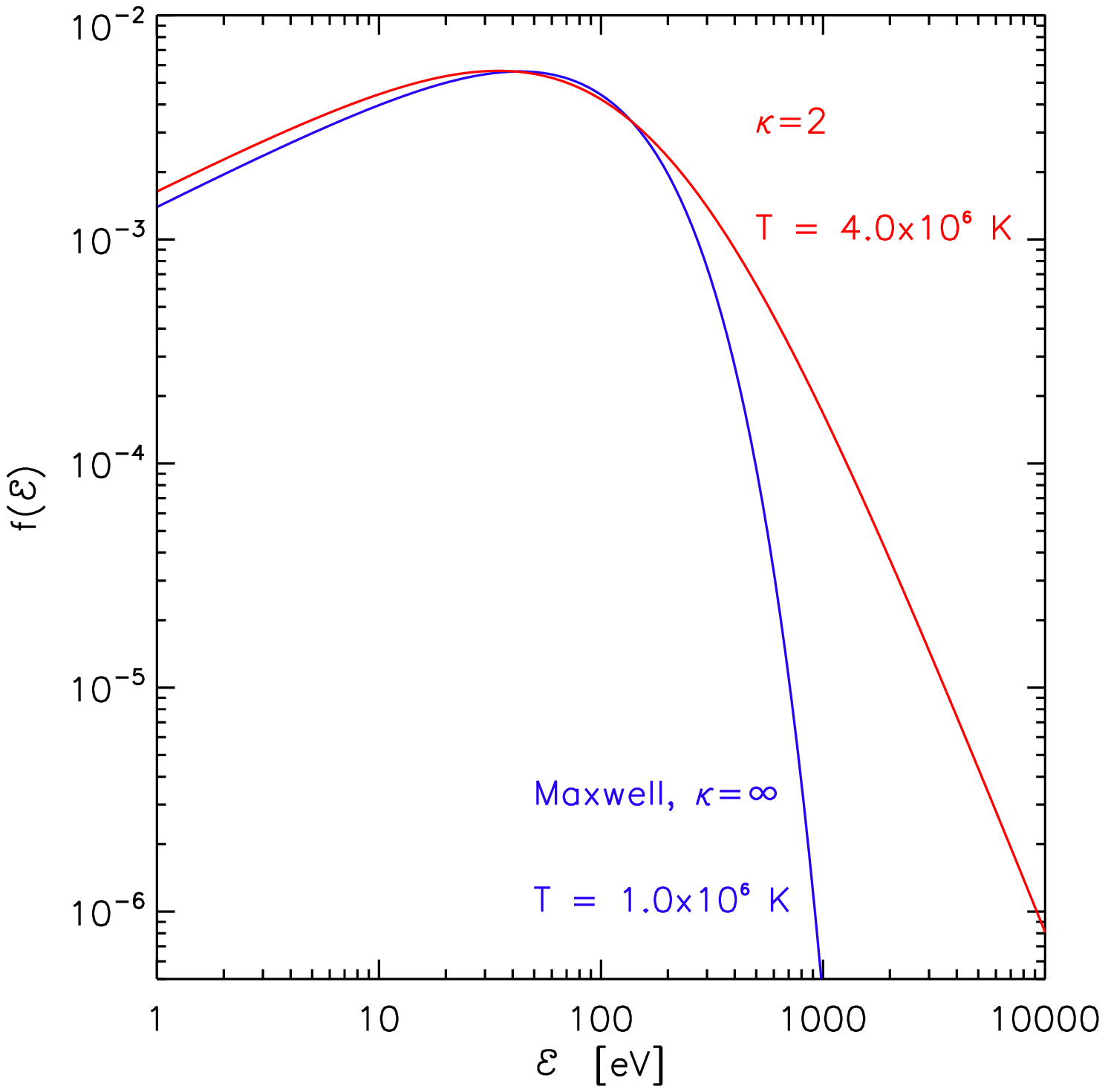}
        \includegraphics[width=5.585cm,clip,bb= 80  0 450 440]{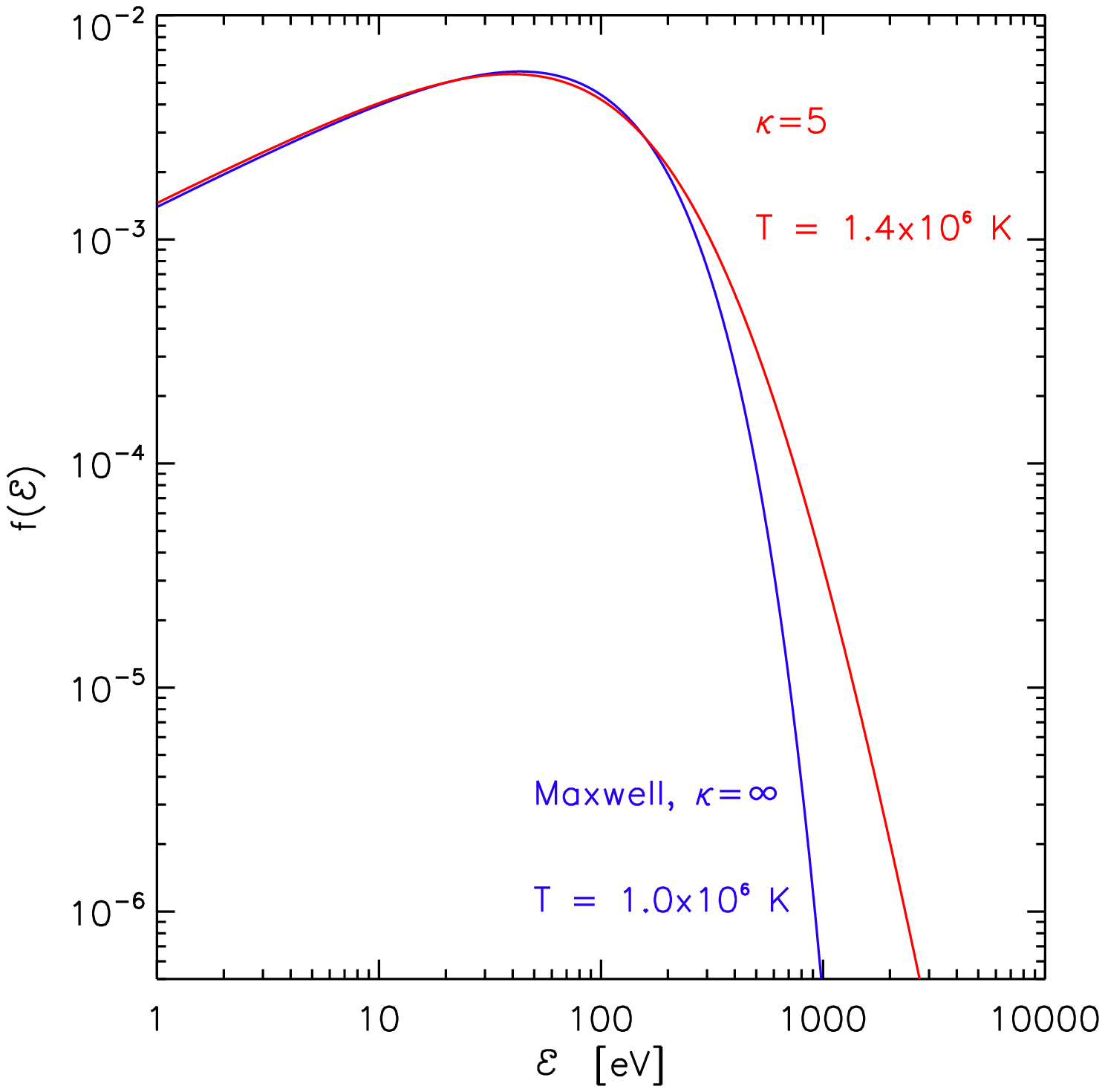}
\caption{Electron energy distributions considered in the model. The Maxwellian distribution with $T$\,=\,1 MK is shown in blue, while the $\kappa$-distributions with $\kappa$\,=\,2 and 5 are shown in the \textit{left} and \textit{right} panels, respectively. The corresponding temperatures of the $\kappa$-distributions are indicated.
\label{Fig:Distribution}}
\end{figure*}
%----------------------------------------------------
%
%________________________________________________________________
\section{Method}
\label{Sect:2}

%----------------------------------------
\subsection{Periodic electron beam}
\label{Sect:2.1}

We consider a simple model where an electron beam passing through the plasma is periodically ``switched on'' and then ``switched off'' with a fixed period $P$. Initially, and when the beam is switched off, the plasma is assumed to have a Maxwellian distribution with a temperature $T$\,=\,1\,MK. This temperature is chosen arbitrarily to represent, for example, a typical temperature of a warm coronal loop as derived from observations \citep[e.g.,][]{DelZanna03,DelZanna13,Warren03,Winebarger03,Aschwanden05a,Aschwanden08,Schmelz09,Schmelz11,Landi10,Brooks11,Brooks12}.

At time $t$\,=\,0, the electron beam is switched on for the duration of a half-period $P/2$. The electron beam is represented by power-law high-energy tail described by a $\kappa$-distribution \citep[e.g.,][]{Vasyliunas68,Owocki83,Dzifcakova15}
\begin{equation}
 f_\kappa(E)dE = A_\kappa C_\kappa \frac{2}{\sqrt{\pi}\left(k_\mathrm{B}T_\kappa\right)^{3/2}} \frac{E^{1/2}dE}{\left(1+\frac{E}{(\kappa-3/2}k_\mathrm{B}T_\kappa\right)^{\kappa+1}}\,,
 \label{Eq:Kappa}
\end{equation}
where $A_\kappa$ is a normalization constant, $k_\mathrm{B}$ is the Boltzmann constant, $C_\kappa$ is a multiplication constant determined by matching the peak of the $\kappa$-distribution to the peak of the Maxwellian (Fig. \ref{Fig:Distribution}), and $T_\kappa$ is the temperature of the $\kappa$-distribution that is in principle different from the chosen $T$\,=\,1\,MK for the Maxwellian distribution. The slope of the power-law tail of the $\kappa$-distribution is given by $\kappa + 1/2$. %The temperature of the $\kappa$-distribution, given by its mean energy \citep[see, e.g.,]{Dzifcakova15} is chosen so that the low-energy bulk of the $\kappa$-distribution is approximately the same as the Maxwellian distribution with $T$\,=\,1\,MK (Fig. \ref{Fig:Distribution}). For $\kappa$\,=\,2, the temperature is 4\,MK, while for $\kappa$\,=\,5, the temperature is 1.4\,MK.

The choice of a $\kappa$-distribution to describe the power-law high-energy tail is made because the $\kappa$-distribution is a distribution with a Maxwellian-like bulk \citep{Oka13} and a strong high-energy tail while still being continuous and described by only  one extra free parameter, $\kappa$. The values of $C_\kappa$ and $T_\kappa$ are then dependent on the value of $\kappa$ chosen, with
\begin{eqnarray}
 T_\kappa &=& \frac{\kappa}{\kappa -3/2}T, \\
 \label{Eq:T_kappa}
 C_\kappa &=& \mathrm{exp}(-1) \frac{\Gamma(\kappa -1/2)}{\Gamma(\kappa+1)} \kappa^{3/2} \left(1 + \frac{1}{\kappa}\right)^{\kappa+1},
 \label{Eq:C_kappa}
\end{eqnarray}
\citep[cf.,][]{Oka13}. For $\kappa$\,=\,2, the temperature $T_\kappa$ is 4\,MK, while for $\kappa$\,=\,5, it is approximately 1.4\,MK. It is not surprising that these $T_\kappa$ values are higher than the Maxwellian $T$, since the switch-on of the electron beam effectively adds energetic particles to the emitting plasma considered. The $\kappa$-distributions with $\kappa$\,=\,5 and 2 adds 19\% and 56\% particles, respectively (see Eq. \ref{Eq:C_kappa}). Most of these particles are added at energies of several keV or below (see Fig. \ref{Fig:Distribution}), i.e., at energies undetectable by RHESSI \citep[cf.,][]{Hannah10}.

We note that we do not consider the evolution of the distribution function because of the collisions of the high-energy electrons with the ambient Maxwellian. This is of course only a crude approximation that is analogous to assuming that the $\kappa$-distribution is being generated for the duration of $P/2$ somewhere else along the emitting loop, passes through the plasma whose radiation is being investigated, exits the region of interest after $P/2$, and is thermalized in the chromosphere. Our aim is only to explore the possible effects of periodic electron beams on the coronal spectra without constructing a too specific coronal loop model including geometry, hydrodynamic evolution, and details of particle acceleration and interaction. %Nevertheless, the thermalization timescales for keV electrons can be quite short, fraction of a second \citep{Che14}, orders of magnitude shorter than the corresponding ionization and recombination timescales at a given density. I.e., the thermalization effects during the assumed abrupt change of the distribution from a $\kappa$-one to a Maxwellian when the beam is switched off should not be important.

%
%---------------------------------------------------- FIGURE 2
\begin{figure}[ht]
        \centering
        \includegraphics[width=6.0cm,clip,bb= 30  0 396 396]{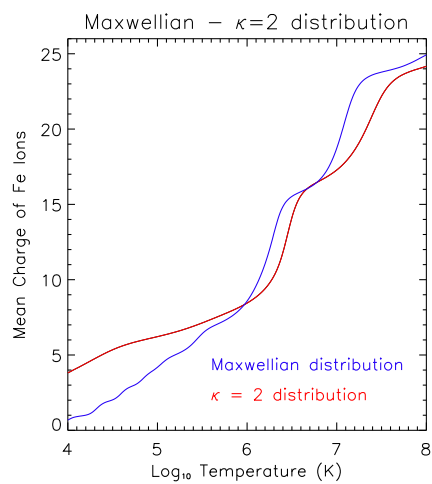}
\caption{ Mean charge of iron in ionization equilibrium as a function of electron temperature for the Maxwellian (blue) and a $\kappa$-distribution with $\kappa$\,=\,2 (red).
\label{Fig:T_eff}}
\end{figure}
%----------------------------------------------------
%
%----------------------------------------
\subsection{Non-equilibrium ionization and the effective temperature}
\label{Sect:2.2}

At $t$\,=\,0, the plasma is assumed to be in  collisional ionization equilibrium corresponding to the Maxwellian with $T$\,=\,1\,MK. The corresponding relative ion abundances and  the ionization and recombination cross-sections are given by the atomic data present in the CHIANTI database, version 7.1 \citep{Dere97,Dere07,Landi13}. At $t$\,=\,0, the beam is switched on, resulting in a change of the electron distribution including $T$\,$\to$\,$T_\kappa$, with the ionization and recombination rates changing correspondingly. If the period $P$ is less than the ionization equilibration timescale \citep[e.g.,][]{Golub89,Reale08,Smith10}, the plasma is no longer in ionization equilibrium. These equilibration timescales are typically tens or hundreds of seconds, depending on the temperature, density, and element considered \citep[see Fig. 1 in][]{Smith10}.

We consider the following equation for ionization non-equilibrium \citep[e.g.,][]{Bradshaw03a,Bradshaw03b,Smith10}
\begin{equation}
 \frac{d Y_i}{d t} = n_\mathrm{e} \left( I_{i-1} Y_{i-1} + R_{i+1} Y_{i+1} - I_i Y_i - R_i Y_i\right)\,,
 \label{Eq:NonEq_Ioniz}
\end{equation}
where $Y_i$ is the relative ion abundance of the $i$-times ionized ion of the element $Y$, with $\sum_i Y_i = 1$, $n_\mathrm{e}$ is electron number density, and $I_i$ and $R_i$ are the total ionization and recombination coefficients \textit{from} the ion stage $+i$. The advective term is not considered as we do not consider any specific hydrodynamic model of a coronal loop.

The ionization and recombination processes considered include the direct impact ionization, autoionization, and  radiative and dielectronic recombination. For the Maxwellian distribution, these rates are obtained from the CHIANTI database, while for the $\kappa$-distributions the rates were obtained by \citet{Dzifcakova13a}.

We note that the electron density $n_\mathrm{e}$ is not constant throughout the period $P$. In the first half-period, where the distribution is a $\kappa$-distribution, the $n_\mathrm{e}$ is $C_\kappa$-times higher than during the second, Maxwellian half-period   because the change of the distribution to a $\kappa$-one effectively adds particles (Fig. \ref{Fig:Distribution}). The first-order differential equation (\ref{Eq:NonEq_Ioniz}) is then solved using the Runge-Kutta method with a sufficiently small time step ensuring the stability of the solution and the condition that $\sum_i Y_i$\,=\,1 is fulfilled at every time step.

To describe the non-equilibrium ionization state, we used the effective ionization temperature. In ionization equilibrium, the mean charge state of an element is a monotonically increasing function of electron temperature. The shape of the curve depends on $\kappa$;  an example for iron is shown in Fig. \ref{Fig:T_eff}. For $\kappa$\,=\,2 and log($T$/K)\,$\lesssim$\,6.2, it is shallower than for the Maxwellian distribution. This is a consequence of the shift of the equilibrium relative ion abundances of Fe towards lower $T$ for low values of $\kappa$ \citep{Dzifcakova13a}. At higher temperatures, the curve is similar to the one for the Maxwellian distribution, but is shifted towards higher $T$ for low values of $\kappa$, again a consequence of the behavior of the ionization equilibrium with $\kappa$.

The monotonic nature of these curves permits us to define the effective temperatures $T_\mathrm{eff}^{\kappa=\mathrm{const.}}$ for a respective $\kappa$-distribution. We define it as the temperature corresponding to a particular mean charge of an element if interpreted in terms of ionization equilibrium for a given value of $\kappa$. We note that in ionization equilibrium, the effective ionization temperature is by definition equal to the electron temperature if a correct value of $\kappa$ is used.

In the remainder of this work, we  study the non-equilibrium ionization of iron under a periodic electron beam (Sect. \ref{Sect:3}). The $T_\mathrm{eff}$ is calculated from the mean Fe charge in the non-equilibrium ionization state under the assumption that the plasma is either Maxwellian or non-Maxwellian with a constant value of $\kappa$, as done when interpreting the coronal observations \citep[see, e.g.,][]{Dudik15}.

%
%----------------------------------------
\subsection{Synthesis of coronal spectra}
\label{Sect:2.3}

Once the relative ion abundances $Y_i$ are obtained, we calculate the synthetic \ion{Fe}{VIII}--\ion{Fe}{XVII} line spectra at wavelengths of 170--290\,\AA, similar to the wavelength range observed by the Extreme-ultraviolet Imaging Spectrometer \citep[EIS,][]{Culhane07} onboard the Hinode satellite \citep{Kosugi07}. The emissivities $\varepsilon_{i,jk}$ of the Fe lines arising due to the transition $j$\,$\to$\,$k$ in the ion $Y_i$ in the optically thin coronal conditions are given by \citep[e.g.,][]{Mason94,Phillips08}
\begin{equation}
 \varepsilon_{i,jk} = \frac{hc}{\lambda_{jk}} \frac{A_{jk}}{n_\mathrm{e}} \frac{n_j}{n_i} Y_i A_\mathrm{Fe} n_\mathrm{e} n_\mathrm{H} = A_\mathrm{Fe} G_{i,jk}(T,n_\mathrm{e},\kappa) n_\mathrm{e} n_\mathrm{H}\,,
 \label{Eq:Emissivities}
\end{equation}
where $\lambda_{jk}$ is the wavelength of the transition, $h$ is the Planck constant, $c$ is the speed of light, $n_\mathrm{H}$ is the hydrogen number density, $n_j$ is the number density of the $i$-times ionized Fe ion with the electron on the excited upper level $j$, $n_i$\,=\,$Y_i n_\mathrm{Fe}$ and $A_\mathrm{Fe}$\,=\,$n_\mathrm{Fe}/n_\mathrm{H}$ is the iron abundance relative to hydrogen, and $G_{i,jk}(T,n_\mathrm{e},\kappa)$ is the line contribution function. The Einstein's coefficients $A_{jk}$ for the spontaneous radiative transition are taken from the CHIANTI database, version 7.1 \citep{Dere97,Landi13}, as are the collisional excitation and deexcitation rates for the Maxwellian distribution. For the $\kappa$-distributions, these collisonal excitation and deexcitation rates are obtained using the approximative method implemented in the KAPPA database \citep{Dzifcakova15}. It was shown therein that the relative accuracy of the rates obtained using this approximative method are typically higher than 5\% when compared to the rates obtained by direct calculations from the respective cross-sections \citep{Dudik14b}. We note however that some of the excitation rates have been recently updated in the version 8 of the CHIANTI database \citep{DelZanna15b}. The influence of these new atomic data on the resulting spectra is discussed in  Appendix \ref{Appendix:A}. Finally, we adopted the ``coronal'' value of $A_\mathrm{Fe}$ based on the abundance measurements of \citet{Feldman92}.

The corresponding line intensities $I_{i,jk}$ are then obtained by using the formula 
\begin{equation}
 I_{i,jk} = A_\mathrm{Fe} \int_l G_{i,jk}(T,n_\mathrm{e},\kappa) n_\mathrm{e} n_\mathrm{H} dl \,,
 \label{Eq:Intensity}
\end{equation}
where $l$ is the observer's line of sight through the optically thin corona. We note that this equation is commonly recast as 
\begin{equation}
 I_{i,jk}= A_\mathrm{Fe} \int_T G_{i,jk}(T,n_\mathrm{e},\kappa) \mathrm{DEM}_\kappa(T) dT\,
 \label{Eq:Intensity_DEM}
\end{equation}
for the purpose of intepretation of observations. Here, the quantity DEM$_\kappa(T) = n_\mathrm{e} n_\mathrm{H} dl/dT$ is the differential emission measure \citep[see, e.g., chapter 4.6 in][]{Phillips08}, generalized for the $\kappa$-distributions by \citet{Mackovjak14}. The corresponding emission measure EM$_{\kappa}(T)$ is then simply given as EM$_\kappa(T)$\,=\,DEM$_\kappa(T)dT$.

%
%---------------------------------------------------- FIGURE 3
\begin{figure*}[ht]
        \centering
        \includegraphics[width=4.74cm,clip,bb=  25  0 390 396]{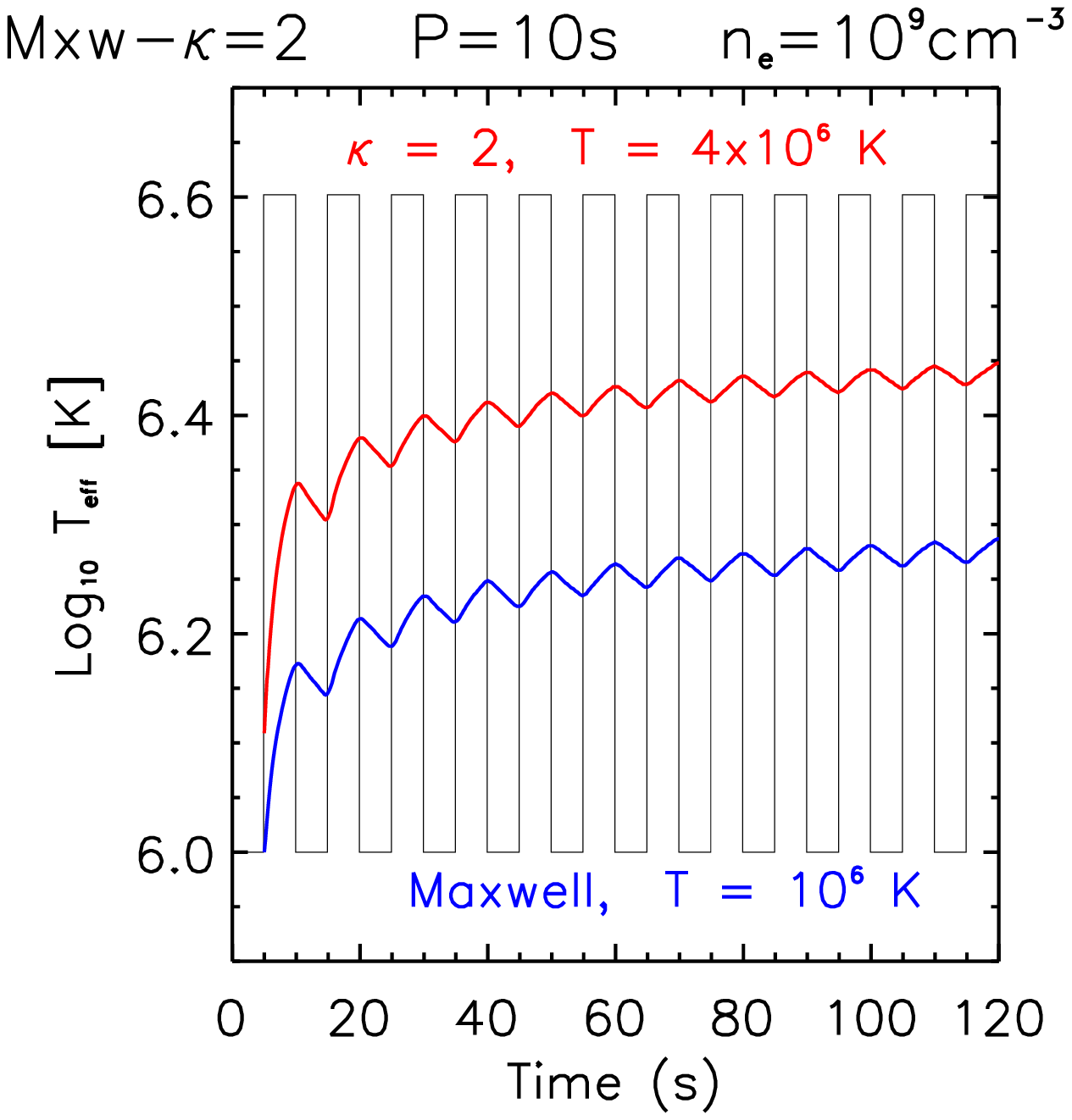}
        \includegraphics[width=6.89cm,clip,bb=  45  0 575 396]{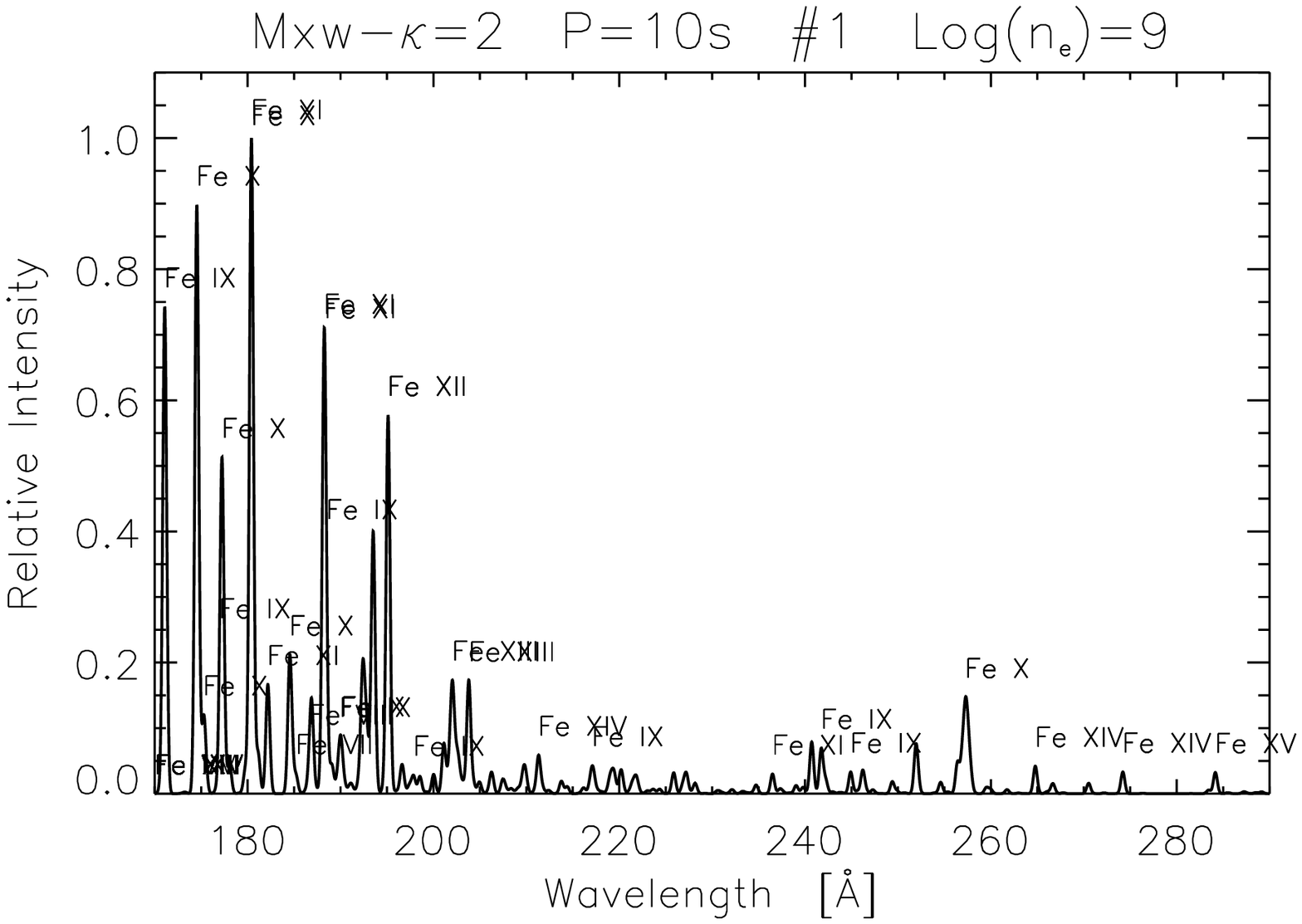}
        \includegraphics[width=6.17cm,clip,bb= 100  0 575 396]{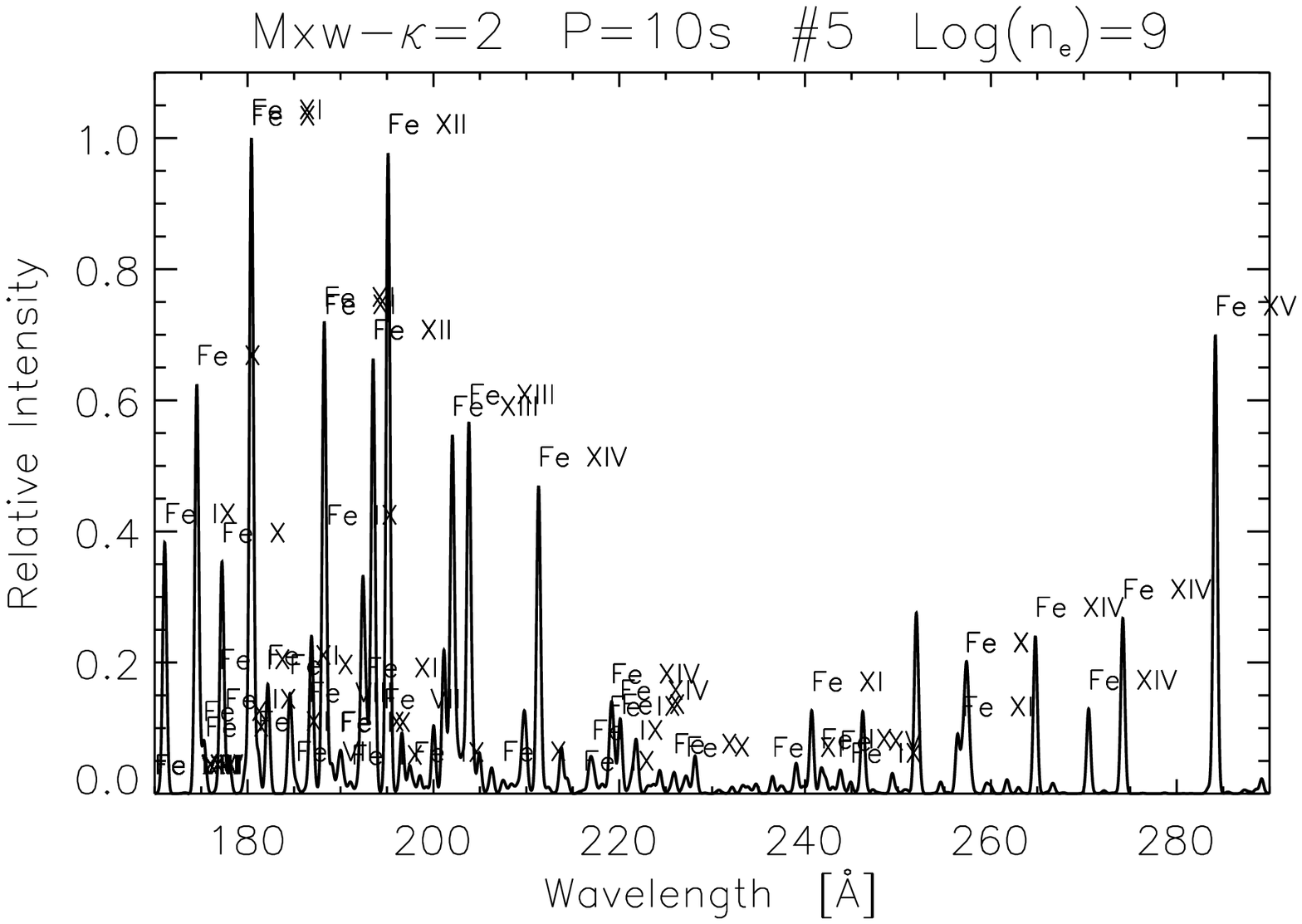}
        \includegraphics[width=4.74cm,clip,bb=  25  0 390 396]{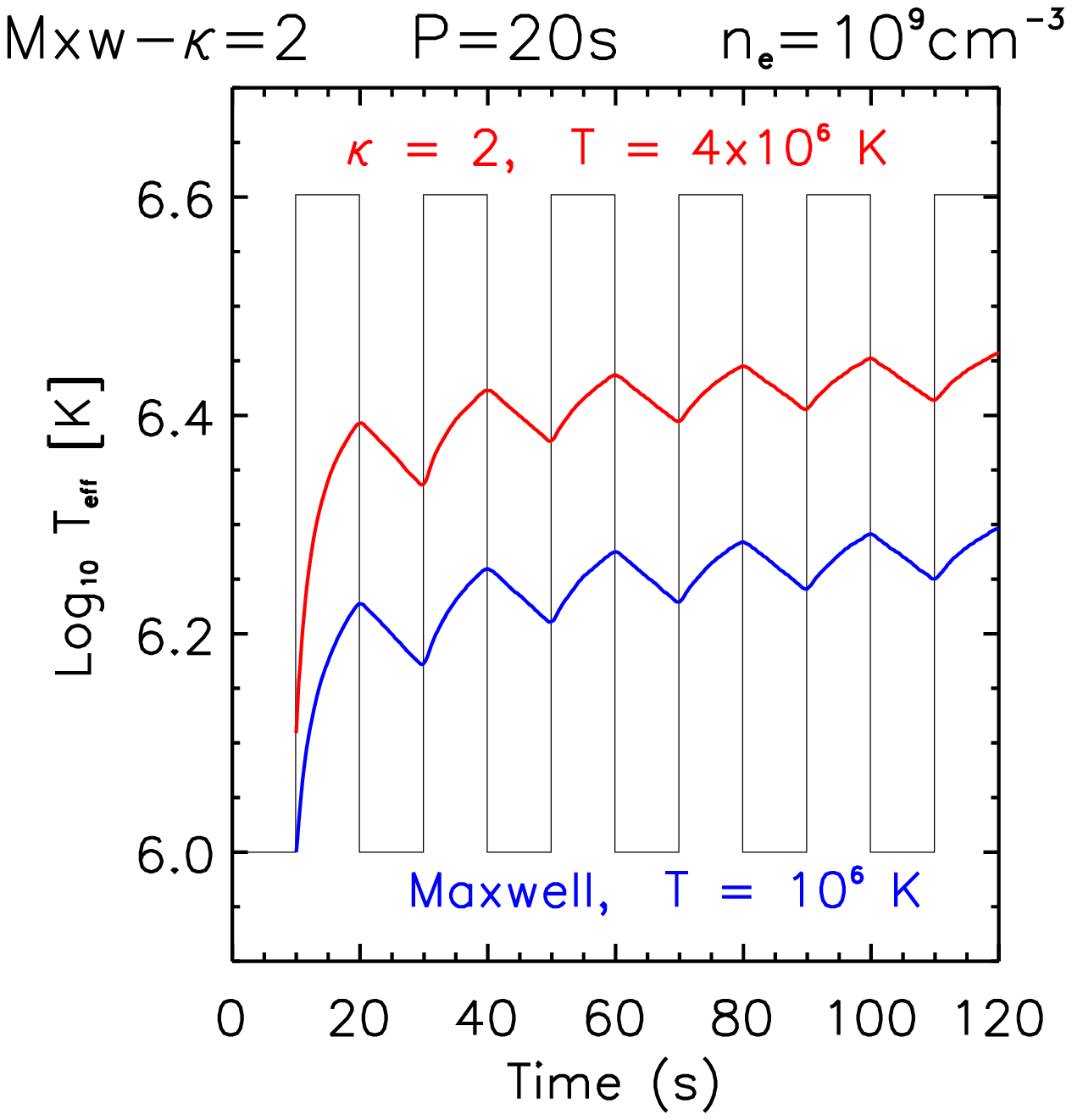}
        \includegraphics[width=6.89cm,clip,bb=  45  0 575 396]{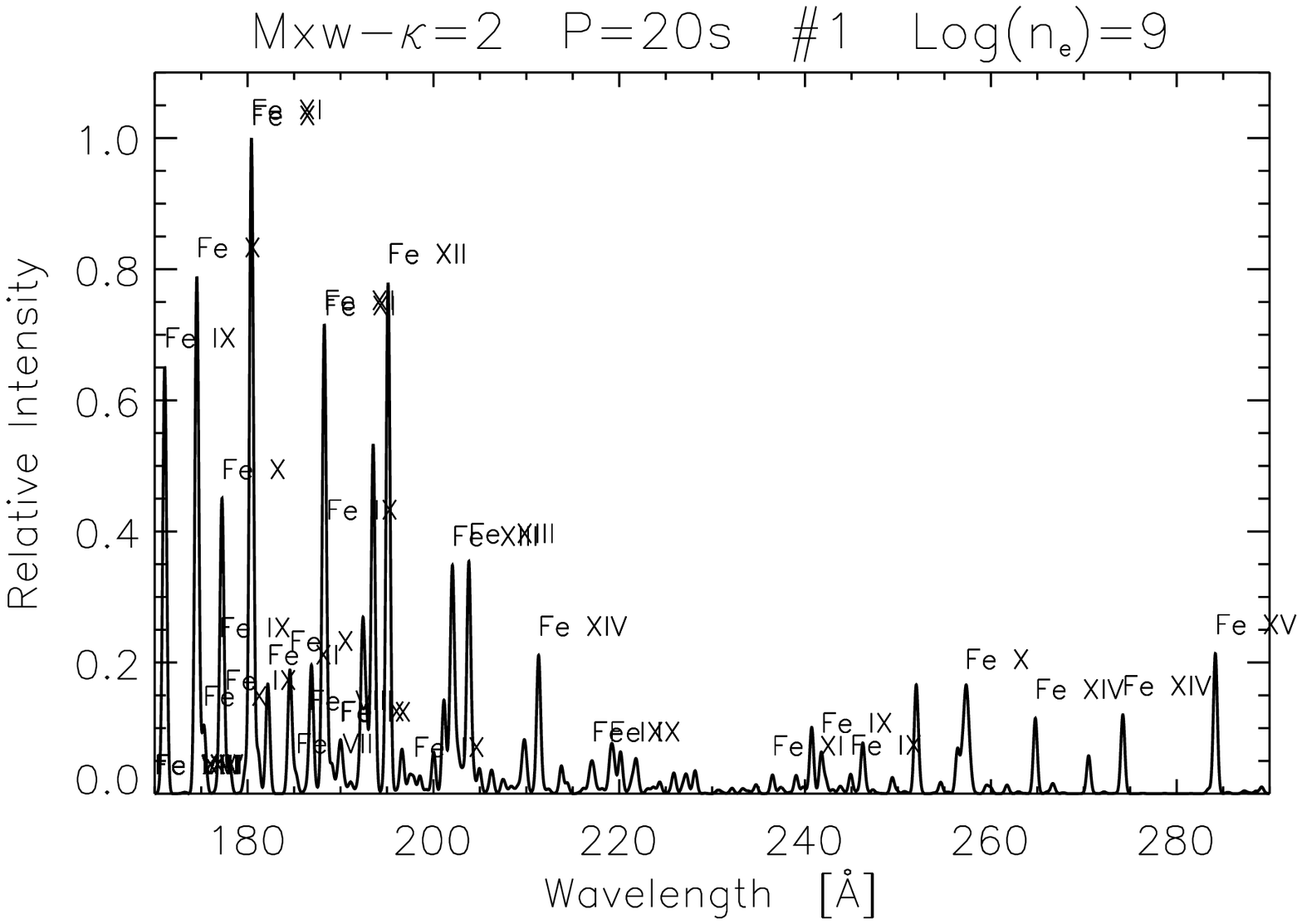}
        \includegraphics[width=6.17cm,clip,bb= 100  0 575 396]{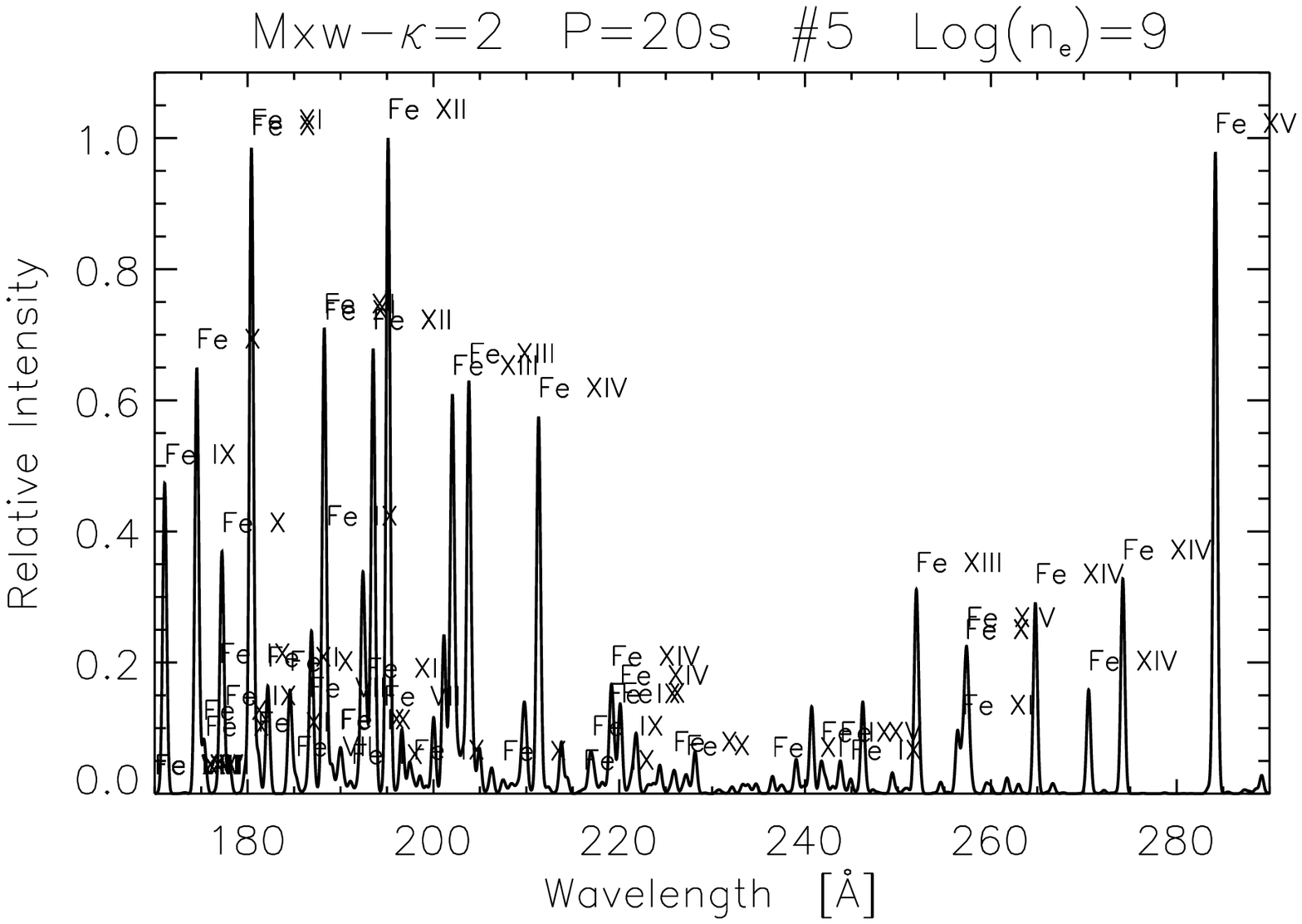}
        \includegraphics[width=4.74cm,clip,bb=  25  0 390 396]{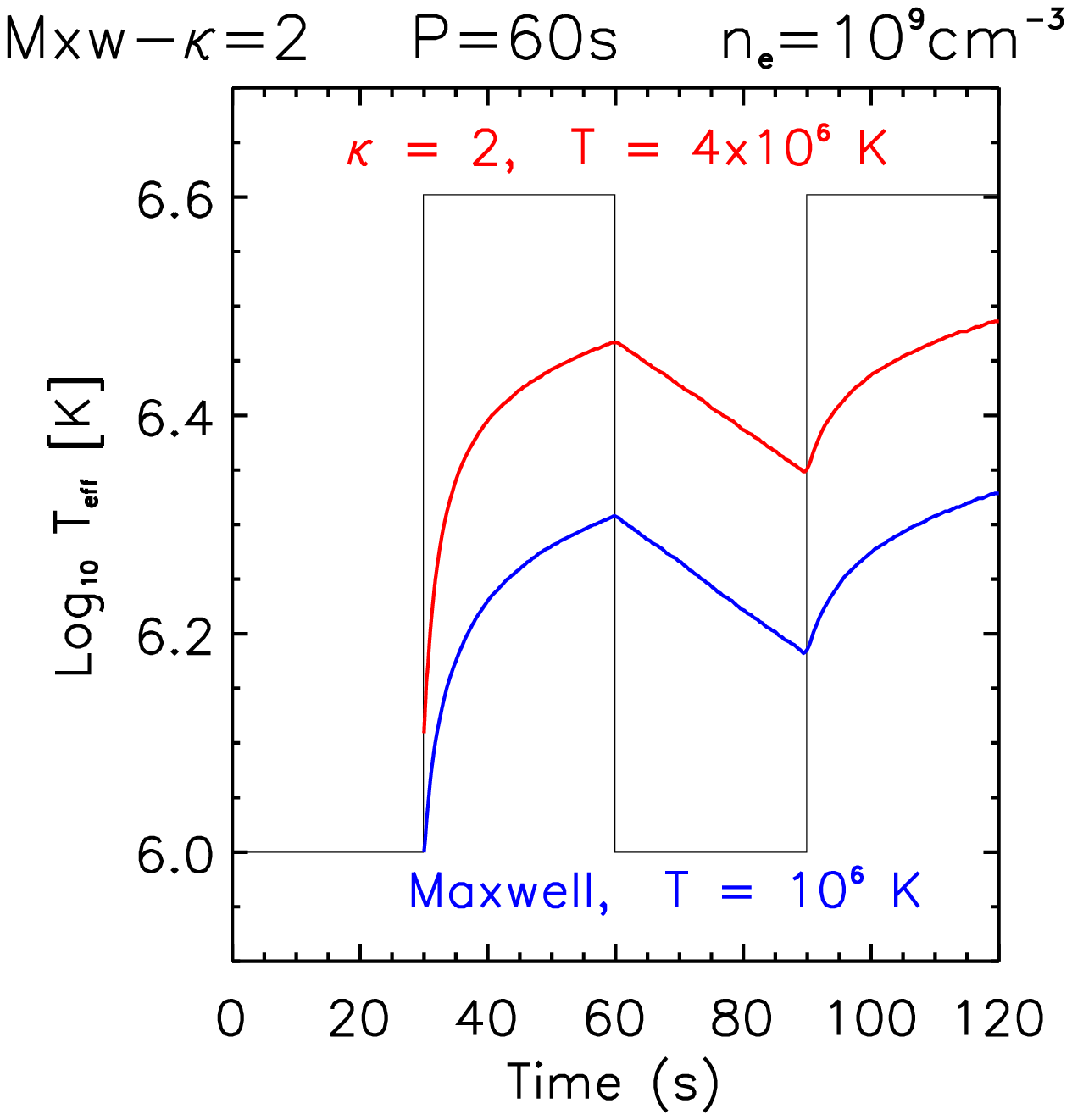}
        \includegraphics[width=6.89cm,clip,bb=  45  0 575 396]{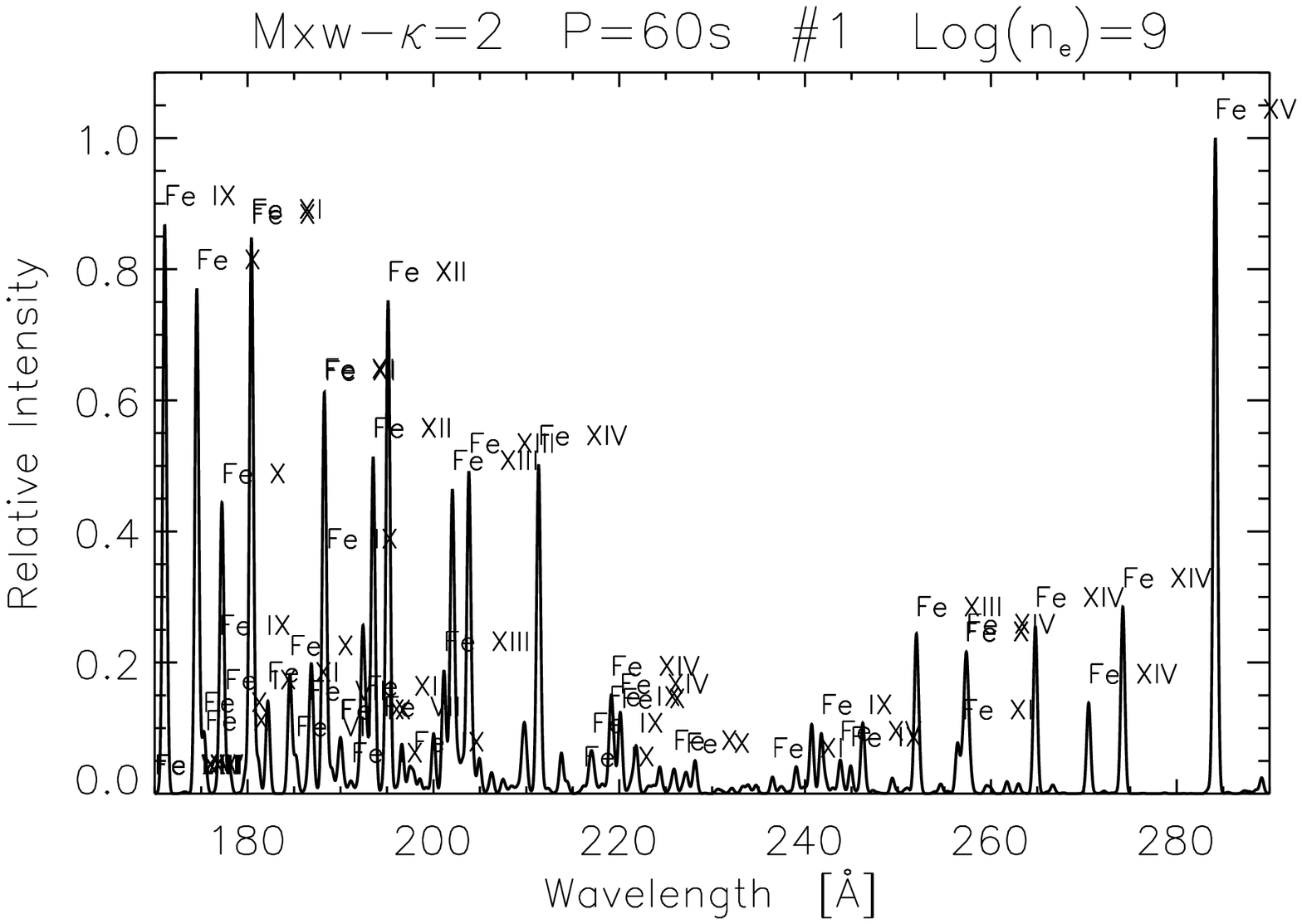}
        \includegraphics[width=6.17cm,clip,bb= 100  0 575 396]{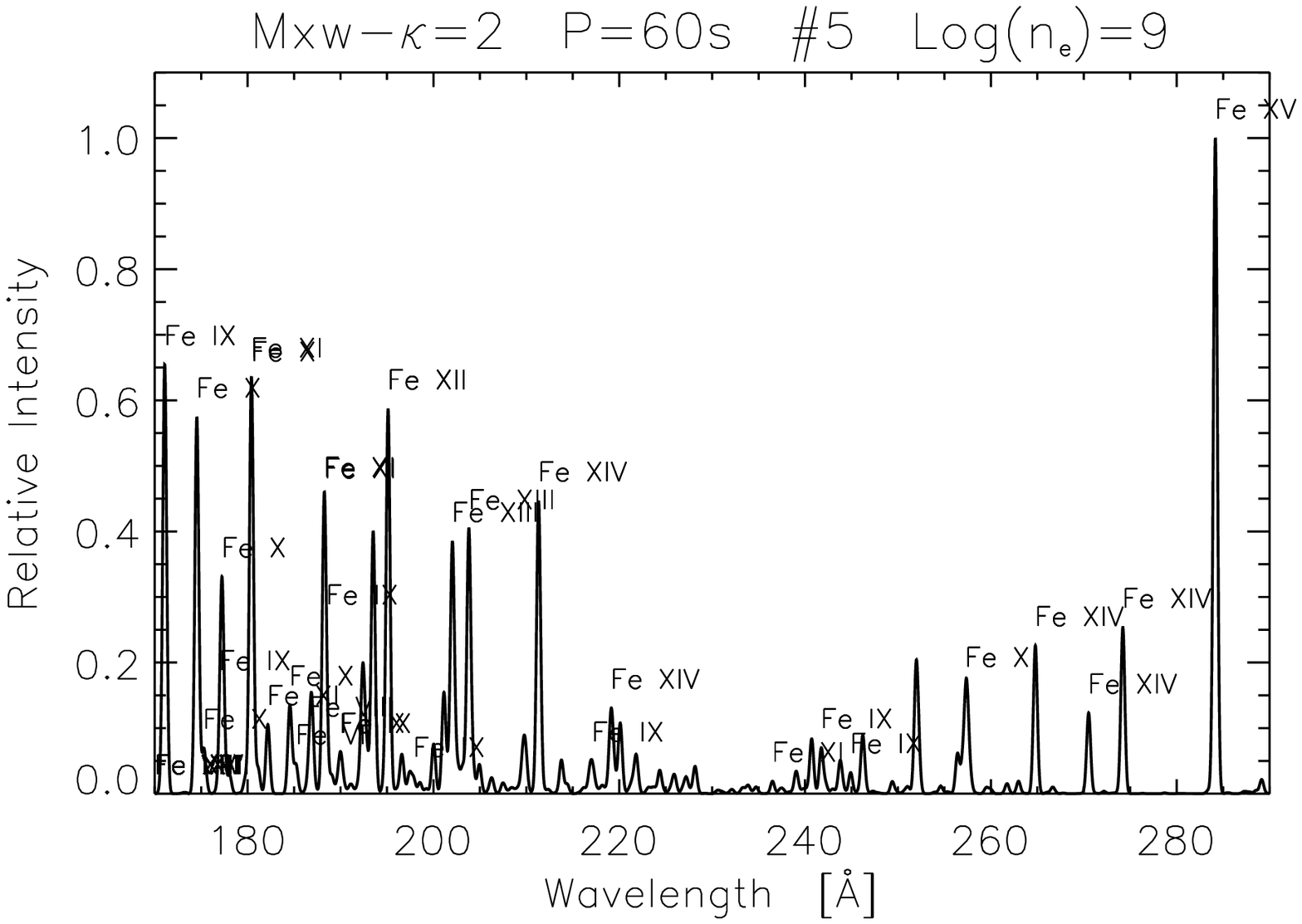}
\caption{Evolution of the effective ionization temperature (\textit{left}) and the corresponding EUV spectra during the first (\textit{center}) and fifth period (\textit{right}). Individual rows stand for different periods. The assumed electron density is log$(n_\mathrm{e}$ [cm$^{-3}$])\,=\,9. Instantaneous spectra at each time step in the calculation are shown in  Movies 1--3.
\label{Fig:Res_lne9_k02}}
\end{figure*}
%----------------------------------------------------
%
%---------------------------------------------------- FIGURE 4
\begin{figure*}[!t]
        \centering
        \includegraphics[width=4.74cm,clip,bb=  25  0 390 396]{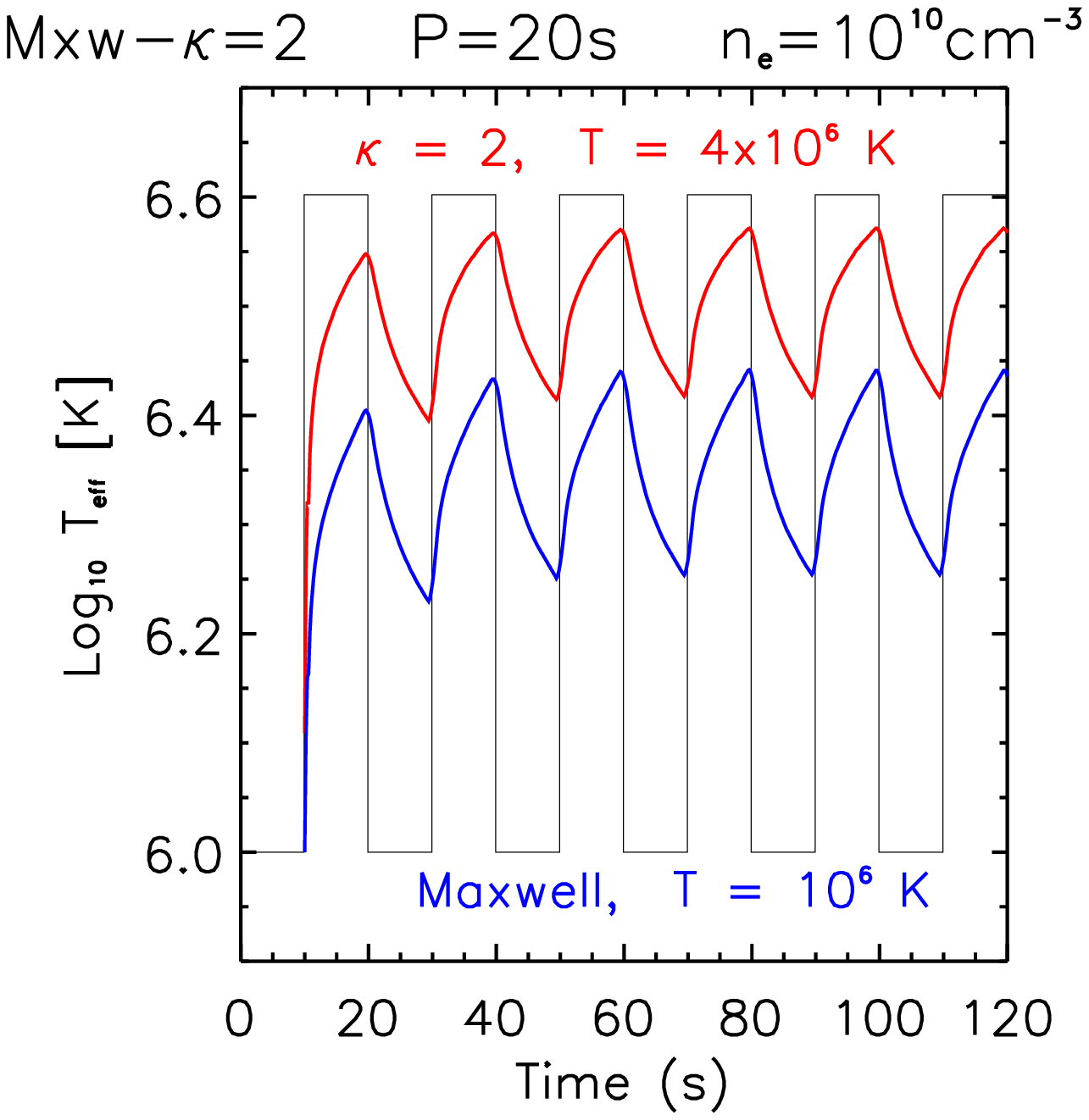}
        \includegraphics[width=6.89cm,clip,bb=  45  0 575 396]{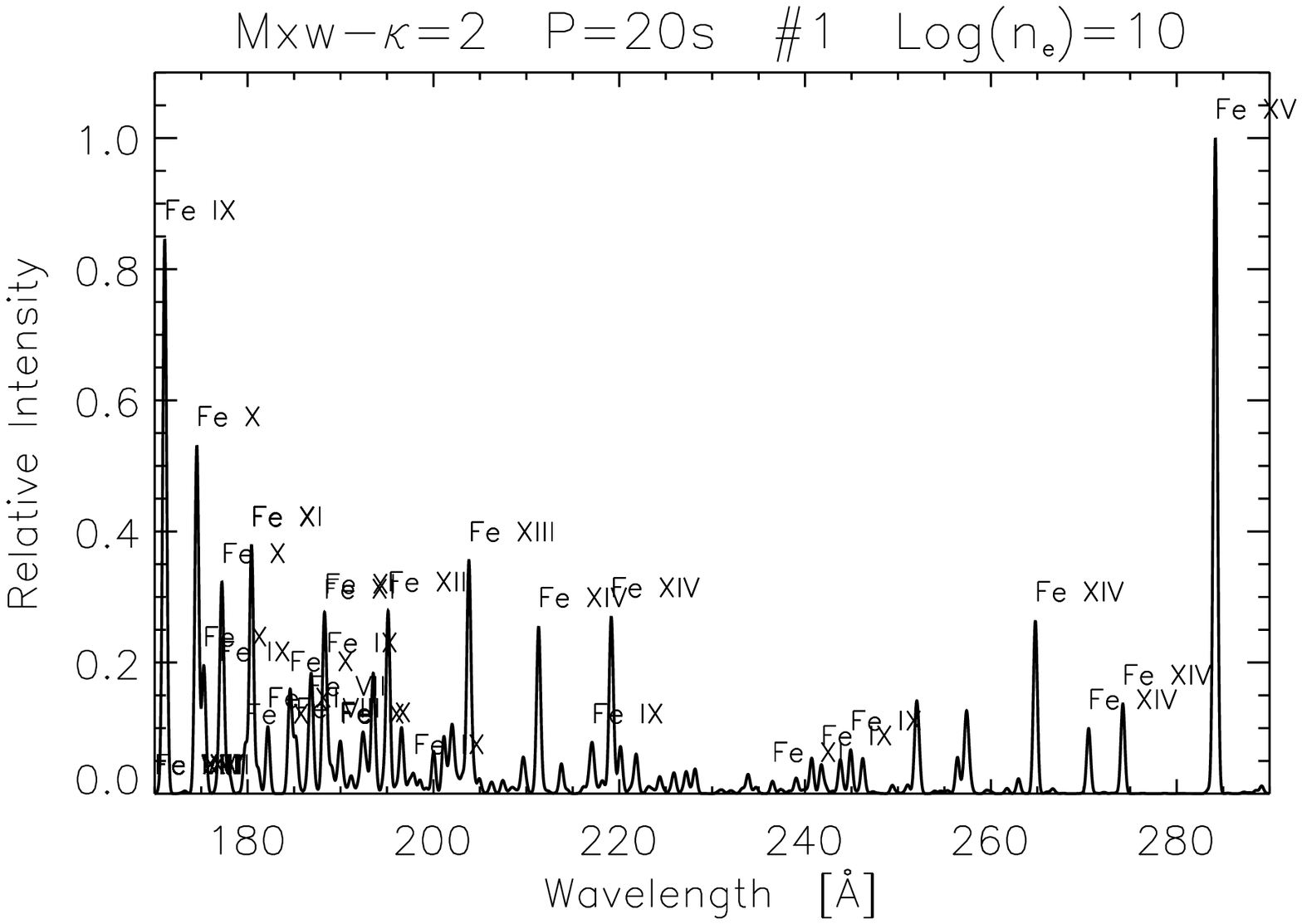}
        \includegraphics[width=6.17cm,clip,bb= 100  0 575 396]{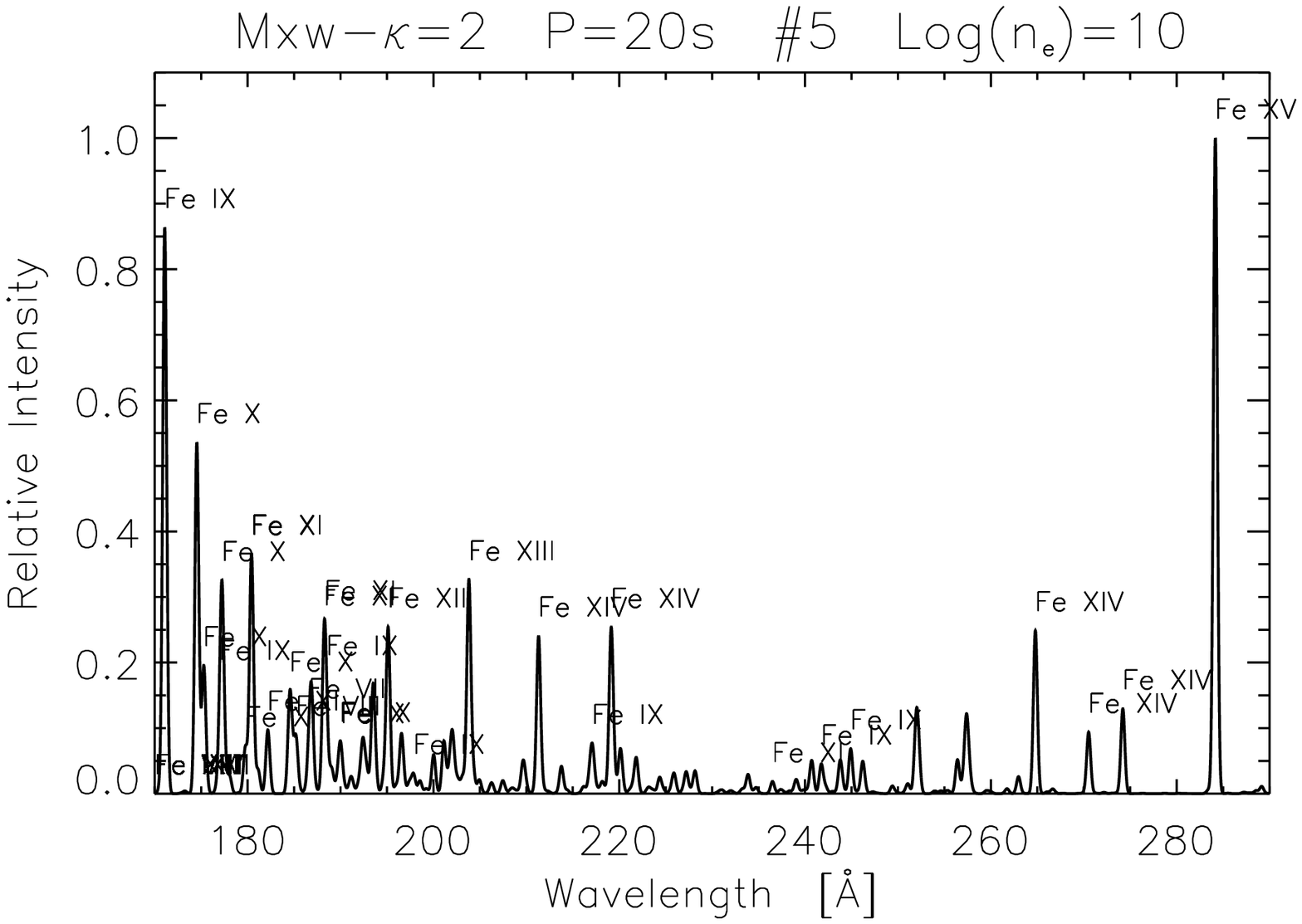}
\caption{Same as in Fig. \ref{Fig:Res_lne9_k02}, but for log$(n_\mathrm{e}$ [cm$^{-3}$])\,=\,10. The corresponding instantaneous spectra are shown in  Movie 4.
\label{Fig:Res_lne10_k02}}
\end{figure*}
%----------------------------------------------------
%
%---------------------------------------------------- FIGURE 5
\begin{figure*}
        \centering
        \includegraphics[width=4.74cm,clip,bb=  25  0 390 396]{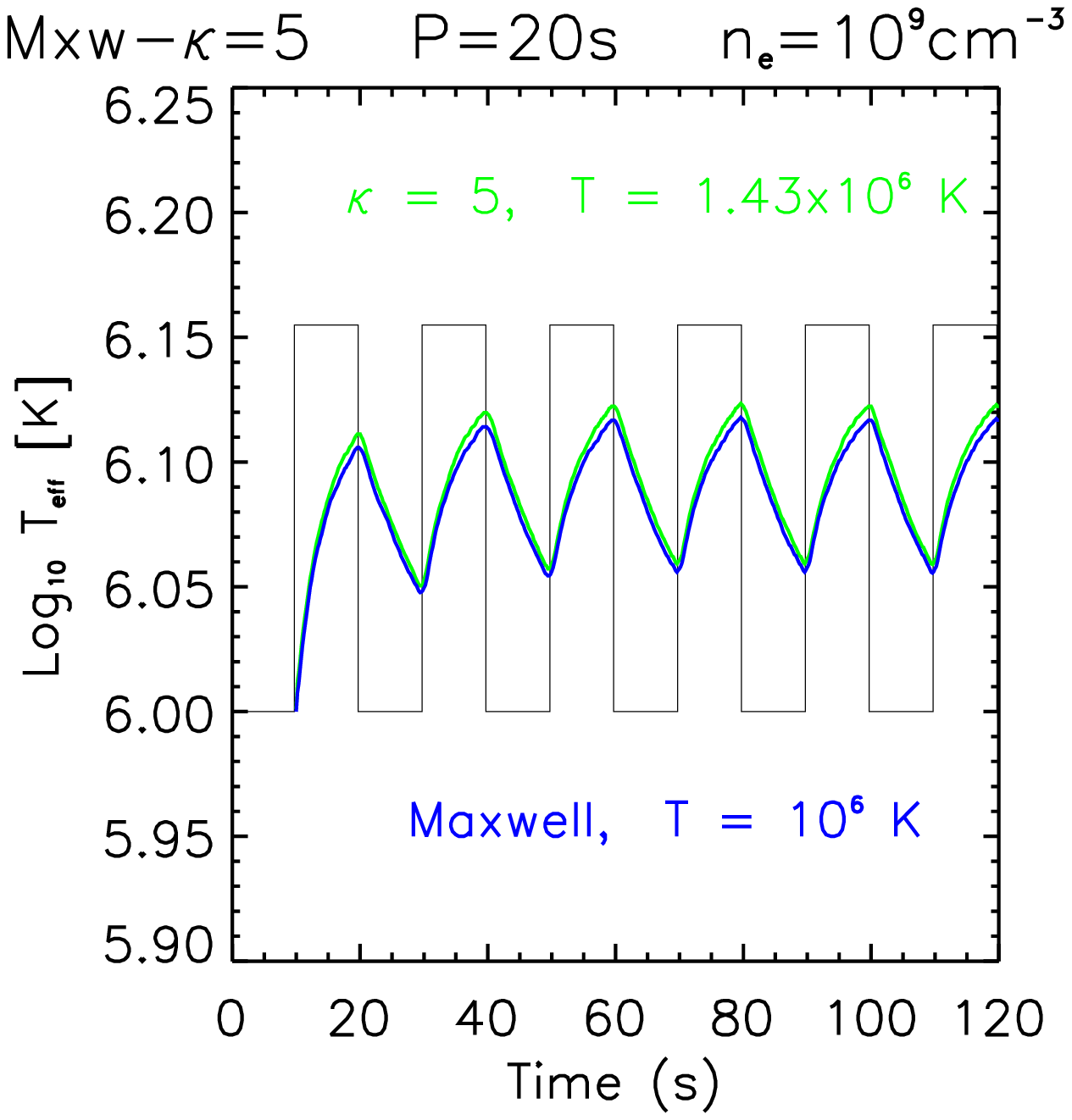}
        \includegraphics[width=6.89cm,clip,bb=  45  0 575 396]{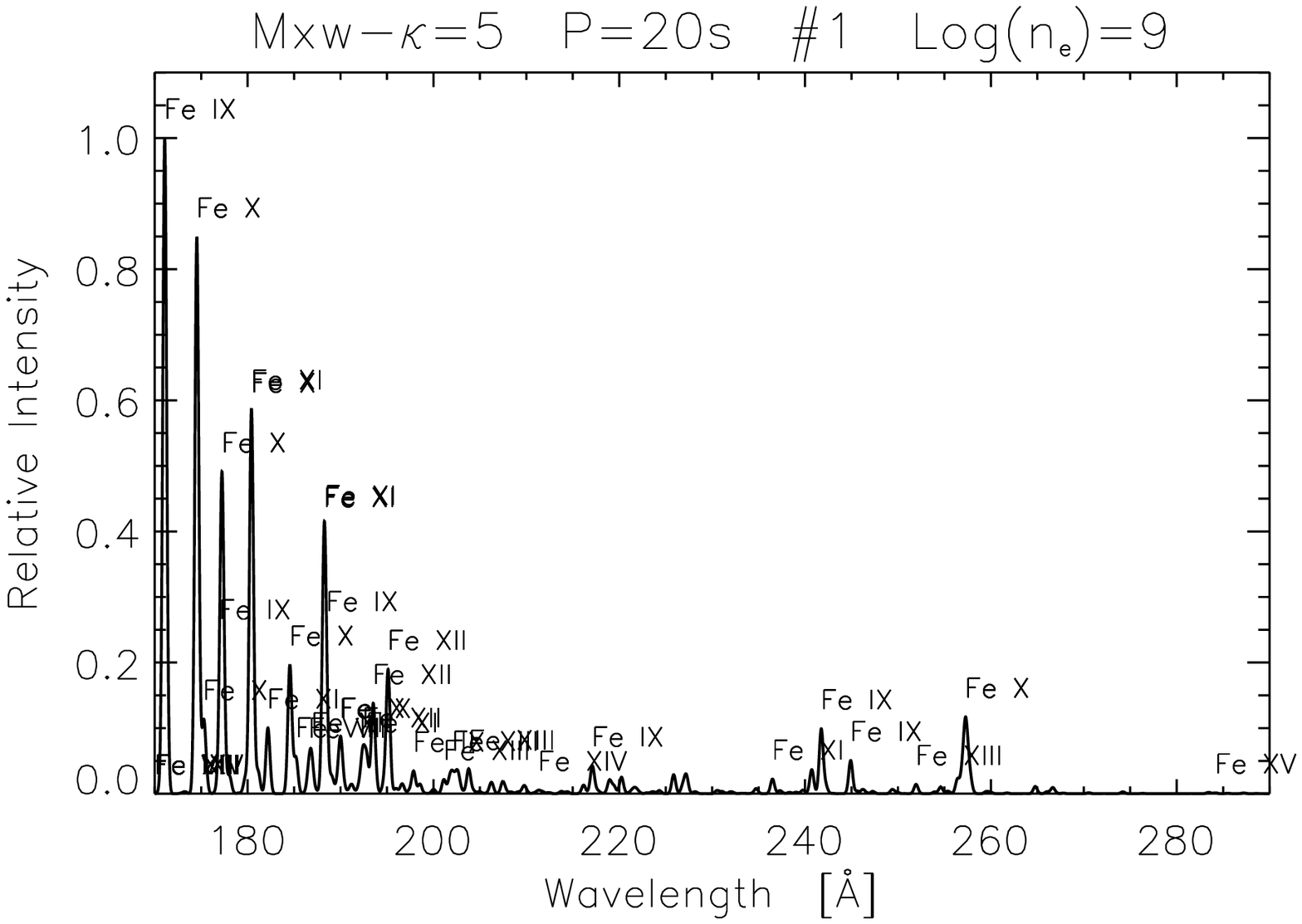}
        \includegraphics[width=6.17cm,clip,bb= 100  0 575 396]{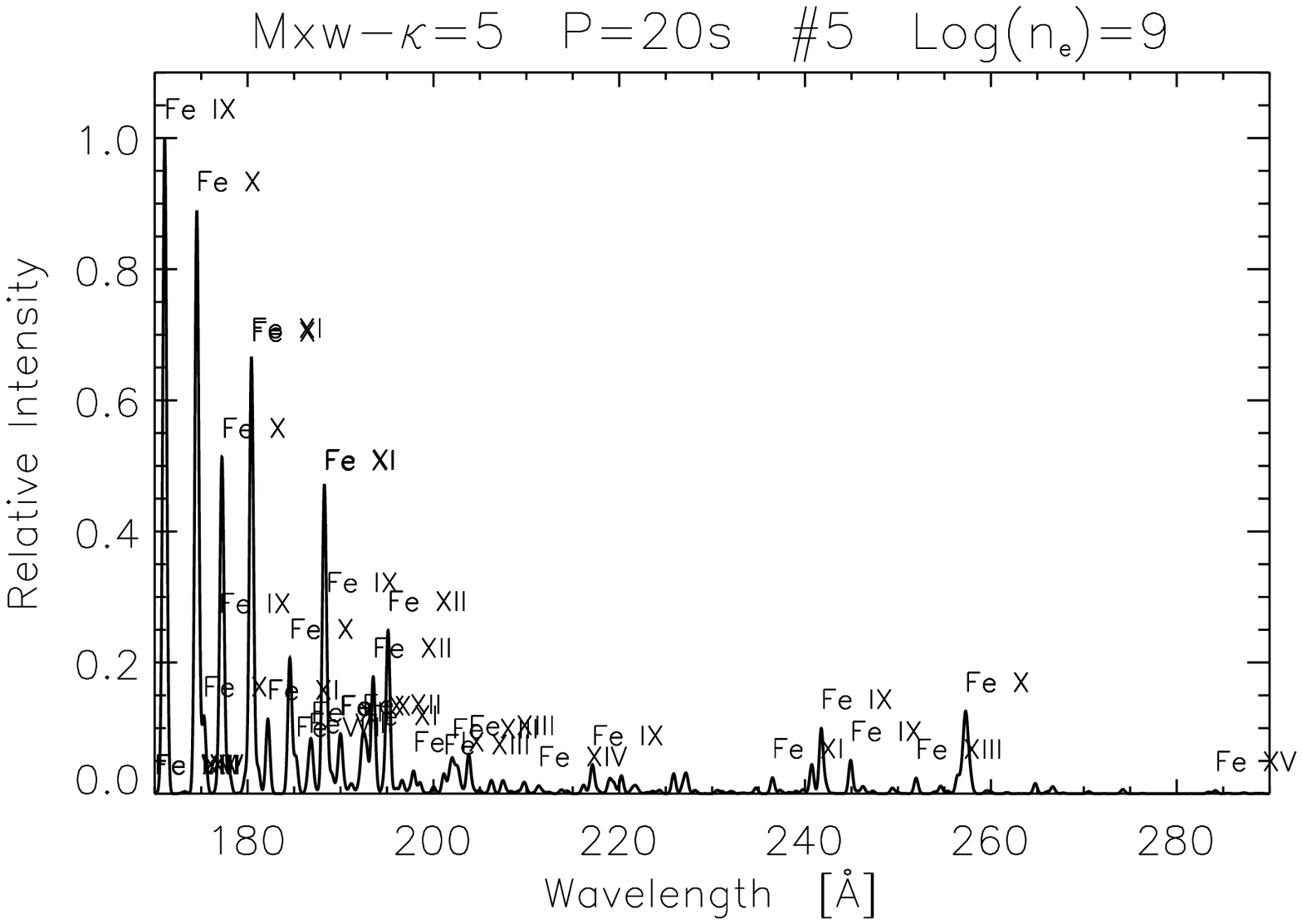}
\caption{Same as in Fig. \ref{Fig:Res_lne9_k02}, but for $\kappa$\,=\,5.  The corresponding instantaneous spectra are shown in  Movie 5.
\label{Fig:Res_lne9_k05}}
\end{figure*}
%----------------------------------------------------
%
%________________________________________________________________
\section{Results}
\label{Sect:3}

\subsection{Results for strong beam}
\label{Sect:3.1}

To explore the consequences of a periodic high-energy electron beam, we studied several cases described by their respective free parameters. First, we chose the electron density to be $n_\mathrm{e}$\,=\,10$^9$\,cm$^{-3}$, a value typical of the solar corona \citep[e.g.,][]{Landi04,Tripathi09,Shestov09,ODwyer11,Dudik15,Gupta15}. This value pertains to the half-period during which the distribution is a $\kappa$-distribution. Next, we selected $\kappa$\,=\,2, which is the most extreme non-Maxwellian value for which the spectra can be synthesized using the KAPPA package \citep{Dzifcakova15}. Finally, we chose  periods on the order of several tens of seconds to be comparable with the typical exposure times and/or cadence of the EUV coronal spectrometers such as Hinode/EIS \citep[except perhaps for observations of brightest lines, see][]{Culhane07} or imagers such as SDO/AIA \citep{Lemen12,Boerner12}. We note that such periods are also of the same order of magnitude as the durations of individual nanoflare heating events \citep[e.g.,][]{Bradshaw03b,Bradshaw06,Taroyan06,Taroyan11,Reale08,Bradshaw09,Bradshaw11,Bradshaw12,Reep13,Klimchuk14,Price15}.

Figure \ref{Fig:Res_lne9_k02} shows the results for $P$\,=\,10, 20, and 60\,s. Although these values are arbitrary, they are sufficient to capture the typical effects of the periodic electron beam on the spectra. In this figure, the left panel shows the stepwise evolution of the distribution function together with the effective ionization temperatures calculated under the assumption of a $\kappa$\,=\,2 distribution and a Maxwellian distribution. The middle and right panels show the corresponding spectra averaged over the first and fifth period, respectively. The respective instantaneous spectra at each time step are shown as animations in  Movies 1-3.

For the shortest period of 10\,s, we find that the $P$ is short enough for the plasma to be out of ionization equilibrium and that it undergoes ionization pumping. This means that during the first half-period $P/2$, when the beam is switched on, the effective ionization temperatures calculated for $\kappa$\,=\,2 and the Maxwellian distribution both spike from the initial value of $T$\,=\,1\,MK and reach log($T_\mathrm{eff}^{\kappa=2}$ [K])\,$\approx$\,6.35 and log($T_\mathrm{eff}^\mathrm{Mxw}$ [K])\,$\approx$\,6.17 (Fig. \ref{Fig:Res_lne9_k02}, top left, and Movie 1). This is followed by a small drop in the second-half period during which the beam is switched off and the distribution is Maxwellian. In the subsequent periods, the effective ionization temperature continues to rise at a slowing rate until it reaches the maximum values of log($T_\mathrm{eff}^{\kappa=2}$ [K])\,$\approx$\,6.45 and log($T_\mathrm{eff}^\mathrm{Mxw}$ [K])\,$\approx$\,6.3. There are, however, small saw-like variations in $T_\mathrm{eff}$ due to the periodic switching on and off of the beam, so that the plasma is never in ionization equilibrium.

The corresponding spectra averaged over the first and fifth periods are shown in the top row of Fig. \ref{Fig:Res_lne9_k02}. During the first period, the average spectrum (black) is dominated by the strong \ion{Fe}{IX}--\ion{Fe}{XII} lines at 171--195\,\AA;  the \ion{Fe}{XI} 180.40\,\AA~line is the strongest one. The higher temperature lines are much weaker, having less than 20\% in terms of intensity than the strongest \ion{Fe}{XI} line. The \ion{Fe}{XV} 284.16\,\AA~line is very weak, having less than about 5\% relative intensity. These intensities are not surprising given the values of $T_\mathrm{eff}$ during the first period, since this quantity is based on the mean ion charge (see Sect. \ref{Sect:2.2}).

In contrast, the average spectrum for the fifth period is significantly different. We note that the fifth period is chosen as a representative one and has an average log($T_\mathrm{eff}$ [K]), which is not significantly different from the near stationary behavior of the ionization composition, which is  lower by less than 0.05 dex. The period-averaged spectrum is now dominated by the \ion{Fe}{XI}--\ion{Fe}{XV} lines (Fig. \ref{Fig:Res_lne9_k02} top right). The \ion{Fe}{XI} 180.40\,\AA~line is still the dominant one, but the intensities of lines belonging to the higher ionization stages have increased significantly, while the \ion{Fe}{IX} and \ion{Fe}{X} lines have decreased. In particular, the relative intensity of the \ion{Fe}{XV} 284.16\,\AA~line is now 0.7. The instantaneous spectra (Movie 1) show fast, subsecond changes especially in the \ion{Fe}{IX} 171.07\,\AA~and \ion{Fe}{XV} 284.16\,\AA~lines during the first period. These changes are slower during the subsequent periods, when the intensity of \ion{Fe}{XV} 284.16\,\AA~line  increases.

For $P$\,=\,20\,s, the situation is similar to that of $P$\,=\,10\,s discussed previously, except that the effect of ionization pumping is now weaker. The log$(T_\mathrm{eff}^{\kappa=2})$ [K]) reaches a maximum of $\approx$6.4 during the first period and 6.45 during the fifth period, while the maxima of the log$(T_\mathrm{eff}^\mathrm{Mxw})$ [K])\,$\approx$\,6.23 and 6.3 for the first and fifth period, respectively. The intensities of the \ion{Fe}{XIII}--\ion{Fe}{XV} lines  increase with respect to the $P$\,=\,10\,s case, with the \ion{Fe}{XV} 284.16\,\AA~line having almost 1.0 relative intensity during the fifth period. The instantaneous spectra show a fast decrease in the \ion{Fe}{IX} 171.07\,\AA~line during the first 5\,seconds, followed by the appearance of the \ion{Fe}{XV} line. The intensity of this line increases on average  from the first half-period, and decreases when the distribution is Maxwellian. In the second and subsequent periods the line remains visible, as can be expected from the behavior of the $T_\mathrm{eff}$.

The $P$\,=\,60\,s period case exhibits a fast increase in $T_\mathrm{eff}$ during the first period to log($T_\mathrm{eff}^{\kappa=2}$ [K]) of about 6.45 and log($T_\mathrm{eff}^{\mathrm{Mxw}}$ [K])\,$\approx$\,6.3, followed by a decrease of about 0.1 dex during the second half-period. In the next periods, the effective temperatures increase to similar maximum values and an oscillatory state is reached within the second period. The averaged spectra are now dominated by the \ion{Fe}{XV} 284.16\,\AA~line, with the intensities of lower ionization stages being weaker during the fifth period than during the first period (see the bottom row of Fig. \ref{Fig:Res_lne9_k02}). The \ion{Fe}{IX} 171.07\,\AA, \ion{Fe}{XI} 180.40\,\AA, and \ion{Fe}{XII} 195.12\,\AA~lines nevertheless remain relatively strong with the relative intensities of about 0.6.

In the instantaneous spectra, however, the \ion{Fe}{IX}--\ion{Fe}{XII} lines are dominant over the first 12\,seconds. During this time, their intensities  decrease, while the intensity of \ion{Fe}{XV}  increases and is dominant approximately in the interval of 12--50\,s. After the first half-period (30\,s), the \ion{Fe}{XV} line intensity decreases and the last 10 seconds of the period are again dominated by lower ionization stages, in particular \ion{Fe}{IX}. These changes reflect again the evolution of $T_\mathrm{eff}$.

In all the cases presented here, the $T_\mathrm{eff}$ is different from the respective $T_\kappa$ or $T$, indicating that the plasma is always out of ionization equilibrium.

\subsection{Higher electron density}
\label{Sect:3.2}

To study the influence of electron density on the evolution, we present the case of $P$\,=\,20\,s calculated assuming an order of magnitude higher density of log($n_\mathrm{e}$\,[cm$^{-3}$])\,=\,10. The results are shown in Fig. \ref{Fig:Res_lne10_k02}.

In this case, the maximum $T_\mathrm{eff}$ reaches higher values during the first period compared to the case with lower density, with log$(T_\mathrm{eff}^{\kappa=2}$ [K])\,=\,6.55 and log$(T_\mathrm{eff}^{\mathrm{Mxw}}$ [K])\,=\,6.40. In particular,  the value of $T_\mathrm{eff}^{\kappa=2}$ is significantly closer to the respective equilibrium value for the first half-period. The enhanced electron density, however, also leads to enhanced recombination. The recombination, dominant in the second half-period, causes a larger drop in $T_\mathrm{eff}$, of about 0.15 dex, than in the case with lower density. In the subsequent periods, the maximum values of $T_\mathrm{eff}^{\kappa=2}$   increase slightly owing to the weak ionization pumping, but never reach the respective equilibrium value (left panel in Fig. \ref{Fig:Res_lne10_k02}).

The average spectra are dominated by \ion{Fe}{XV}, in agreement with the values of $T_\mathrm{eff}$, but show significant \ion{Fe}{IX} 171.07\,\AA~intensities as well. The intensity of this \ion{Fe}{IX} spikes and becomes temporarily dominant in the second half-period for the last 6\,seconds owing to the enhanced recombination.

Increasing the density by another order of magnitude to 10$^{11}$\,cm$^{-3}$ would cause the $T_\mathrm{eff}^{\kappa=2}$ to temporarily reach its equilibrium value of 4\,MK for nearly the entire second quarter of a period, while the $T_\mathrm{eff}^{\mathrm{Mxw}}$ would approach its equilibrium value of 1\,MK towards the end of the first period. This means that a sufficient increase in electron density would lead to the plasma being in ionization equilibrium for a non-negligible portion of the half-period, characterized by its respective distribution.

\subsection{A weaker beam}
\label{Sect:3.3}

To study the effects of weaker beam, we selected $\kappa$\,=\,5, which is an intermediate value of $\kappa$ in terms of the influence of stationary $\kappa$-distributions on spectra \citep{Dzifcakova13a,Dudik14a,Dudik14b}. The other parameters are set to be log($n_\mathrm{e}$\,[cm$^{-3}$])\,=\,9 and $P$\,=\,20\,s.

Owing to the lower number of high-energy particles in a $\kappa$\,=\,5 distribution, the $T_\mathrm{eff}$ does not attain values larger than 6.12 in the log. The values of $T_\mathrm{eff}^{\kappa=5}$ and $T_\mathrm{eff}^{\mathrm{Mxw}}$ are only marginally different, which is likely due to the behavior of the ionization equilibrium \citep{Dzifcakova13a} and thus the mean ion charge used to define the $T_\mathrm{eff}$. The averaged spectra are dominated by the \ion{Fe}{IX} line;  the \ion{Fe}{XIII}--\ion{Fe}{XV} lines are strongly suppressed corresponding to the relatively low values of $T_\mathrm{eff}$. The spectra of the first and fifth period are similar, which is expected given the absence of a strong ionization pumping. In the instantaneous spectra (Movie 5), the \ion{Fe}{IX} 171.07\,\AA~line intensity oscillates by more than a factor of two, while the intensity of the \ion{Fe}{X} 174.53\,\AA~line exhibits much weaker changes on the order of 20\%.

%---------------------------------------------------- FIGURE 6
\begin{figure*}
% \sidecaption
%
        \centering
        \includegraphics[width=5.48cm,clip,bb= 40 70 725 555]{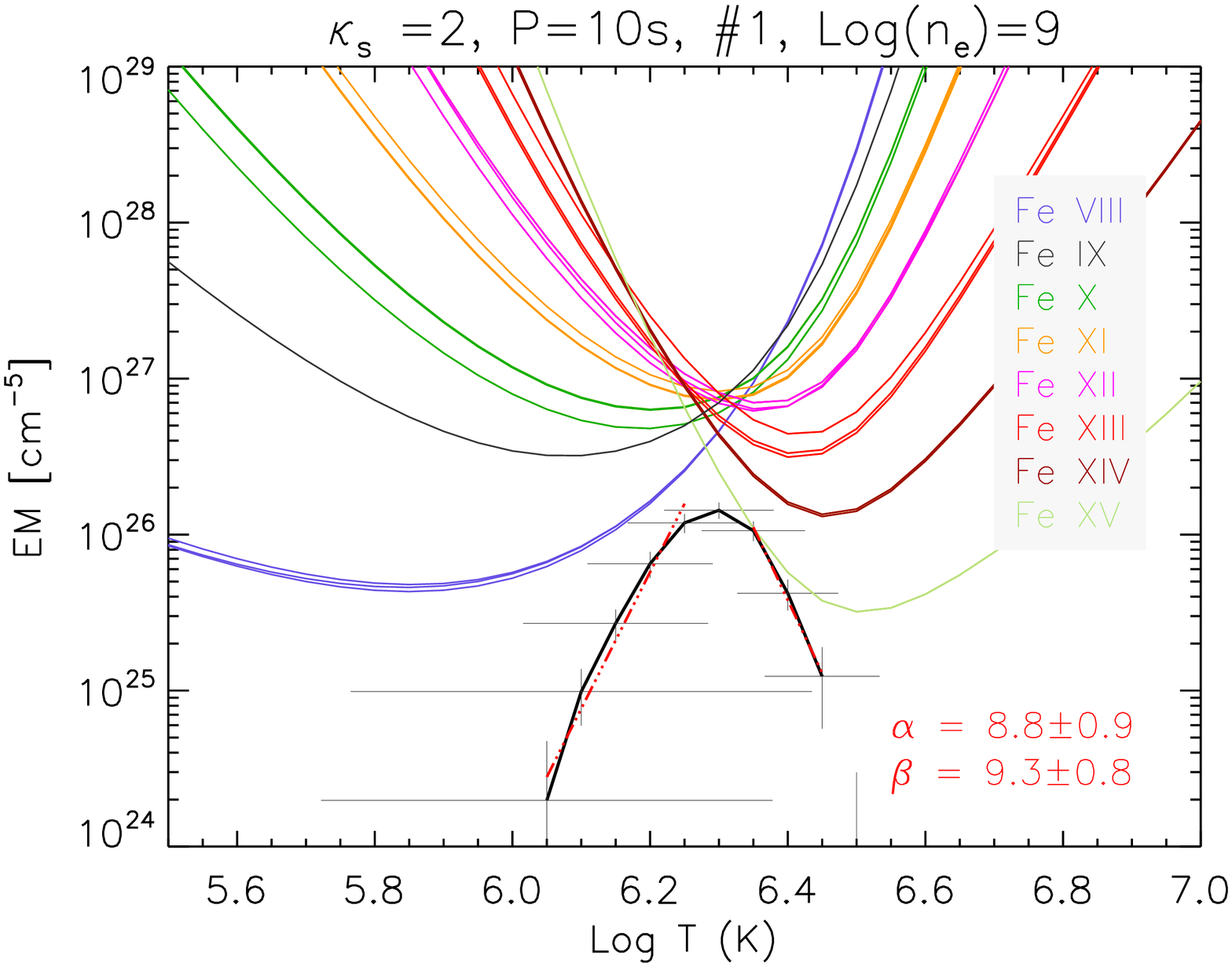}
        \includegraphics[width=4.76cm,clip,bb=130 70 725 555]{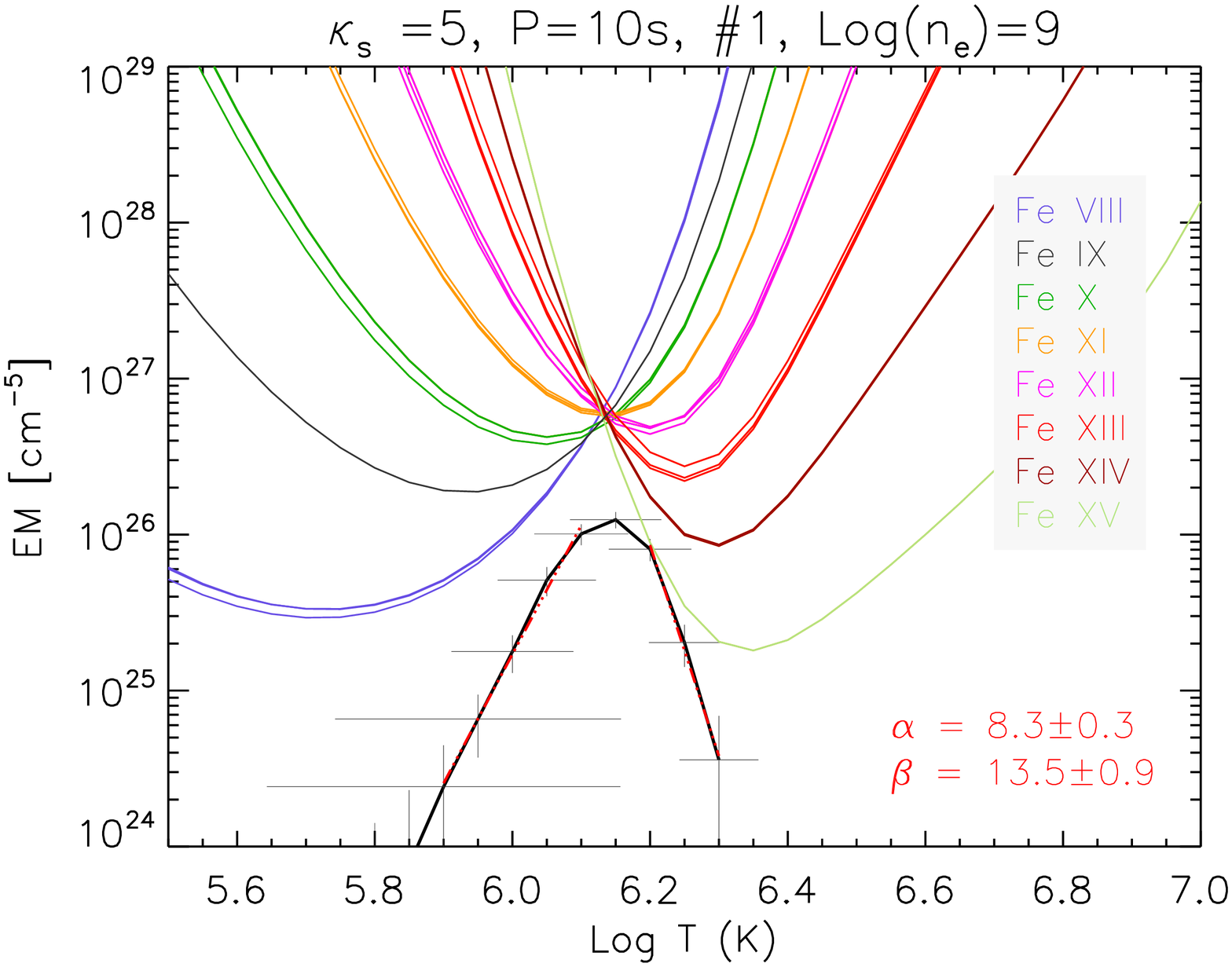}
        \includegraphics[width=4.76cm,clip,bb=130 70 725 555]{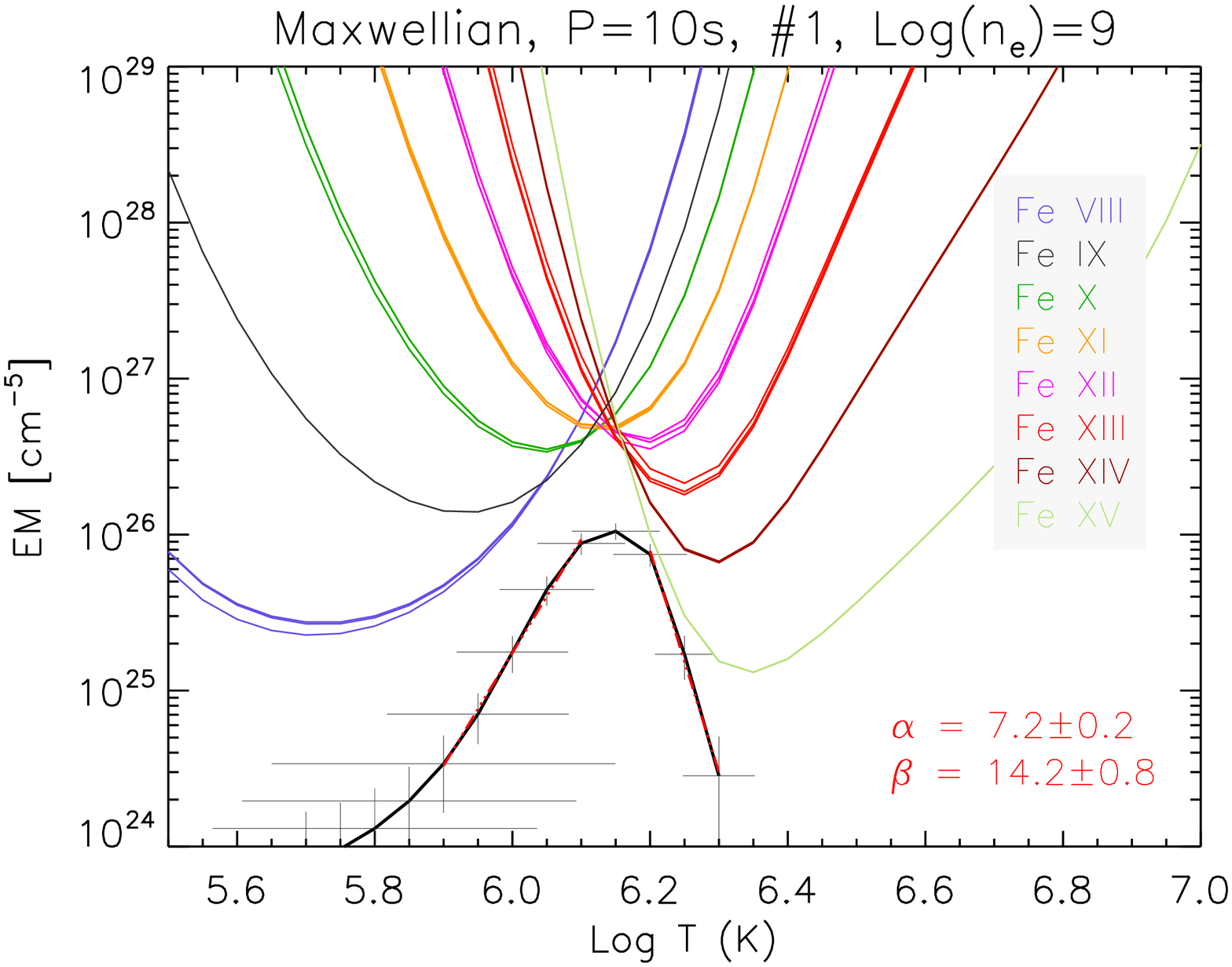}
        \includegraphics[width=5.48cm,clip,bb= 40 70 725 555]{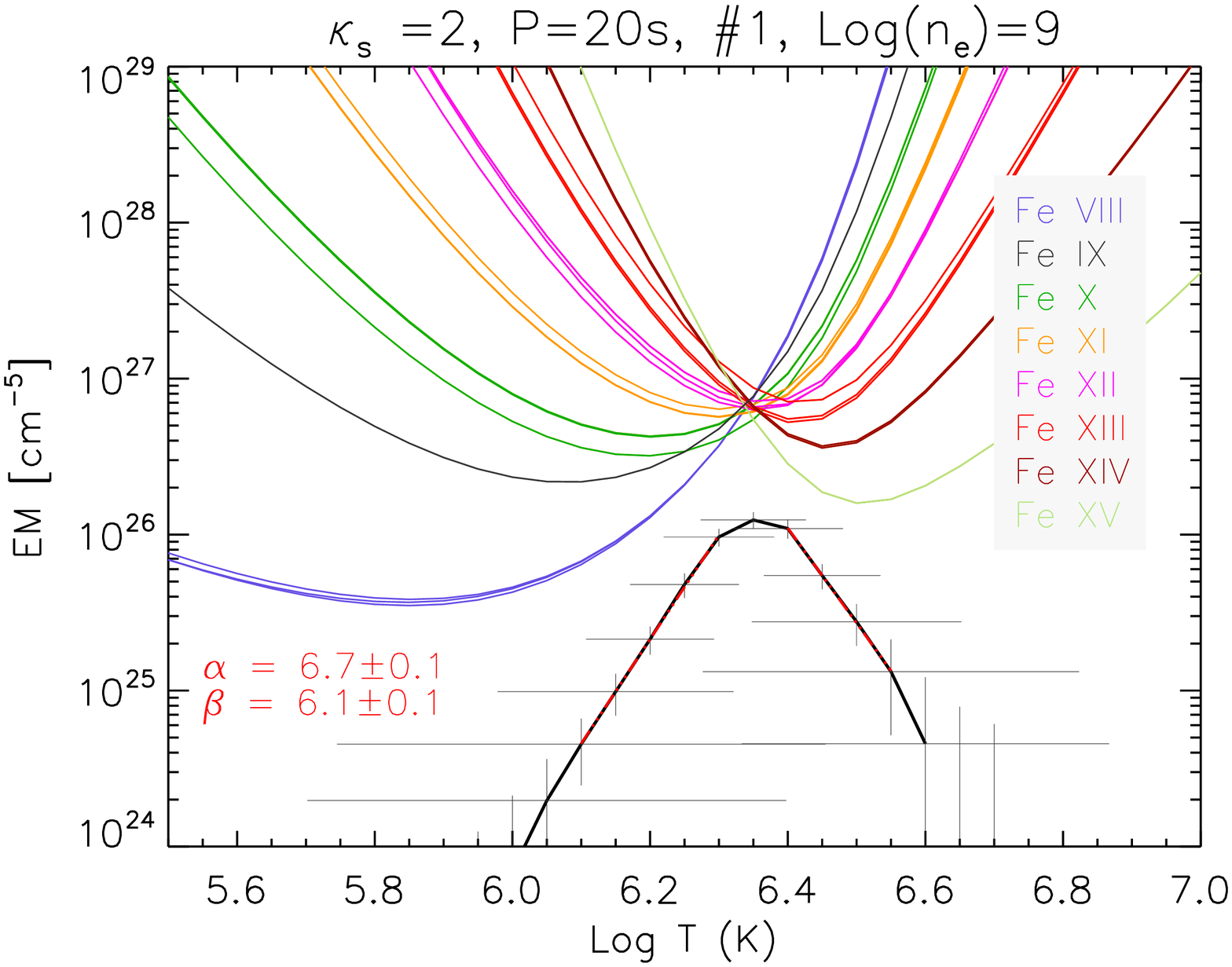}
        \includegraphics[width=4.76cm,clip,bb=130 70 725 555]{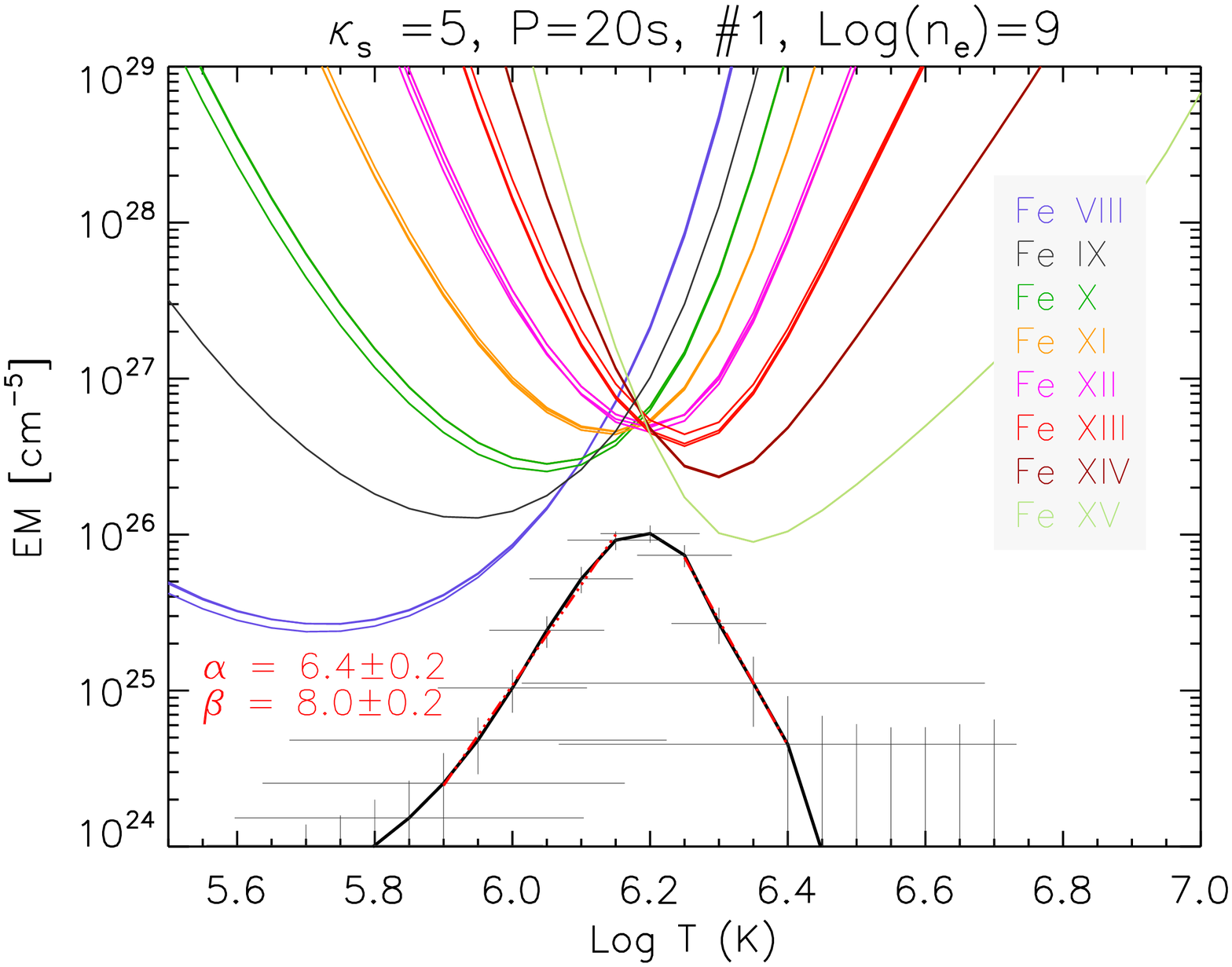}
        \includegraphics[width=4.76cm,clip,bb=130 70 725 555]{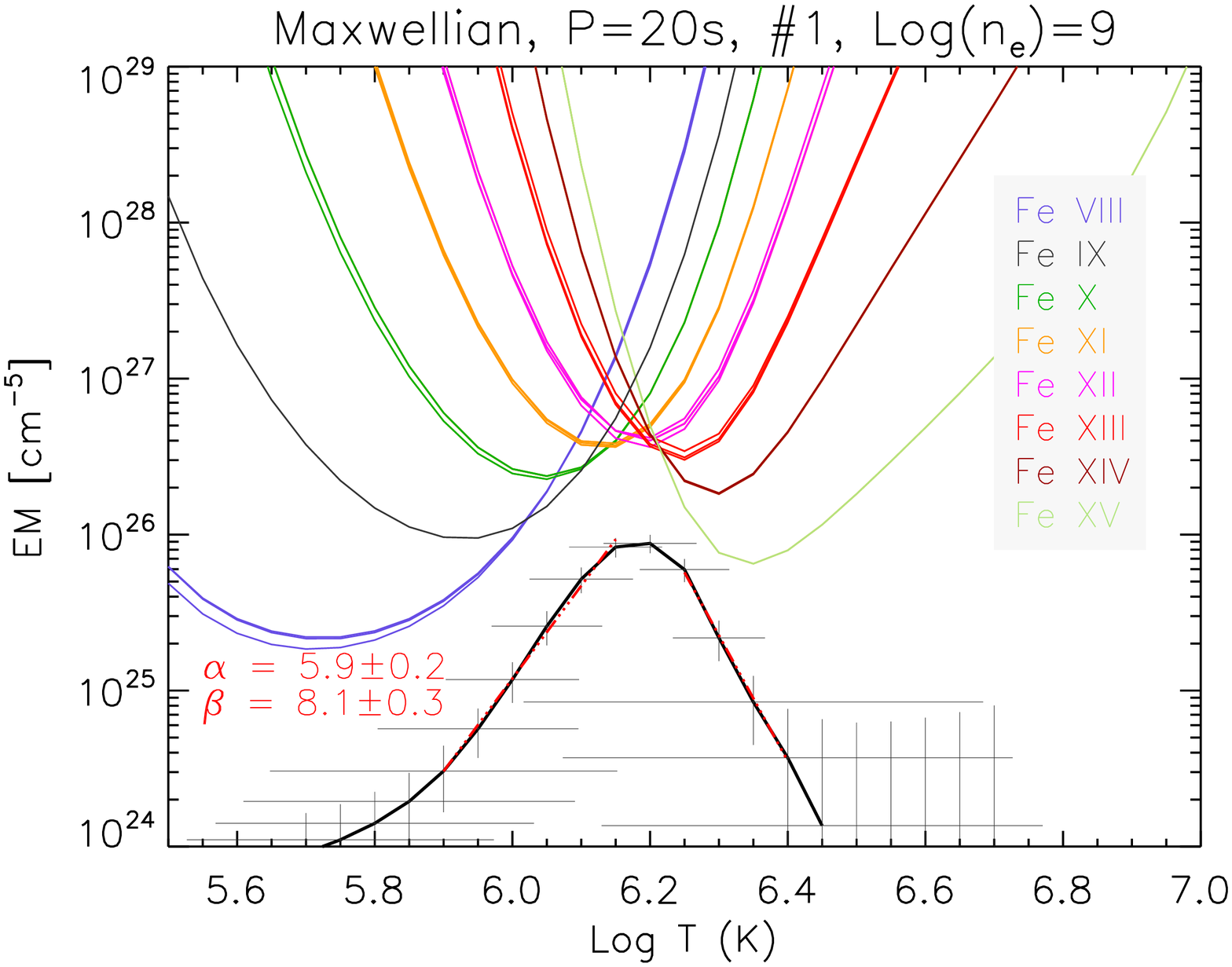}
        \includegraphics[width=5.48cm,clip,bb= 40 15 725 555]{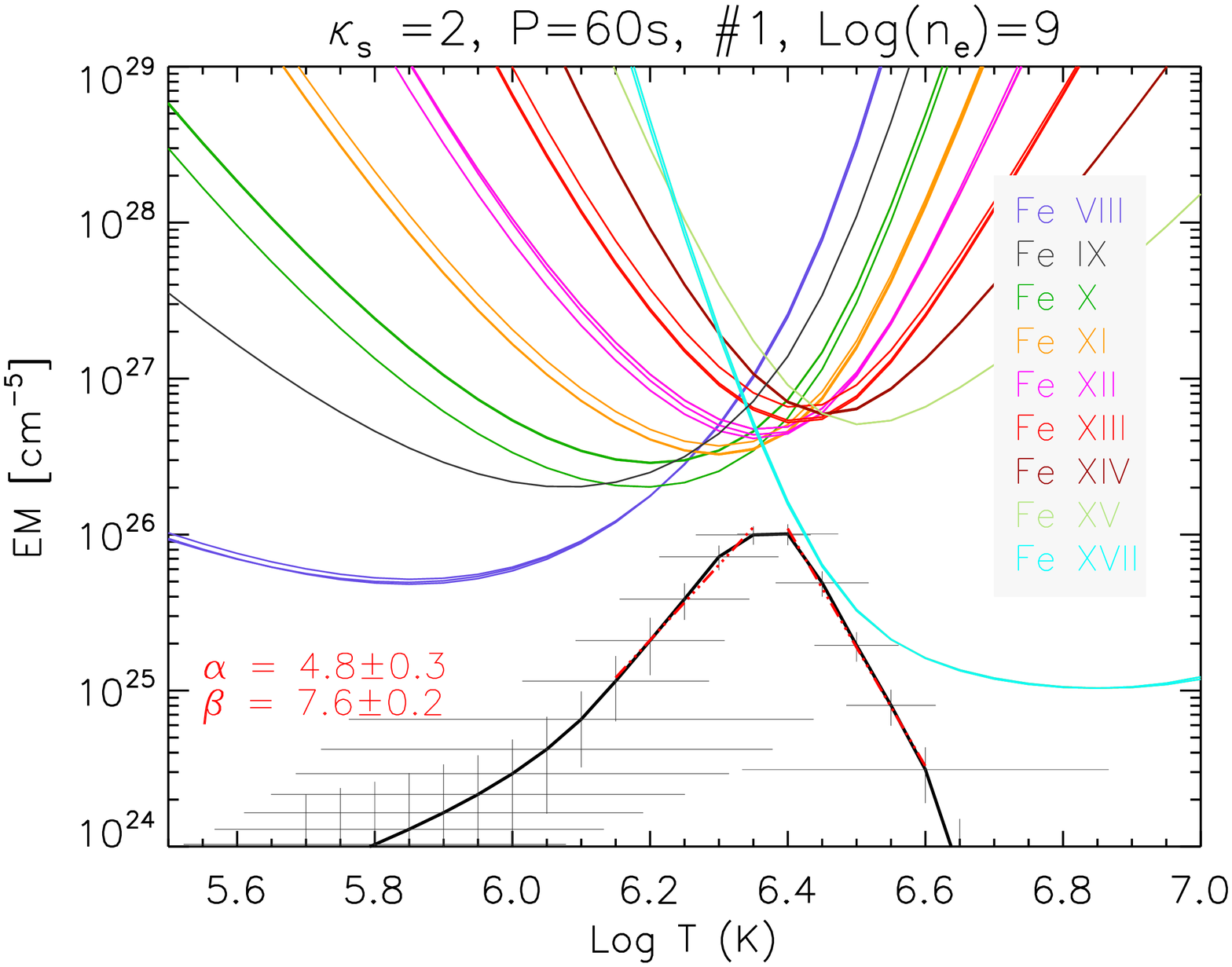}
        \includegraphics[width=4.76cm,clip,bb=130 15 725 555]{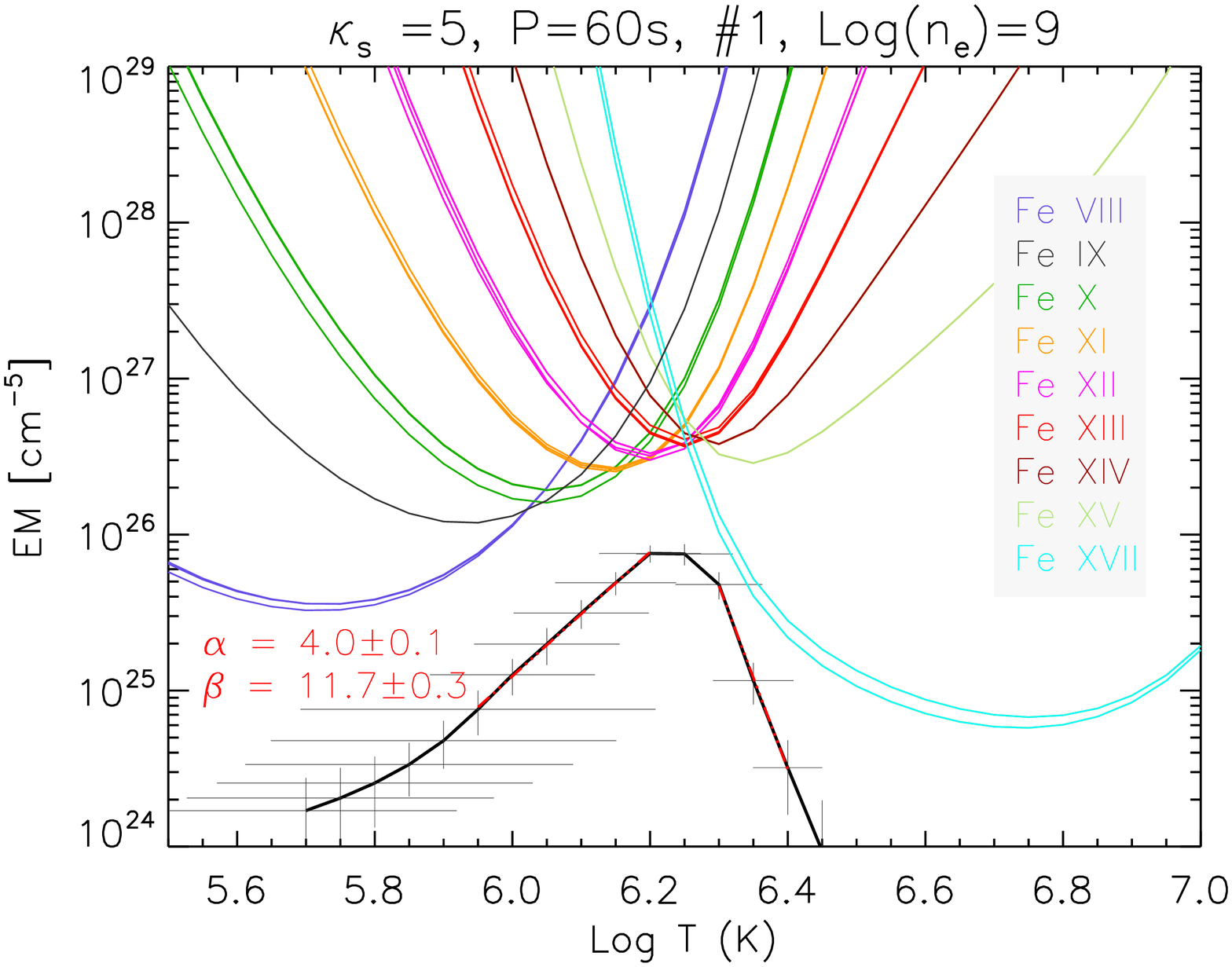}
        \includegraphics[width=4.76cm,clip,bb=130 15 725 555]{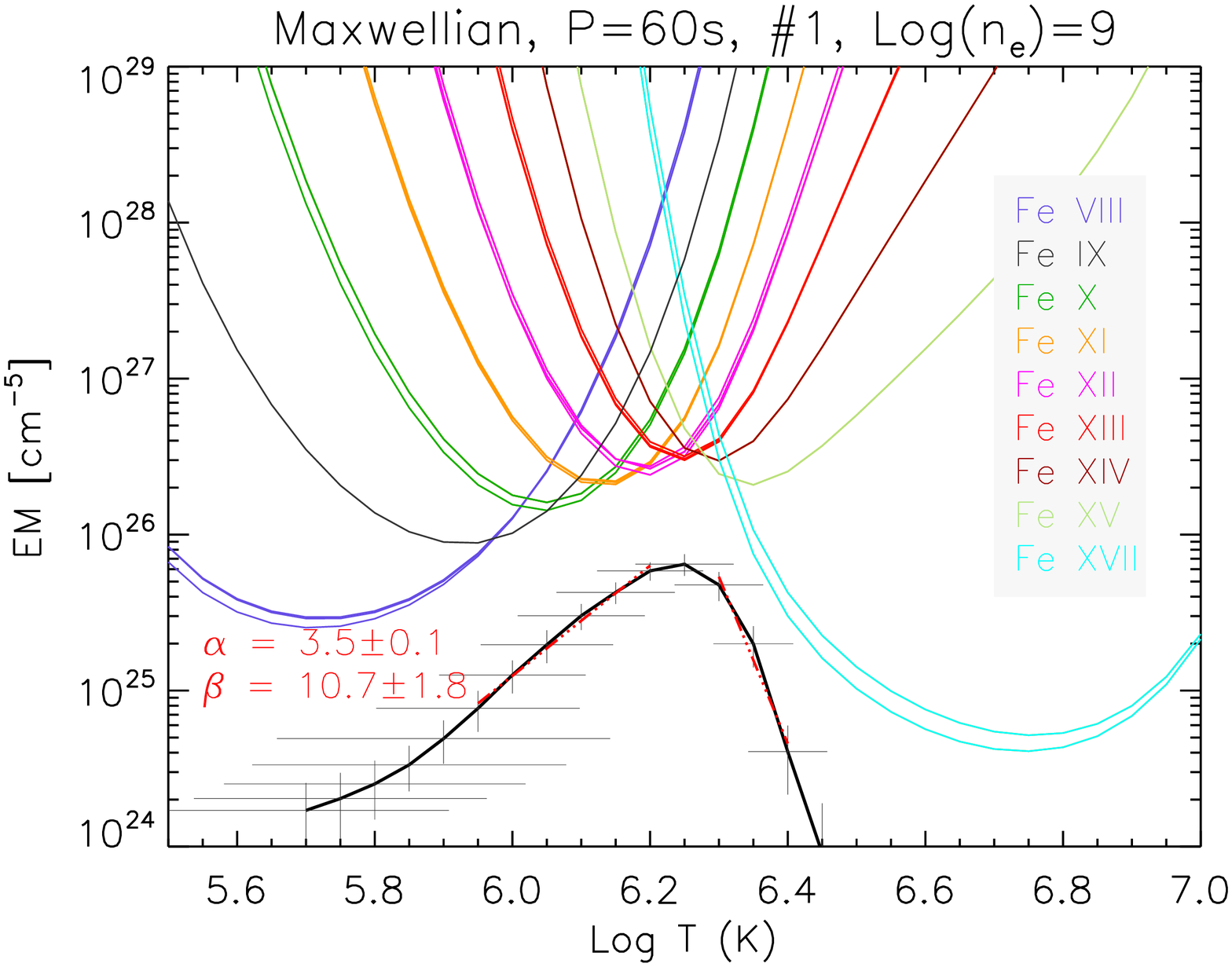}
\caption{Emission measure distributions EM($T$) derived under the assumption of a constant $\kappa_\mathrm{s}$ using the intensities averaged over the first period for $P$\,=\,10\,s (\textit{top}), 20\,s (\textit{middle}), and 60\,s (\textit{bottom}). The spectra correspond to log$(n_\mathrm{e}$ [cm$^{-3}$])\,=\,9. Low-$T$ and high-$T$ power-law slopes of the EM($T$) functions are denoted  $\alpha$ and $\beta$, respectively.
\label{Fig:EMs_1}}
\end{figure*}
%----------------------------------------------------
%
%---------------------------------------------------- FIGURE 7
\begin{figure*}
% \sidecaption
%
        \centering
        \includegraphics[width=5.48cm,clip,bb= 40 70 725 555]{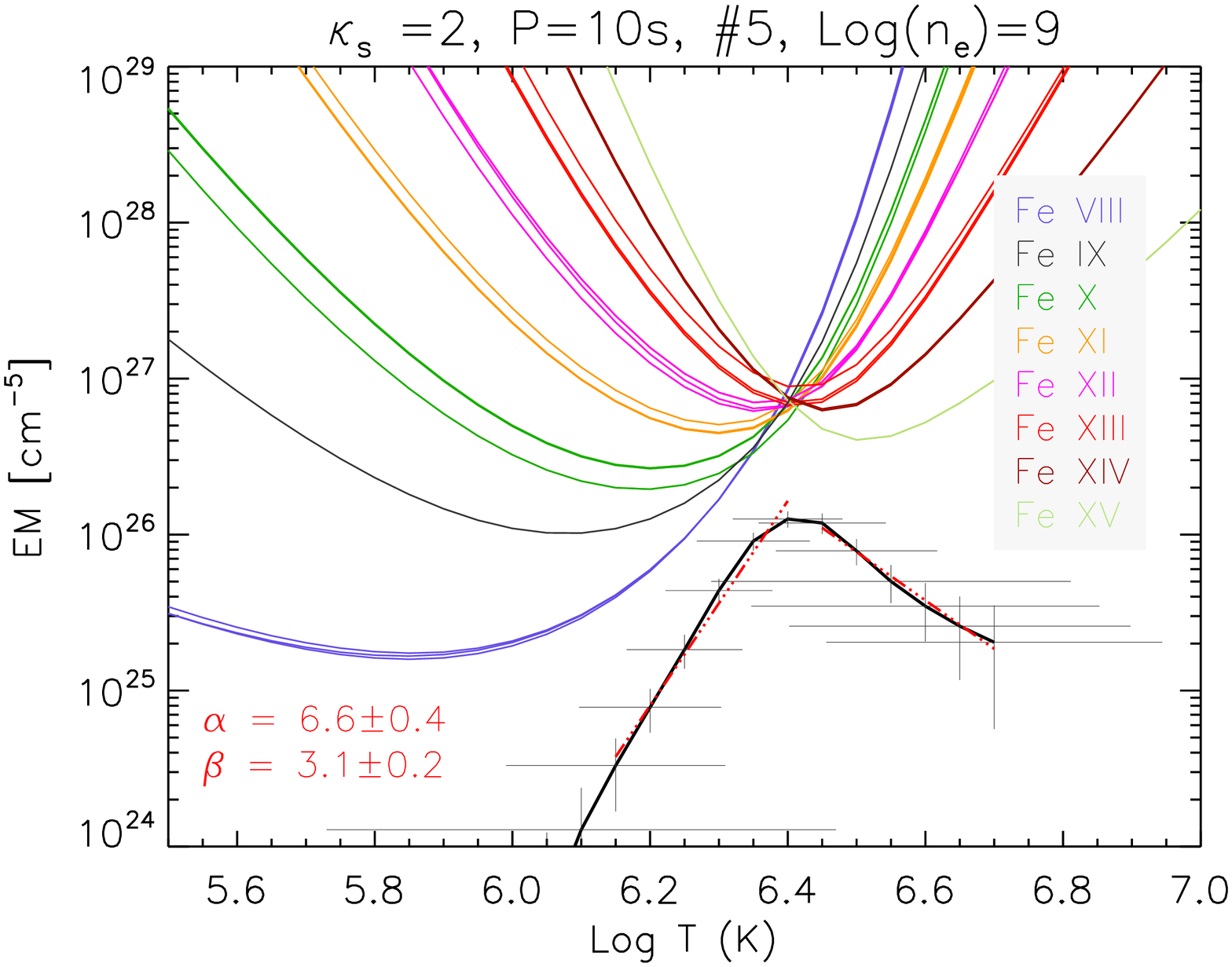}
        \includegraphics[width=4.76cm,clip,bb=130 70 725 555]{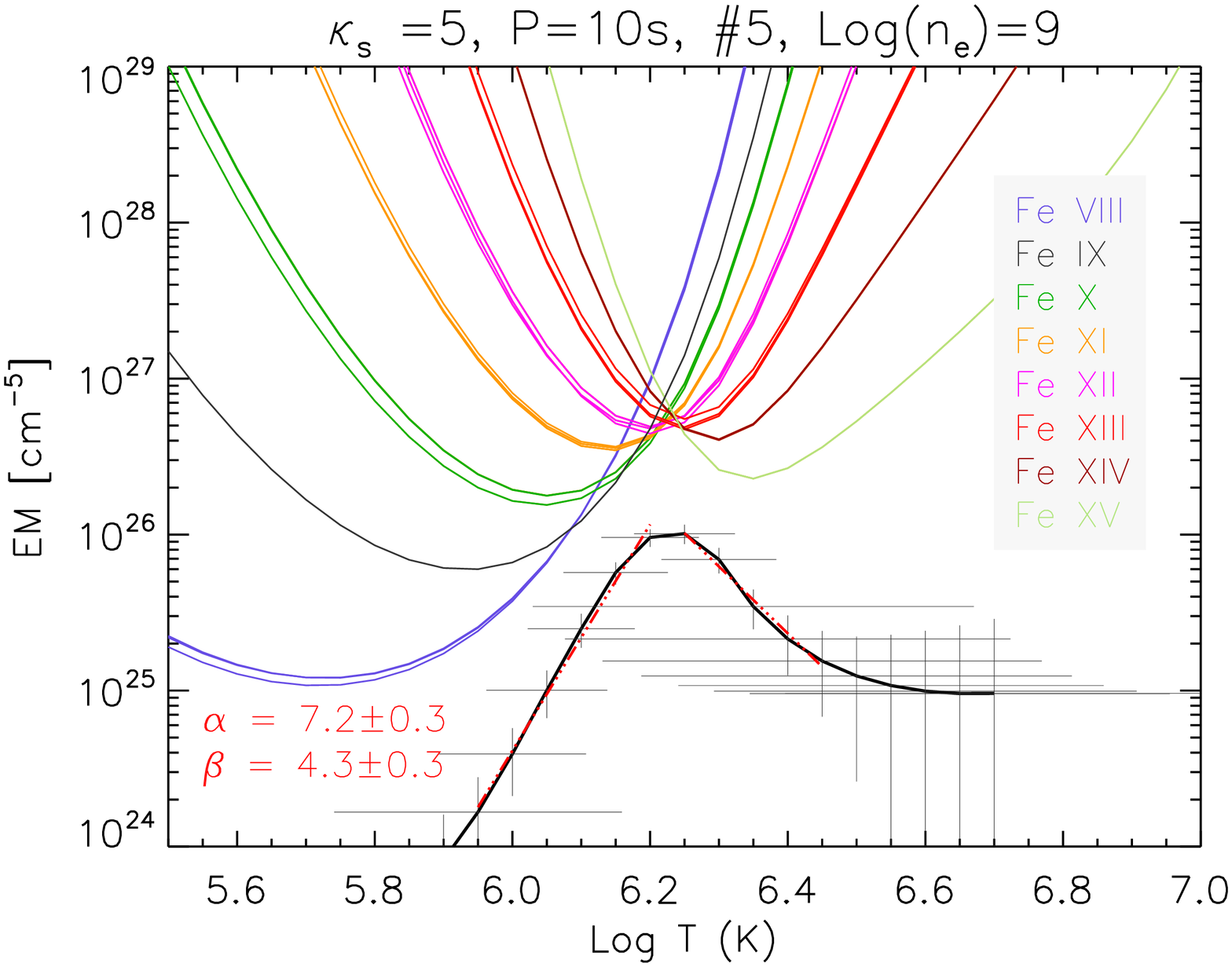}
        \includegraphics[width=4.76cm,clip,bb=130 70 725 555]{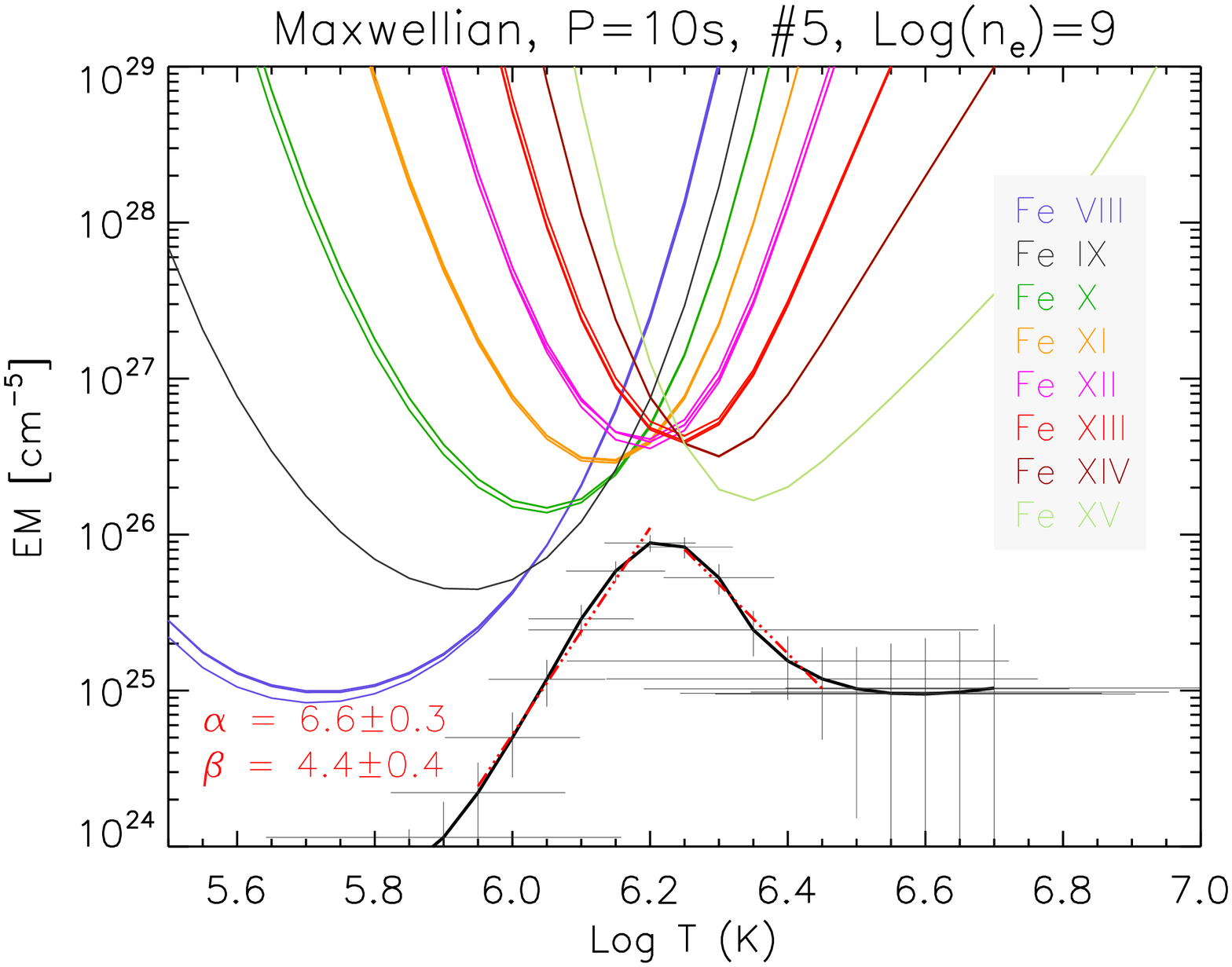}
        \includegraphics[width=5.48cm,clip,bb= 40 70 725 555]{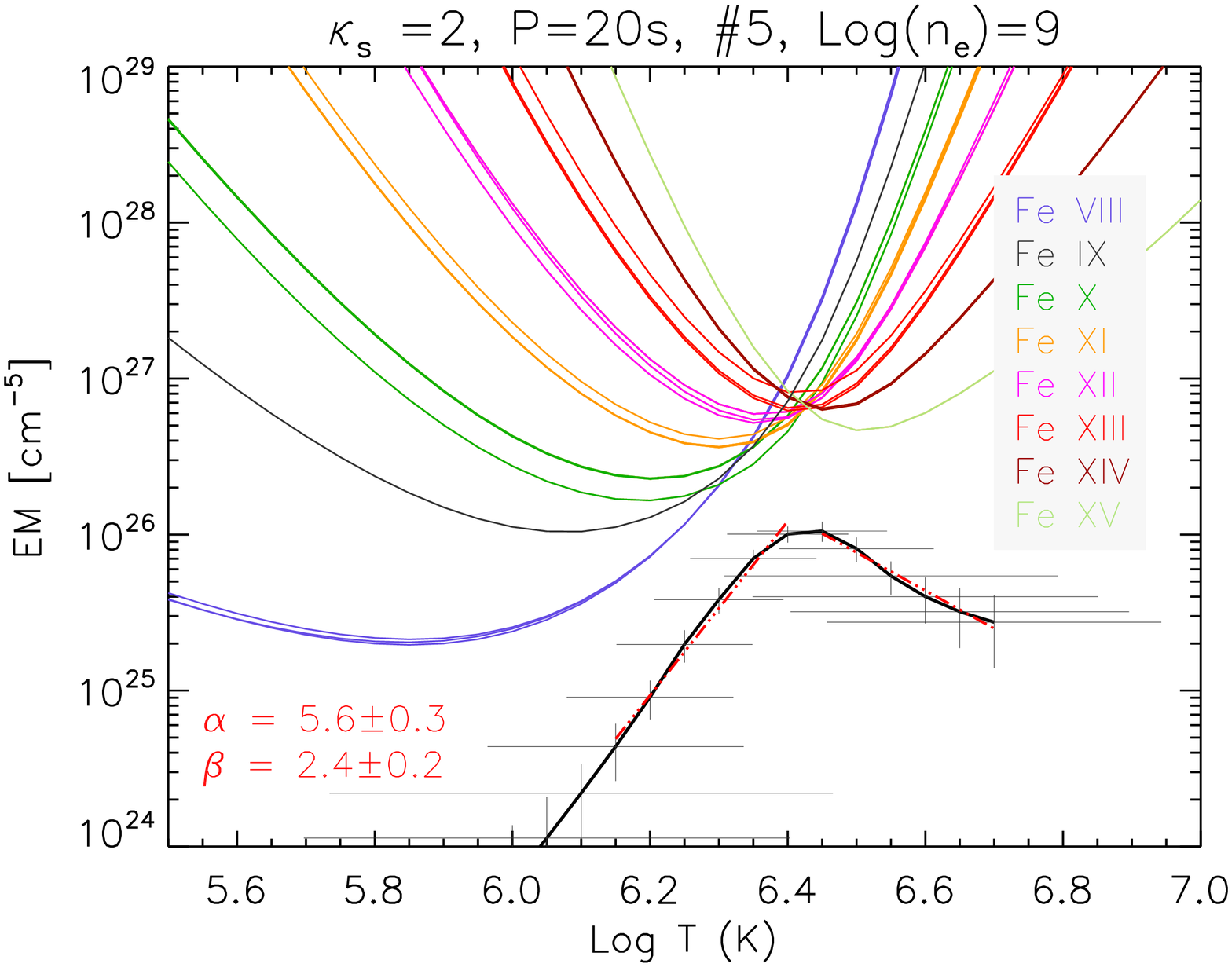}
        \includegraphics[width=4.76cm,clip,bb=130 70 725 555]{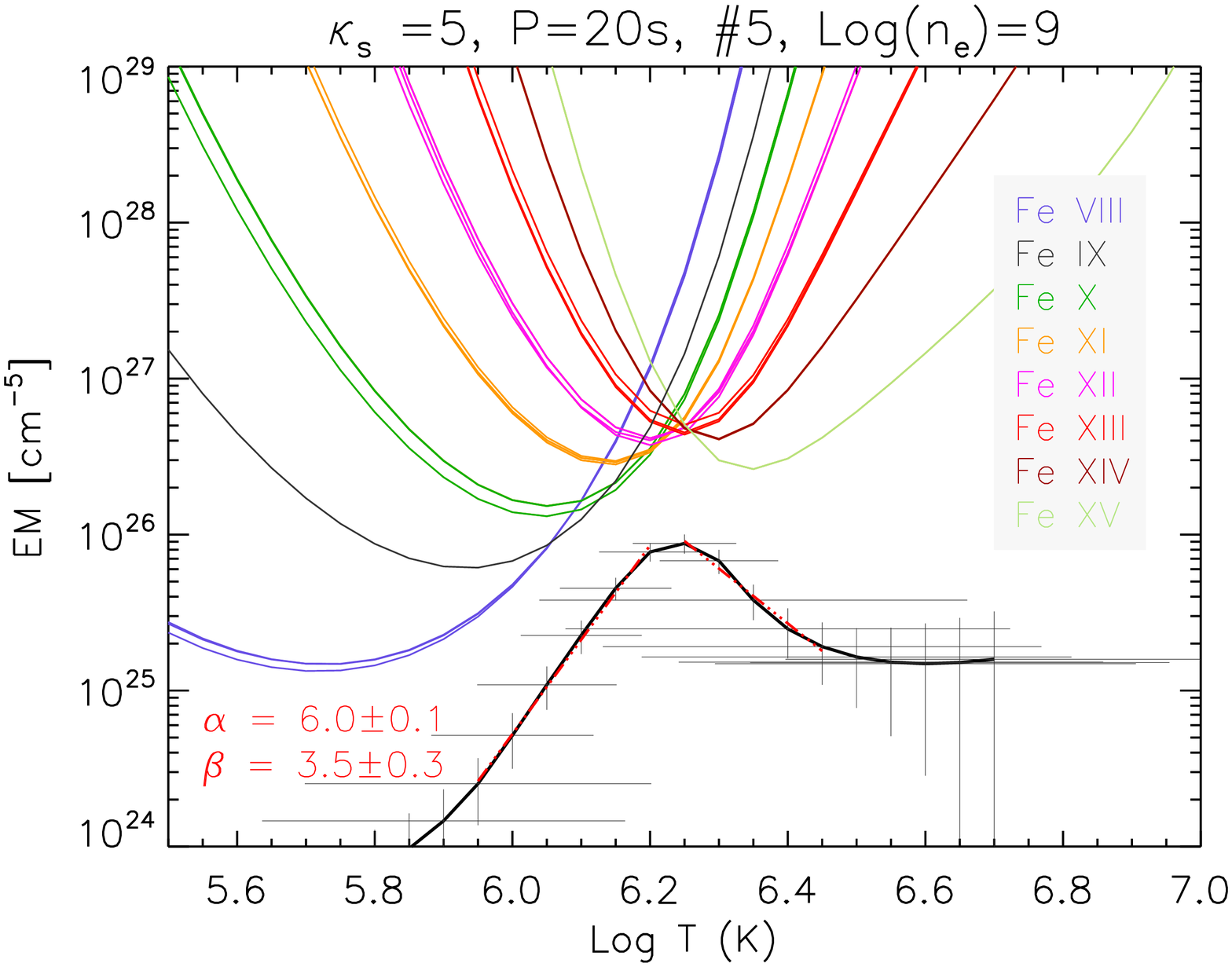}
        \includegraphics[width=4.76cm,clip,bb=130 70 725 555]{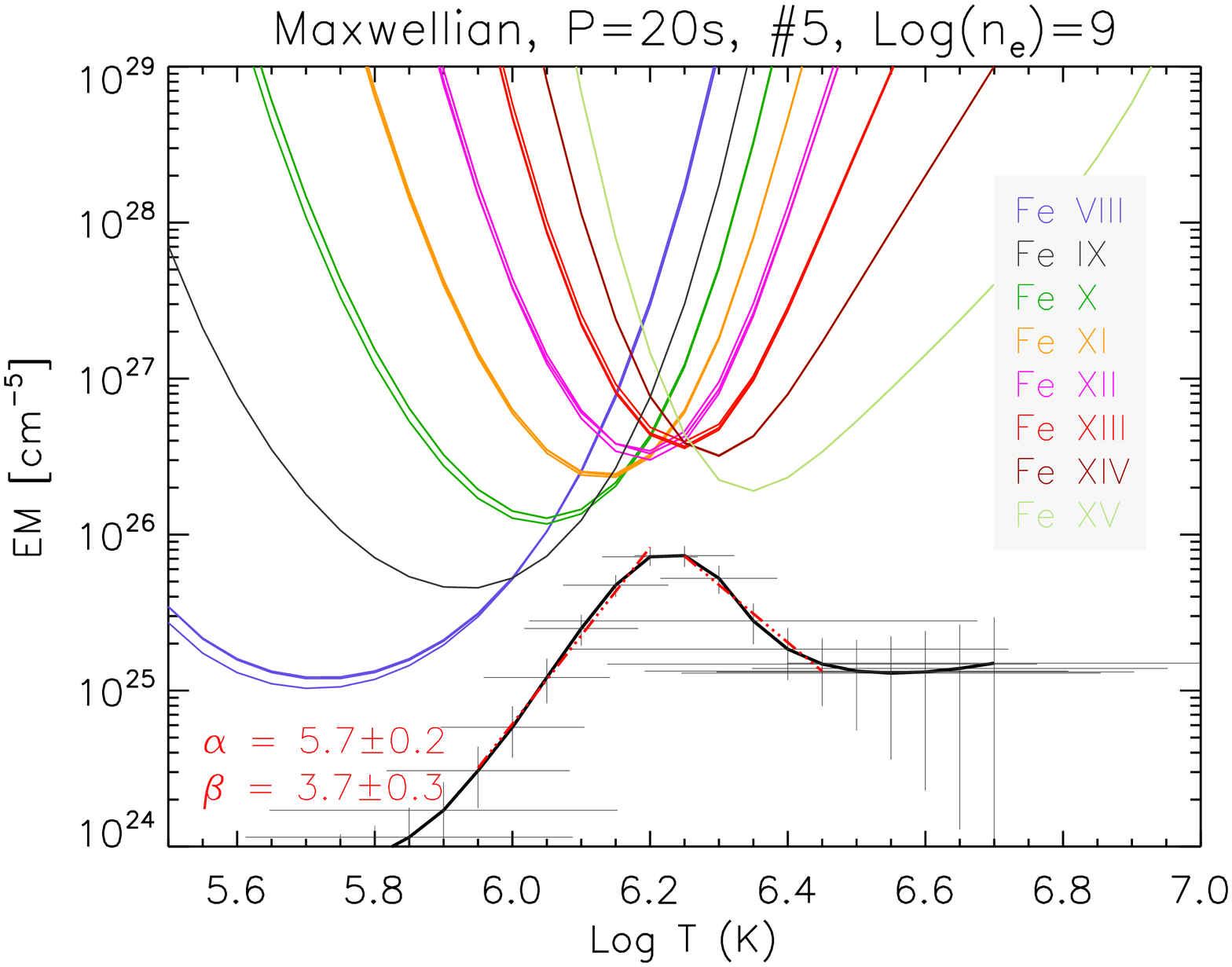}
        \includegraphics[width=5.48cm,clip,bb= 40 15 725 555]{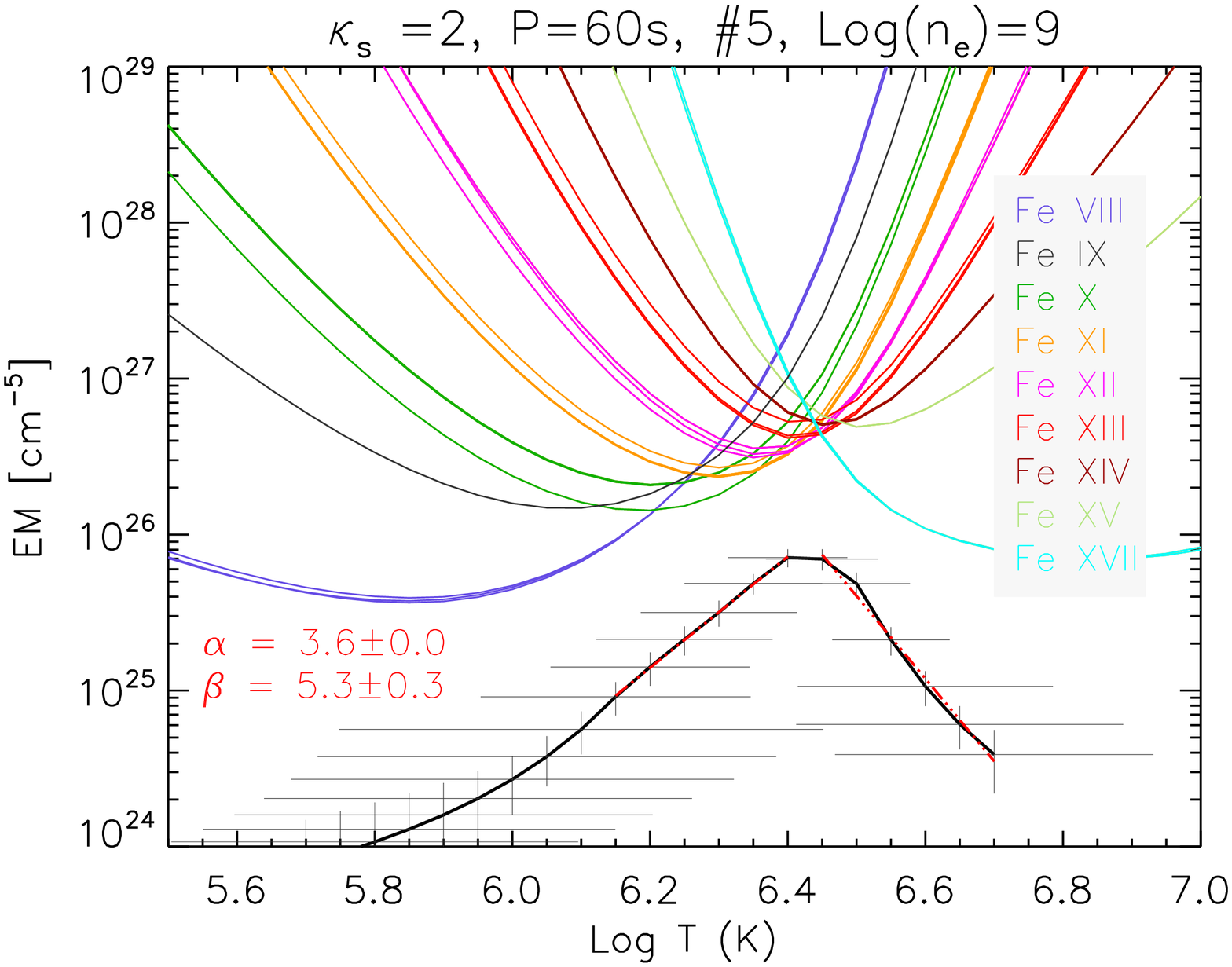}
        \includegraphics[width=4.76cm,clip,bb=130 15 725 555]{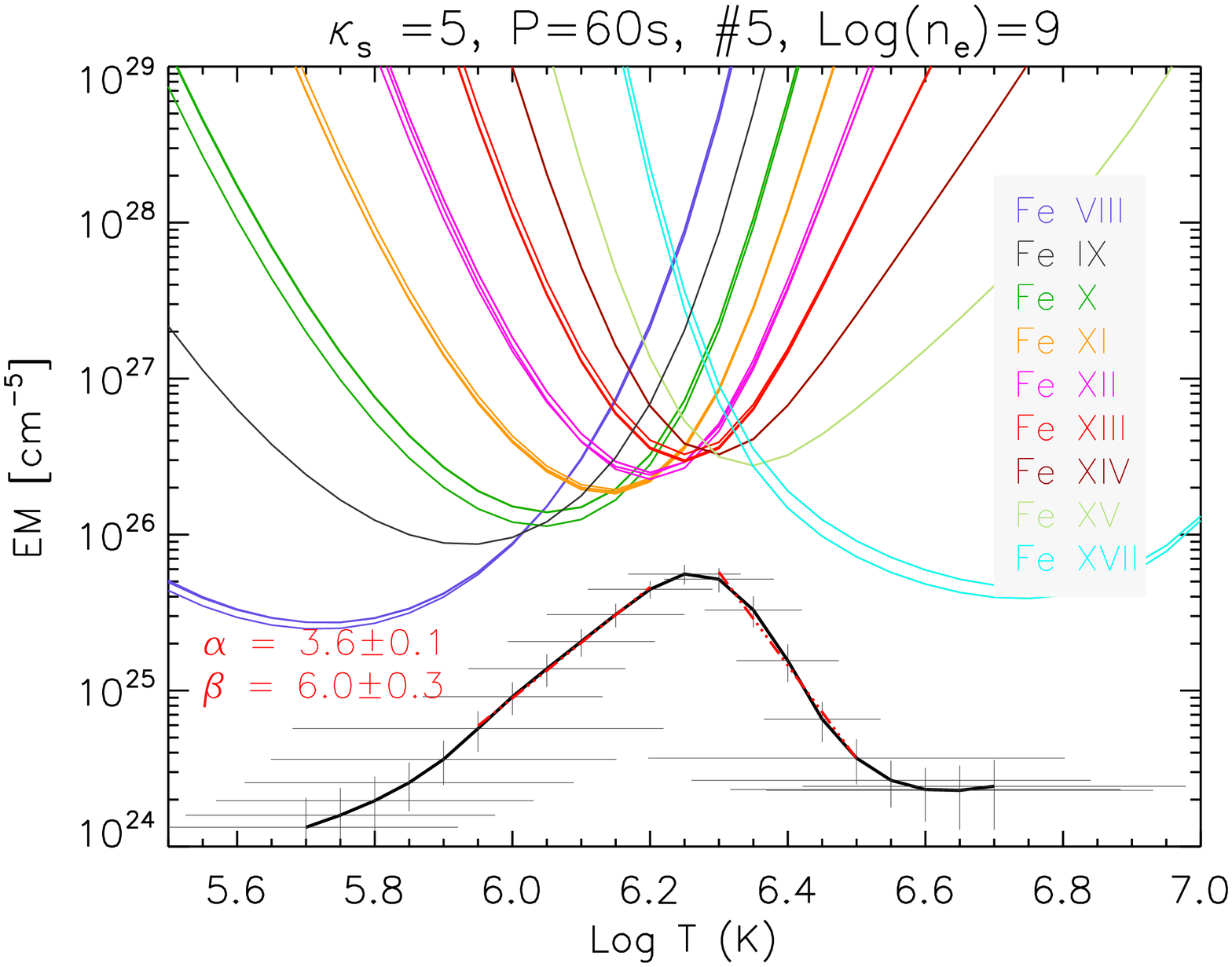}
        \includegraphics[width=4.76cm,clip,bb=130 15 725 555]{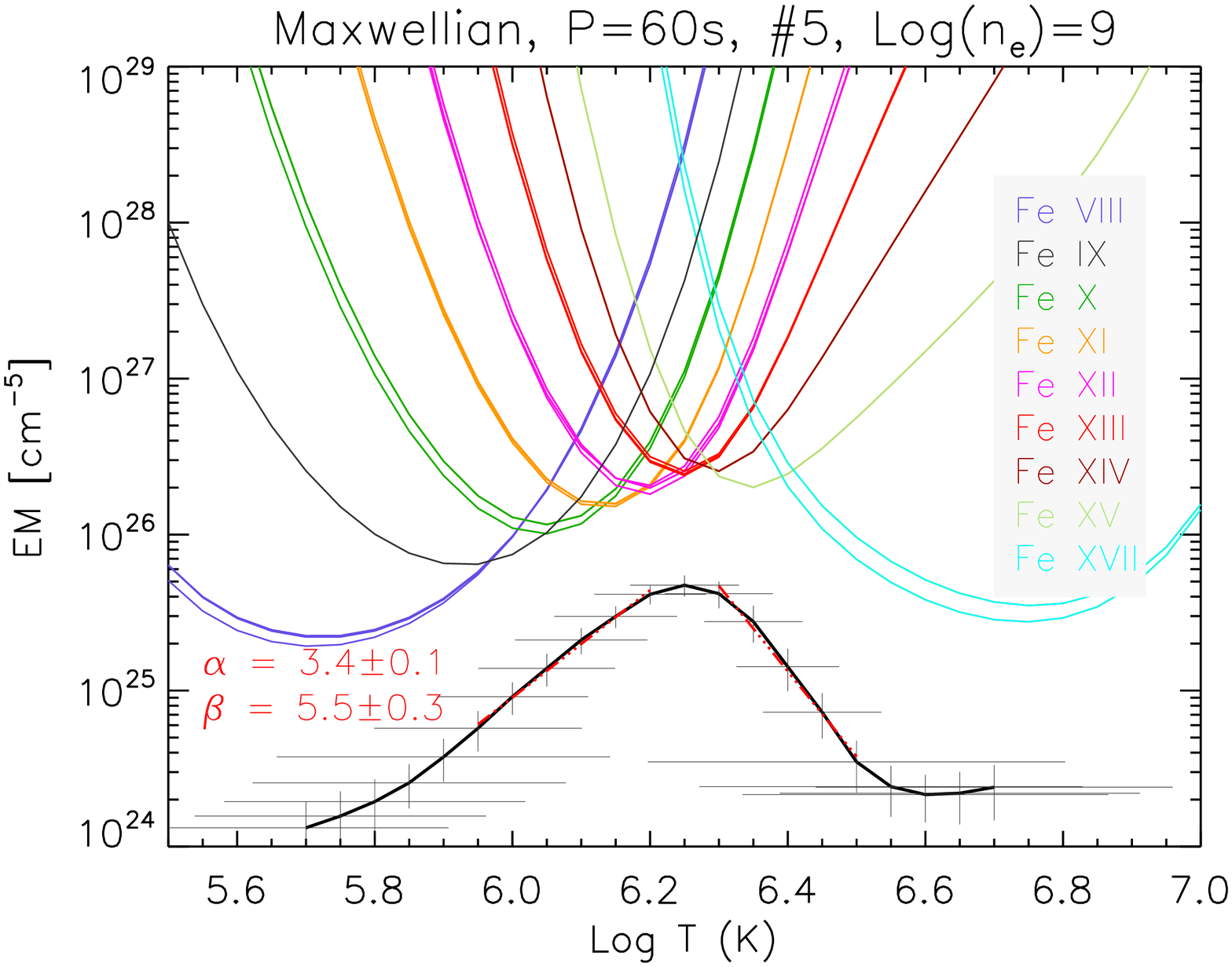}
\caption{Same as in Fig. \ref{Fig:EMs_1}, but for the fifth period.
\label{Fig:EMs_5}}
\end{figure*}
%----------------------------------------------------
%
%---------------------------------------------------- FIGURE 8
\begin{figure*}
\sidecaption
        \centering
        \includegraphics[width=5.80cm]{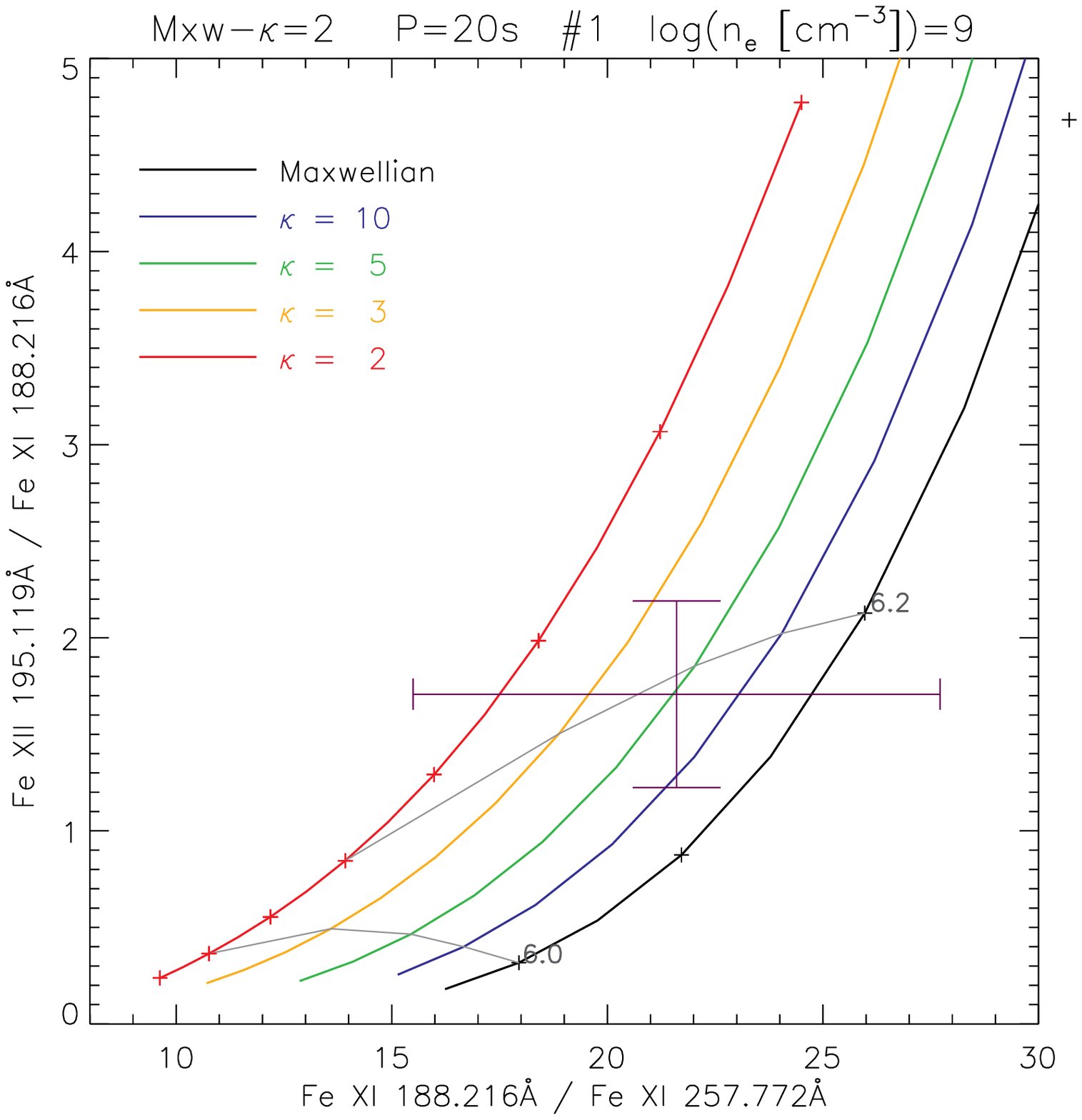}
        \includegraphics[width=5.80cm]{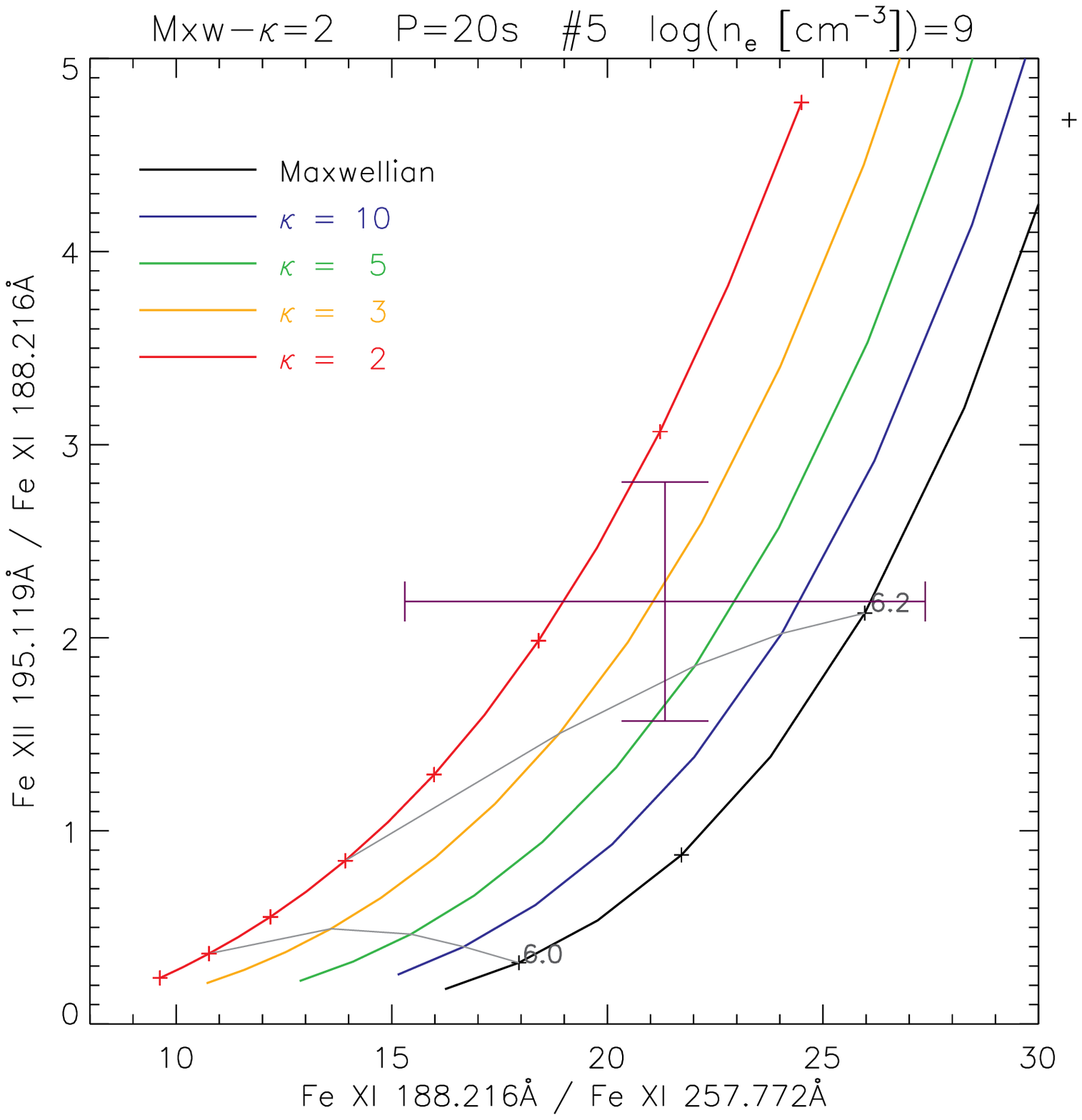}
\caption{Attempted diagnostics of a stationary $\kappa$-distribution using the period-averaged spectra for the first period (left) and the fifth period (right). Individual colors stand for different values of the stationary $\kappa$-distribution, while the violet cross represents the line intensity ratios with their assumed uncertainties.
\label{Fig:Diag}}
\end{figure*}
%----------------------------------------------------

%
%________________________________________________________________
\section{Consequences for interpretation of observations}
\label{Sect:4}

In Sect. \ref{Sect:3} it was found that all the studied cases are out of ionization equilibrium at all times, and that the respective period-averaged spectra show the presence of lines from multiple ionization stages. We now focus on the interpretation of such spectra, both in terms of plasma multithermality, i.e., using DEM techniques, as well as in terms of diagnosing possible departures from the Maxwellian distribution.

\subsection{Multithermality}
\label{Sect:4.1}

We first used the period-averaged synthetic intensities to calculate the respective DEMs as a function of $\kappa_\mathrm{s}$. Here, we use $\kappa_\mathrm{s}$ to denote to a stationary $\kappa$-distribution, i.e., a distribution of electron energies with a constant value of $\kappa_\mathrm{s}$, assumed by the prospective observer in analysis of the spectra. This analysis was used by \citet{Mackovjak14} on the spectra of active region cores and quiet Sun presented by \citet{Warren12} and \citet{Landi10}, respectively. \citet{Mackovjak14} found that the slopes of the EM($T$) for the multithermal active region cores did not change with $\kappa_\mathrm{s}$, while the quiet Sun spectra were multithermal with a decreasing degree of multithermality with decreasing $\kappa_\mathrm{s}$.

In Figs. \ref{Fig:EMs_1} and \ref{Fig:EMs_5}, we present the EM($T$) distributions calculated using the synthetic intensities from the synthetic spectra presented in Fig. \ref{Fig:Res_lne9_k02}, i.e., a periodic beam with $\kappa$\,=\,2 and log($n_\mathrm{e}$ [cm$^{-3}$])\,=\,9. The EM($T$) distributions presented in Fig. \ref{Fig:EMs_1} correspond to the intensities averaged over the first period, reflecting the case of a single pulse, while those in Fig. \ref{Fig:EMs_5} correspond to the intensities averaged over the fifth period, representing a near-stationary oscillatory state. The EM($T$) distributions shown in Figs. \ref{Fig:EMs_1} and \ref{Fig:EMs_5} were obtained from the respective DEM($T$) calculated using the regularized inversion method of \citet{Hannah12} under the assumption of a constant $\kappa_\mathrm{s}$ value. We note that the $\kappa_\mathrm{s}$ influences the line contribution functions, which are calculated using the method described in \citet{Dzifcakova15} under the assumption of the ionization equlibrium. The values of $\kappa_\mathrm{s}$ used for the construction of these EM$(T)$ distributions are $\kappa_\mathrm{s}$\,=\,2 (\textit{left column}), 5 (\textit{middle}), and $\infty$, i.e., a Maxwellian distribution (\textit{right column}).

We find that for nearly all the cases presented in Figs. \ref{Fig:EMs_1} and \ref{Fig:EMs_5}, the plasma is multithermal. The low-temperature and high-temperature slopes, $\alpha$ and $\beta$, respectively, are indicated in each panel of Figs. \ref{Fig:EMs_1} and \ref{Fig:EMs_5}. These slopes and their respective errors are obtained by linear least-square fitting of the log(EM) as a function of log($T$\,[K]). For $P$\,=\,10, the EM$(T)$ distributions are near-isothermal \citep[cf.,][Fig. 4 therein]{Warren14}, with $\alpha$ being higher than 7.2 independently of $\kappa_\mathrm{s}$. However, for $\kappa_\mathrm{s}$\,=\,5 and the first period, the EM-loci curves \citep[see, e.g.,][]{Strong78,Veck84,DelZanna03} have an almost isothermal crossing point at EM\,$\approx$\,6\,$\times$10$^{26}$\,cm$^{-5}$ and log($T$\,[K])\,$\approx$\,6.15. The corresponding spectrum can then be interpreted, within the limit of observational uncertainties, as isothermal for $\kappa_\mathrm{s}$\,=\,5, even though the spectrum originates in an out-of-equilibrium plasma heated by an electron beam.

A similar situation arises for the $P$\,=\,10\,s and the fifth period (Fig. \ref{Fig:EMs_5}, top); however, the EM-loci curves now indicate a near-isothermal plasma for $\kappa_\mathrm{s}$\,=\,2, with the crossing point at EM\,$\approx$\,7\,$\times$10$^{26}$\,cm$^{-5}$ and log($T$\,[K])\,$\approx$\,6.40. If this spectrum were interpreted as arising in a Maxwellian plasma, as is commonly done, the EM-loci curves would indicate a multithermal plasma with EM$(T)$ somewhat similar to that of the quiet Sun \citep{Landi10,Mackovjak14}.

With  increasing $P$, the spectra become progressively more multithermal, and the value of $\alpha$ decreases with increasing $P$. Furthermore, the spectra averaged over the fifth period are more multithermal than the spectra averaged over the first period. For $P$\,=\,60\,s and $\kappa_\mathrm{s}$ corresponding to the Maxwellian distribution, we find $\alpha$\,$\approx$\,3.5 with a somewhat higher value for lower $\kappa_\mathrm{s}$. These EM$(T)$ distributions are similar to those reported for coronal loops \citep[see, e.g.,][]{DelZanna03,Brooks11,Brooks12,Schmelz09,Schmelz13a,Schmelz13b,Schmelz14,Subramanian14,Dudik15}, although some of these were derived using different observed lines or different DEM methods, which makes the direct comparison difficult. Nevertheless, it is not the purpose of this paper to directly model the observed EM$(T)$ distributions; rather, we emphasize the finding that {out-of-equilibrium plasma will nearly always resemble multithermal plasma}. This is especially true if the periodic electron beam, or other agent causing the departures from equilibrium ionization, occurs on periods shorter than or comparable to that of the typical integration times of EUV spectrometers such as \textit{Hinode}/EIS. Finally, we note that the multithermal representation of the period-averaged synthetic spectra is made possible by the changes in the ionization state of the plasma, and that different excitation datasets have only a small effect on the resulting EM-loci and DEMs (see Appendix \ref{Appendix:A}).

\subsection{Diagnostics of a stationary $\kappa$-distribution}
\label{Sect:4.2}

We next attempted to diagnose the $\kappa_\mathrm{s}$ from the period-averaged spectra using the method presented in \citet{Dudik15}. There, the observed \textit{Hinode}/EIS spectra of a transient coronal loop were interpreted in terms of a stationary $\kappa$-distribution with a constant value of $\kappa_\mathrm{s}$. It was found that the loop is consistent with a very low value of $\kappa_\mathrm{s}$\,$\lesssim$\,2. This diagnostics was performed using the ratio-ratio technique, where two line ratios are combined to simultaneously diagnose the temperature $T_\kappa$ and the value of $\kappa$. Simultaneous diagnostics is necessary since $T_\kappa$ and $\kappa$ are both parameters of the distribution (Eq. \ref{Eq:Kappa}). The line ratios used by \citet{Dudik15} involved one ratio from ions in the neighboring ionization stages, \ion{Fe}{XI} and \ion{Fe}{XII}, which is sensitive to $\kappa_\mathrm{s}$ and strongly sensitive to $T_\kappa$. The other line ratio was a ratio of two \ion{Fe}{XI} lines separated in wavelength, one from the short-wavelength channel of EIS and the other  in the long-wavelength EIS channel. The difference in excitation energy thresholds of these two lines causes the ratio to become sensitive to the shape of the distribution, i.e., to both $T_\kappa$ and $\kappa_\mathrm{s}$.

Here, we use this ratio-ratio technique involving the same lines as \citet{Dudik15}. In Fig. \ref{Fig:Diag}, individual colors are used to distinguish the dependence of the \ion{Fe}{XII} 195.119\,\AA / \ion{Fe}{XI} 188.216\,\AA~ratio on the \ion{Fe}{XI} 188.216\,\AA\,/\,257.772\,\AA~ratio as a function of the assumed value of $\kappa_\mathrm{s}$. Maxwellian is in black, while red stands for $\kappa_\mathrm{s}$\,=\,2. We see that the individual ratio-ratio curves as a function of $\kappa_\mathrm{s}$ are separated and do not cross each other. Subsequently, the period-averaged intensities were used to calculate the values of individual line ratios, and the results are plotted in violet. The errorbars are calculated assuming a 20\% uncertainty in line intensities, a value typical for EIS observations \citep{Culhane07,DelZanna13b}.

The ratios with their respective error-bars shown in Fig. \ref{Fig:Diag} correspond to the case with $P$\,=\,20\,s, log($n_\mathrm{e}$\,[cm$^{-3}$])\,=\,9, and the beam described with $\kappa$\,=\,2. We see that the crosses indicate a moderate value of $\kappa_\mathrm{s}$\,=\,5 for the first period, and $\kappa_\mathrm{s}$\,=\,3 for the fifth period. Unfortunately, the errorbars, resulting from the use of a weak \ion{Fe}{XI} 257.772\,\AA~line, are too large to unambiguously diagnose the value of $\kappa_\mathrm{s}$. This is the case for all beam parameters investigated in Sect. \ref{Sect:3}. The only measurable quantity using the ratio-ratio technique is $T_\kappa$, ranging from $\approx$6.15 to 6.35, with lower values for the Maxwellian distribution. These values are slightly lower than the respective $T_\mathrm{eff}$ (Fig. \ref{Fig:Res_lne9_k02}, \textit{middle row}). 

Although the value of $\kappa_\mathrm{s}$ cannot be diagnosed using the application of this technique on the intensities obtained in Sect. \ref{Sect:3}, this result implies that the case of $\kappa_\mathrm{s}$\,$\lesssim$\,2 diagnosed by \citet{Dudik15} likely corresponds to a nearly time-independent $\kappa$-distribution rather than periodic high-energy beam.

%
%________________________________________________________________
\section{Summary}
\label{Sect:5}

We investigated the effect of a periodic, high-energy, power-law, electron beam on the coronal spectra of Fe. To do this, a simple model was employed, assuming that the distribution of electron energies switches periodically from a Maxwellian to a $\kappa$-distribution and back every half-period $P/2$. The $\kappa$-distribution was assumed to have nearly the same core and low-energy part as the Maxwellian distribution with the pre-set temperature of $T$\,=\,1 MK. This situation mimics, for example, the behavior of plasma in a portion of a warm coronal loop being periodically crossed by an electron beam being generated elsewhere along the loop, e.g., during periodically recurring nanoflare-like events.

We found that the periodic occurrence of the electron beam drives the plasma out of ionization equilibrium. The studied periods, $P$\,=\,10, 20, and 60\,s, are on the order of the typical spectrometer exposure times or the nanoflare duration. For all the cases studied here, the plasma is out of ionization equilibrium at all times, unless a sufficiently high electron density, on the order of 10$^{11}$ cm$^{-3}$, is assumed. The presence of the high-energy beam described by a $\kappa$-distribution also leads to an increase in the effective ionization temperature of the plasma, a quantity calculated from the mean charge state of the plasma.

The period-averaged and instantaneous spectra were calculated in order to investigate the possible observables. The instantaneous spectra show fast changes in lines of ions that are formed in equilibrium at temperatures outside of the effective ionization temperature; in our case these lines were typically those of \ion{Fe}{IX} and \ion{Fe}{XV}. The period-averaged spectra, calculated to approximate the observables obtained by a spectrometer with similar exposure times, were interpreted using the regularized inversion DEM technique under an assumption of a stationary Maxwellian or a $\kappa$-distribution, as is commonly done for the coronal observations. It was found that the out-of-equilibrium plasma nearly always resembles multithermal plasma. However, for some combination of parameters including $\kappa_\mathrm{s}$, the EM-loci curves showed an almost isothermal crossing point, which can be misinterpreted as an isothermal plasma with a constant value of $\kappa_\mathrm{s}$, even though the periodic occurrence of an electron beam with a $\kappa$-distribution drives the plasma out of ionization equilibrium.

We also attempted to perform a diagnostics of a stationary $\kappa$-distribution from the period-averaged spectra. Moderate departures from the Maxwellian were found; however, taking the realistic observational uncertainties into account can effectively prevent the determination of the $\kappa_\mathrm{s}$ using the ratio-ratio technique. This is due to the large error-bars compared to the sensitivity of the ratio-ratio curves to $\kappa_\mathrm{s}$. 

In summary, we showed that the periodic presence of energetic electrons, possibly created during  recurring nanoflare-type reconnection events in a given strand in the solar corona, could be at least a partial explanation for the presence of single structures observed in the solar corona that appear multithermal even after background subtraction, such as some coronal loops.  It is likely that multithermality also arises owing to the optically thin nature of the solar corona; however,  our results imply that the effects of energetic particles and the resulting out-of-equilibrium ionization on the spectra cannot be discounted when interpreting the coronal spectra and/or modeling the observables produced by nanoflare heating events. Clearly, identifying these non-equilibrium processes --  the energetic particles, the non-equilibrium ionization, and their combination -- deserve further study. In the meantime, we suggest that fast changes in some of the lines being formed outside of the peak of the respective emission measure distributions could be a potential observable for the presence of a periodic electron beam. Transition-region lines, where fast changes in line intensities are often detected \citep[][]{Doyle06,Testa13,Regnier14,Huang14,Vissers15}, will be the subject of a following paper.

\begin{acknowledgements}
The authors thank the referee for the careful reading of the manuscript and comments that helped to improve it. J.D. and E.Dz. acknowledge the Grant P209/12/1652 of the Grant Agency of the Czech Republic. J.D. also acknowledges support from the Royal Society via the Newton Alumni Programme. CHIANTI is a collaborative project involving George Mason University, the University of Michigan (USA), and the University of Cambridge (UK). Hinode is a Japanese mission developed and launched by ISAS/JAXA, with NAOJ as domestic partner and NASA and STFC (UK) as international partners. It is operated by these agencies in cooperation with ESA and NSC (Norway).
\end{acknowledgements}

%-------------------------------------------------------------------

\bibliographystyle{aa}         % A&A style
\bibliography{NonEq_Ioniz}   % name your BibTeX data base

\begin{thebibliography}{104}
\expandafter\ifx\csname natexlab\endcsname\relax\def\natexlab#1{#1}\fi

\bibitem[{{Aschwanden} \& {Nightingale}(2005)}]{Aschwanden05a}
{Aschwanden}, M.~J. \& {Nightingale}, R.~W. 2005, \apj, 633, 499

\bibitem[{{Aschwanden} {et~al.}(2008){Aschwanden}, {Nitta}, {Wuelser}, \&
  {Lemen}}]{Aschwanden08}
{Aschwanden}, M.~J., {Nitta}, N.~V., {Wuelser}, J.-P., \& {Lemen}, J.~R. 2008,
  \apj, 680, 1477

\bibitem[{{Bian} {et~al.}(2014){Bian}, {Emslie}, {Stackhouse}, \&
  {Kontar}}]{Bian14}
{Bian}, N.~H., {Emslie}, A.~G., {Stackhouse}, D.~J., \& {Kontar}, E.~P. 2014,
  \apj, 796, 142

\bibitem[{{Boerner} {et~al.}(2012){Boerner}, {Edwards}, {Lemen}, {Rausch},
  {Schrijver}, {Shine}, {Shing}, {Stern}, {Tarbell}, {Title}, {Wolfson},
  {Soufli}, {Spiller}, {Gullikson}, {McKenzie}, {Windt}, {Golub}, {Podgorski},
  {Testa}, \& {Weber}}]{Boerner12}
{Boerner}, P., {Edwards}, C., {Lemen}, J., {et~al.} 2012, \solphys, 275, 41

\bibitem[{{Bradshaw}(2009)}]{Bradshaw09}
{Bradshaw}, S.~J. 2009, \aap, 502, 409

\bibitem[{{Bradshaw} \& {Cargill}(2006)}]{Bradshaw06}
{Bradshaw}, S.~J. \& {Cargill}, P.~J. 2006, \aap, 458, 987

\bibitem[{{Bradshaw} \& {Klimchuk}(2011)}]{Bradshaw11}
{Bradshaw}, S.~J. \& {Klimchuk}, J.~A. 2011, \apjs, 194, 26

\bibitem[{{Bradshaw} {et~al.}(2012){Bradshaw}, {Klimchuk}, \&
  {Reep}}]{Bradshaw12}
{Bradshaw}, S.~J., {Klimchuk}, J.~A., \& {Reep}, J.~W. 2012, \apj, 758, 53

\bibitem[{{Bradshaw} \& {Mason}(2003{\natexlab{a}})}]{Bradshaw03a}
{Bradshaw}, S.~J. \& {Mason}, H.~E. 2003{\natexlab{a}}, \aap, 401, 699

\bibitem[{{Bradshaw} \& {Mason}(2003{\natexlab{b}})}]{Bradshaw03b}
{Bradshaw}, S.~J. \& {Mason}, H.~E. 2003{\natexlab{b}}, \aap, 407, 1127

\bibitem[{{Brooks} {et~al.}(2012){Brooks}, {Warren}, \&
  {Ugarte-Urra}}]{Brooks12}
{Brooks}, D.~H., {Warren}, H.~P., \& {Ugarte-Urra}, I. 2012, \apjl, 755, L33

\bibitem[{{Brooks} {et~al.}(2011){Brooks}, {Warren}, \& {Young}}]{Brooks11}
{Brooks}, D.~H., {Warren}, H.~P., \& {Young}, P.~R. 2011, \apj, 730, 85

\bibitem[{{Burge} {et~al.}(2014){Burge}, {MacKinnon}, \& {Petkaki}}]{Burge14}
{Burge}, C.~A., {MacKinnon}, A.~L., \& {Petkaki}, P. 2014, \aap, 561, A107

\bibitem[{{Cargill}(1994)}]{Cargill94}
{Cargill}, P.~J. 1994, \apj, 422, 381

\bibitem[{{Cargill}(2014)}]{Cargill14}
{Cargill}, P.~J. 2014, \apj, 784, 49

\bibitem[{{Cargill} \& {Klimchuk}(2004)}]{Cargill04}
{Cargill}, P.~J. \& {Klimchuk}, J.~A. 2004, \apj, 605, 911

\bibitem[{{Cargill} {et~al.}(2012){Cargill}, {Vlahos}, {Baumann}, {Drake}, \&
  {Nordlund}}]{Cargill12}
{Cargill}, P.~J., {Vlahos}, L., {Baumann}, G., {Drake}, J.~F., \& {Nordlund},
  {\AA}. 2012, \ssr, 173, 223

\bibitem[{{Che} \& {Goldstein}(2014)}]{Che14}
{Che}, H. \& {Goldstein}, M.~L. 2014, \apjl, 795, L38

\bibitem[{{Cirtain}(2005)}]{Cirtain05}
{Cirtain}, J.~W. 2005, PhD thesis, Montana State University, Montana, USA

\bibitem[{{Culhane} {et~al.}(2007){Culhane}, {Harra}, {James}, {Al-Janabi},
  {Bradley}, {Chaudry}, {Rees}, {Tandy}, {Thomas}, {Whillock}, {Winter},
  {Doschek}, {Korendyke}, {Brown}, {Myers}, {Mariska}, {Seely}, {Lang}, {Kent},
  {Shaughnessy}, {Young}, {Simnett}, {Castelli}, {Mahmoud}, {Mapson-Menard},
  {Probyn}, {Thomas}, {Davila}, {Dere}, {Windt}, {Shea}, {Hagood}, {Moye},
  {Hara}, {Watanabe}, {Matsuzaki}, {Kosugi}, {Hansteen}, \&
  {Wikstol}}]{Culhane07}
{Culhane}, J.~L., {Harra}, L.~K., {James}, A.~M., {et~al.} 2007, \solphys, 243,
  19

\bibitem[{{Curdt} {et~al.}(2004){Curdt}, {Landi}, \& {Feldman}}]{Curdt04}
{Curdt}, W., {Landi}, E., \& {Feldman}, U. 2004, \aap, 427, 1045

\bibitem[{{Del Zanna}(2013{\natexlab{a}})}]{DelZanna13b}
{Del Zanna}, G. 2013{\natexlab{a}}, \aap, 555, A47

\bibitem[{{Del Zanna}(2013{\natexlab{b}})}]{DelZanna13}
{Del Zanna}, G. 2013{\natexlab{b}}, \aap, 558, A73

\bibitem[{{Del Zanna} {et~al.}(2015){Del Zanna}, {Dere}, {Young}, {Landi}, \&
  {Mason}}]{DelZanna15b}
{Del Zanna}, G., {Dere}, K.~P., {Young}, P.~R., {Landi}, E., \& {Mason}, H.~E.
  2015, \aap, 582, A56

\bibitem[{{Del Zanna} \& {Mason}(2003)}]{DelZanna03}
{Del Zanna}, G. \& {Mason}, H.~E. 2003, \aap, 406, 1089

\bibitem[{{Del Zanna} {et~al.}(2012){Del Zanna}, {Storey}, {Badnell}, \&
  {Mason}}]{DelZanna12}
{Del Zanna}, G., {Storey}, P.~J., {Badnell}, N.~R., \& {Mason}, H.~E. 2012,
  \aap, 543, A139

\bibitem[{{Dere}(2007)}]{Dere07}
{Dere}, K.~P. 2007, \aap, 466, 771

\bibitem[{{Dere} {et~al.}(1997){Dere}, {Landi}, {Mason}, {Monsignori Fossi}, \&
  {Young}}]{Dere97}
{Dere}, K.~P., {Landi}, E., {Mason}, H.~E., {Monsignori Fossi}, B.~C., \&
  {Young}, P.~R. 1997, \aaps, 125, 149

\bibitem[{{Doyle} {et~al.}(2006){Doyle}, {Ishak}, {Madjarska}, {O'Shea}, \&
  {Dzif{\'c}{\'a}kov{\'a}}}]{Doyle06}
{Doyle}, J.~G., {Ishak}, B., {Madjarska}, M.~S., {O'Shea}, E., \&
  {Dzif{\'c}{\'a}kov{\'a}}, E. 2006, \aap, 451, L35

\bibitem[{{Dud{\'{\i}}k} {et~al.}(2014{\natexlab{a}}){Dud{\'{\i}}k}, {Del
  Zanna}, {Dzif{\v c}{\'a}kov{\'a}}, {Mason}, \& {Golub}}]{Dudik14a}
{Dud{\'{\i}}k}, J., {Del Zanna}, G., {Dzif{\v c}{\'a}kov{\'a}}, E., {Mason},
  H.~E., \& {Golub}, L. 2014{\natexlab{a}}, \apjl, 780, L12

\bibitem[{{Dud{\'{\i}}k} {et~al.}(2014{\natexlab{b}}){Dud{\'{\i}}k}, {Del
  Zanna}, {Mason}, \& {Dzif{\v c}{\'a}kov{\'a}}}]{Dudik14b}
{Dud{\'{\i}}k}, J., {Del Zanna}, G., {Mason}, H.~E., \& {Dzif{\v
  c}{\'a}kov{\'a}}, E. 2014{\natexlab{b}}, \aap, 570, A124

\bibitem[{{Dud{\'{\i}}k} {et~al.}(2015){Dud{\'{\i}}k}, {Mackovjak}, {Dzif{\v
  c}{\'a}kov{\'a}}, {Del Zanna}, {Williams}, {Karlick{\'y}}, {Mason},
  {L{\"o}rin{\v c}{\'{\i}}k}, {Kotr{\v c}}, {F{\'a}rn{\'{\i}}k}, \&
  {Zemanov{\'a}}}]{Dudik15}
{Dud{\'{\i}}k}, J., {Mackovjak}, {\v S}., {Dzif{\v c}{\'a}kov{\'a}}, E.,
  {et~al.} 2015, \apj, 807, 123

\bibitem[{{Dzif{\v c}{\'a}kov{\'a}} \& {Dud{\'{\i}}k}(2013)}]{Dzifcakova13a}
{Dzif{\v c}{\'a}kov{\'a}}, E. \& {Dud{\'{\i}}k}, J. 2013, Astrophys. J. Suppl.
  Ser., 206, 6

\bibitem[{{Dzif{\v c}{\'a}kov{\'a}} {et~al.}(2015){Dzif{\v c}{\'a}kov{\'a}},
  {Dud{\'{\i}}k}, {Kotr{\v c}}, {F{\'a}rn{\'{\i}}k}, \&
  {Zemanov{\'a}}}]{Dzifcakova15}
{Dzif{\v c}{\'a}kov{\'a}}, E., {Dud{\'{\i}}k}, J., {Kotr{\v c}}, P.,
  {F{\'a}rn{\'{\i}}k}, F., \& {Zemanov{\'a}}, A. 2015, \apjs, 217, 14

\bibitem[{{Dzif{\v c}{\'a}kov{\'a}} {et~al.}(2011){Dzif{\v c}{\'a}kov{\'a}},
  {Homola}, \& {Dud{\'{\i}}k}}]{Dzifcakova11}
{Dzif{\v c}{\'a}kov{\'a}}, E., {Homola}, M., \& {Dud{\'{\i}}k}, J. 2011, \aap,
  531, A111

\bibitem[{{Feldman} {et~al.}(1992){Feldman}, {Mandelbaum}, {Seely}, {Doschek},
  \& {Gursky}}]{Feldman92}
{Feldman}, U., {Mandelbaum}, P., {Seely}, J.~F., {Doschek}, G.~A., \& {Gursky},
  H. 1992, Astrophys. J. Suppl. Ser., 81, 387

\bibitem[{{Fletcher} {et~al.}(2011){Fletcher}, {Dennis}, {Hudson}, {Krucker},
  {Phillips}, {Veronig}, {Battaglia}, {Bone}, {Caspi}, {Chen}, {Gallagher},
  {Grigis}, {Ji}, {Liu}, {Milligan}, \& {Temmer}}]{Fletcher11}
{Fletcher}, L., {Dennis}, B.~R., {Hudson}, H.~S., {et~al.} 2011, Space Sci.
  Rev., 159, 19

\bibitem[{{Golub} {et~al.}(1989){Golub}, {Hartquist}, \& {Quillen}}]{Golub89}
{Golub}, L., {Hartquist}, T.~W., \& {Quillen}, A.~C. 1989, \solphys, 122, 245

\bibitem[{{Gontikakis} {et~al.}(2013){Gontikakis}, {Patsourakos},
  {Efthymiopoulos}, {Anastasiadis}, \& {Georgoulis}}]{Gontikakis13}
{Gontikakis}, C., {Patsourakos}, S., {Efthymiopoulos}, C., {Anastasiadis}, A.,
  \& {Georgoulis}, M.~K. 2013, \apj, 771, 126

\bibitem[{{Gordovskyy} {et~al.}(2013){Gordovskyy}, {Browning}, {Kontar}, \&
  {Bian}}]{Gordovskyy13}
{Gordovskyy}, M., {Browning}, P.~K., {Kontar}, E.~P., \& {Bian}, N.~H. 2013,
  \solphys, 284, 489

\bibitem[{{Gordovskyy} {et~al.}(2014){Gordovskyy}, {Browning}, {Kontar}, \&
  {Bian}}]{Gordovskyy14}
{Gordovskyy}, M., {Browning}, P.~K., {Kontar}, E.~P., \& {Bian}, N.~H. 2014,
  \aap, 561, A72

\bibitem[{{Gupta} {et~al.}(2015){Gupta}, {Tripathi}, \& {Mason}}]{Gupta15}
{Gupta}, G.~R., {Tripathi}, D., \& {Mason}, H.~E. 2015, \apj, 800, 140

\bibitem[{{Hannah} {et~al.}(2010){Hannah}, {Hudson}, {Hurford}, \&
  {Lin}}]{Hannah10}
{Hannah}, I.~G., {Hudson}, H.~S., {Hurford}, G.~J., \& {Lin}, R.~P. 2010, \apj,
  724, 487

\bibitem[{{Hannah} \& {Kontar}(2012)}]{Hannah12}
{Hannah}, I.~G. \& {Kontar}, E.~P. 2012, \aap, 539, A146

\bibitem[{{Hasegawa} {et~al.}(1985){Hasegawa}, {Mima}, \&
  {Duong-van}}]{Hasegawa85}
{Hasegawa}, A., {Mima}, K., \& {Duong-van}, M. 1985, Physical Review Letters,
  54, 2608

\bibitem[{{Huang} {et~al.}(2014){Huang}, {Madjarska}, {Xia}, {Doyle},
  {Galsgaard}, \& {Fu}}]{Huang14}
{Huang}, Z., {Madjarska}, M.~S., {Xia}, L., {et~al.} 2014, \apj, 797, 88

\bibitem[{{Klimchuk}(2006)}]{Klimchuk06}
{Klimchuk}, J.~A. 2006, \solphys, 234, 41

\bibitem[{{Klimchuk}(2015)}]{Klimchuk15}
{Klimchuk}, J.~A. 2015, Royal Society of London Philosophical Transactions
  Series A, 373, 40256

\bibitem[{{Klimchuk} \& {Bradshaw}(2014)}]{Klimchuk14}
{Klimchuk}, J.~A. \& {Bradshaw}, S.~J. 2014, \apj, 791, 60

\bibitem[{{Klimchuk} {et~al.}(2010){Klimchuk}, {Karpen}, \&
  {Antiochos}}]{Klimchuk10}
{Klimchuk}, J.~A., {Karpen}, J.~T., \& {Antiochos}, S.~K. 2010, \apj, 714, 1239

\bibitem[{{Kosugi} {et~al.}(2007){Kosugi}, {Matsuzaki}, {Sakao}, {Shimizu},
  {Sone}, {Tachikawa}, {Hashimoto}, {Minesugi}, {Ohnishi}, {Yamada}, {Tsuneta},
  {Hara}, {Ichimoto}, {Suematsu}, {Shimojo}, {Watanabe}, {Shimada}, {Davis},
  {Hill}, {Owens}, {Title}, {Culhane}, {Harra}, {Doschek}, \&
  {Golub}}]{Kosugi07}
{Kosugi}, T., {Matsuzaki}, K., {Sakao}, T., {et~al.} 2007, \solphys, 243, 3

\bibitem[{{Laming} \& {Lepri}(2007)}]{Laming07}
{Laming}, J.~M. \& {Lepri}, S.~T. 2007, \apj, 660, 1642

\bibitem[{{Landi} {et~al.}(2002){Landi}, {Feldman}, \& {Dere}}]{Landi02}
{Landi}, E., {Feldman}, U., \& {Dere}, K.~P. 2002, Astrophys. J. Suppl. Ser.,
  139, 281

\bibitem[{{Landi} \& {Landini}(2004)}]{Landi04}
{Landi}, E. \& {Landini}, M. 2004, \apj, 608, 1133

\bibitem[{{Landi} \& {Young}(2009)}]{Landi09}
{Landi}, E. \& {Young}, P.~R. 2009, \apj, 706, 1

\bibitem[{{Landi} \& {Young}(2010)}]{Landi10}
{Landi}, E. \& {Young}, P.~R. 2010, \apj, 714, 636

\bibitem[{{Landi} {et~al.}(2013){Landi}, {Young}, {Dere}, {Del Zanna}, \&
  {Mason}}]{Landi13}
{Landi}, E., {Young}, P.~R., {Dere}, K.~P., {Del Zanna}, G., \& {Mason}, H.~E.
  2013, \apj, 763, 86

\bibitem[{{Lemen} {et~al.}(2012){Lemen}, {Title}, {Akin}, {Boerner}, {Chou},
  {Drake}, {Duncan}, {Edwards}, {Friedlaender}, {Heyman}, {Hurlburt}, {Katz},
  {Kushner}, {Levay}, {Lindgren}, {Mathur}, {McFeaters}, {Mitchell}, {Rehse},
  {Schrijver}, {Springer}, {Stern}, {Tarbell}, {Wuelser}, {Wolfson}, {Yanari},
  {Bookbinder}, {Cheimets}, {Caldwell}, {Deluca}, {Gates}, {Golub}, {Park},
  {Podgorski}, {Bush}, {Scherrer}, {Gummin}, {Smith}, {Auker}, {Jerram},
  {Pool}, {Soufli}, {Windt}, {Beardsley}, {Clapp}, {Lang}, \&
  {Waltham}}]{Lemen12}
{Lemen}, J.~R., {Title}, A.~M., {Akin}, D.~J., {et~al.} 2012, \solphys, 275, 17

\bibitem[{{Mackovjak} {et~al.}(2014){Mackovjak}, {Dzif{\v c}{\'a}kov{\'a}}, \&
  {Dud{\'{\i}}k}}]{Mackovjak14}
{Mackovjak}, {\v S}., {Dzif{\v c}{\'a}kov{\'a}}, E., \& {Dud{\'{\i}}k}, J.
  2014, \aap, 564, A130

\bibitem[{{Mason} \& {Mosignori-Fossi}(1994)}]{Mason94}
{Mason}, H.~E. \& {Mosignori-Fossi}, B.~C. 1994, \aapr, 6, 123

\bibitem[{{O'Dwyer} {et~al.}(2011){O'Dwyer}, {Del Zanna}, {Mason}, {Sterling},
  {Tripathi}, \& {Young}}]{ODwyer11}
{O'Dwyer}, B., {Del Zanna}, G., {Mason}, H.~E., {et~al.} 2011, \aap, 525, A137

\bibitem[{{Oka} {et~al.}(2013){Oka}, {Ishikawa}, {Saint-Hilaire}, {Krucker}, \&
  {Lin}}]{Oka13}
{Oka}, M., {Ishikawa}, S., {Saint-Hilaire}, P., {Krucker}, S., \& {Lin}, R.~P.
  2013, \apj, 764, 6

\bibitem[{{Olluri} {et~al.}(2013{\natexlab{a}}){Olluri}, {Gudiksen}, \&
  {Hansteen}}]{Olluri13b}
{Olluri}, K., {Gudiksen}, B.~V., \& {Hansteen}, V.~H. 2013{\natexlab{a}}, \apj,
  767, 43

\bibitem[{{Olluri} {et~al.}(2013{\natexlab{b}}){Olluri}, {Gudiksen}, \&
  {Hansteen}}]{Olluri13a}
{Olluri}, K., {Gudiksen}, B.~V., \& {Hansteen}, V.~H. 2013{\natexlab{b}}, \aj,
  145, 72

\bibitem[{{Olluri} {et~al.}(2015){Olluri}, {Gudiksen}, {Hansteen}, \& {De
  Pontieu}}]{Olluri15}
{Olluri}, K., {Gudiksen}, B.~V., {Hansteen}, V.~H., \& {De Pontieu}, B. 2015,
  \apj, 802, 5

\bibitem[{{Owocki} \& {Scudder}(1983)}]{Owocki83}
{Owocki}, S.~P. \& {Scudder}, J.~D. 1983, \apj, 270, 758

\bibitem[{{Parker}(1988)}]{Parker88}
{Parker}, E.~N. 1988, \apj, 330, 474

\bibitem[{{Phillips} {et~al.}(2008){Phillips}, {Feldman}, \&
  {Landi}}]{Phillips08}
{Phillips}, K.~J.~H., {Feldman}, U., \& {Landi}, E. 2008, {Ultraviolet and
  X-ray Spectroscopy of the Solar Atmosphere} (Cambridge University Press)

\bibitem[{{Price} {et~al.}(2015){Price}, {Taroyan}, {Innes}, \&
  {Bradshaw}}]{Price15}
{Price}, D.~J., {Taroyan}, Y., {Innes}, D.~E., \& {Bradshaw}, S.~J. 2015,
  \solphys, 290, 1931

\bibitem[{{Reale} \& {Orlando}(2008)}]{Reale08}
{Reale}, F. \& {Orlando}, S. 2008, \apj, 684, 715

\bibitem[{{Reep} {et~al.}(2013){Reep}, {Bradshaw}, \& {Klimchuk}}]{Reep13}
{Reep}, J.~W., {Bradshaw}, S.~J., \& {Klimchuk}, J.~A. 2013, \apj, 764, 193

\bibitem[{{R{\'e}gnier} {et~al.}(2014){R{\'e}gnier}, {Alexander}, {Walsh},
  {Winebarger}, {Cirtain}, {Golub}, {Korreck}, {Mitchell}, {Platt}, {Weber},
  {De Pontieu}, {Title}, {Kobayashi}, {Kuzin}, \& {DeForest}}]{Regnier14}
{R{\'e}gnier}, S., {Alexander}, C.~E., {Walsh}, R.~W., {et~al.} 2014, \apj,
  784, 134

\bibitem[{{Schmelz} {et~al.}(2013{\natexlab{a}}){Schmelz}, {Jenkins}, \&
  {Pathak}}]{Schmelz13b}
{Schmelz}, J.~T., {Jenkins}, B.~S., \& {Pathak}, S. 2013{\natexlab{a}}, \apj,
  770, 14

\bibitem[{{Schmelz} {et~al.}(2011){Schmelz}, {Jenkins}, {Worley}, {Anderson},
  {Pathak}, \& {Kimble}}]{Schmelz11}
{Schmelz}, J.~T., {Jenkins}, B.~S., {Worley}, B.~T., {et~al.} 2011, \apj, 731,
  49

\bibitem[{{Schmelz} {et~al.}(2009){Schmelz}, {Nasraoui}, {Rightmire}, {Kimble},
  {del Zanna}, {Cirtain}, {DeLuca}, \& {Mason}}]{Schmelz09}
{Schmelz}, J.~T., {Nasraoui}, K., {Rightmire}, L.~A., {et~al.} 2009, \apj, 691,
  503

\bibitem[{{Schmelz} {et~al.}(2014){Schmelz}, {Pathak}, {Dhaliwal}, {Christian},
  \& {Fair}}]{Schmelz14}
{Schmelz}, J.~T., {Pathak}, S., {Dhaliwal}, R.~S., {Christian}, G.~M., \&
  {Fair}, C.~B. 2014, \apj, 795, 139

\bibitem[{{Schmelz} {et~al.}(2013{\natexlab{b}}){Schmelz}, {Winebarger},
  {Kimble}, {Pathak}, {Golub}, {Jenkins}, \& {Worley}}]{Schmelz13a}
{Schmelz}, J.~T., {Winebarger}, A.~R., {Kimble}, J.~A., {et~al.}
  2013{\natexlab{b}}, \apj, 770, 160

\bibitem[{{Shestov} {et~al.}(2009){Shestov}, {Urnov}, {Kuzin}, {Zhitnik}, \&
  {Bogachev}}]{Shestov09}
{Shestov}, S.~V., {Urnov}, A.~M., {Kuzin}, S.~V., {Zhitnik}, I.~A., \&
  {Bogachev}, S.~A. 2009, Astronomy Letters, 35, 45

\bibitem[{{Smith} \& {Hughes}(2010)}]{Smith10}
{Smith}, R.~K. \& {Hughes}, J.~P. 2010, \apj, 718, 583

\bibitem[{{Strong}(1978)}]{Strong78}
{Strong}, K.~T. 1978, PhD thesis, , Univ.~College London, (1978)

\bibitem[{{Subramanian} {et~al.}(2014){Subramanian}, {Tripathi}, {Klimchuk}, \&
  {Mason}}]{Subramanian14}
{Subramanian}, S., {Tripathi}, D., {Klimchuk}, J.~A., \& {Mason}, H.~E. 2014,
  \apj, 795, 76

\bibitem[{{Taroyan} {et~al.}(2006){Taroyan}, {Bradshaw}, \&
  {Doyle}}]{Taroyan06}
{Taroyan}, Y., {Bradshaw}, S.~J., \& {Doyle}, J.~G. 2006, \aap, 446, 315

\bibitem[{{Taroyan} {et~al.}(2011){Taroyan}, {Erd{\'e}lyi}, \&
  {Bradshaw}}]{Taroyan11}
{Taroyan}, Y., {Erd{\'e}lyi}, R., \& {Bradshaw}, S.~J. 2011, \solphys, 269, 295

\bibitem[{{Teriaca} {et~al.}(2012){Teriaca}, {Warren}, \& {Curdt}}]{Teriaca12}
{Teriaca}, L., {Warren}, H.~P., \& {Curdt}, W. 2012, \apjl, 754, L40

\bibitem[{{Testa} {et~al.}(2014){Testa}, {De Pontieu}, {Allred}, {Carlsson},
  {Reale}, {Daw}, {Hansteen}, {Martinez-Sykora}, {Liu}, {DeLuca}, {Golub},
  {McKillop}, {Reeves}, {Saar}, {Tian}, {Lemen}, {Title}, {Boerner}, {Hulburt},
  {Tarbell}, {Wuelser}, {Kleint}, {Kankelborg}, \& {Jaeggli}}]{Testa14}
{Testa}, P., {De Pontieu}, B., {Allred}, J., {et~al.} 2014, Science, 346,
  1255724

\bibitem[{{Testa} {et~al.}(2013){Testa}, {De Pontieu},
  {Mart{\'{\i}}nez-Sykora}, {DeLuca}, {Hansteen}, {Cirtain}, {Winebarger},
  {Golub}, {Kobayashi}, {Korreck}, {Kuzin}, {Walsh}, {DeForest}, {Title}, \&
  {Weber}}]{Testa13}
{Testa}, P., {De Pontieu}, B., {Mart{\'{\i}}nez-Sykora}, J., {et~al.} 2013,
  \apjl, 770, L1

\bibitem[{{Tripathi} {et~al.}(2009){Tripathi}, {Mason}, {Dwivedi}, {del Zanna},
  \& {Young}}]{Tripathi09}
{Tripathi}, D., {Mason}, H.~E., {Dwivedi}, B.~N., {del Zanna}, G., \& {Young},
  P.~R. 2009, \apj, 694, 1256

\bibitem[{{Tripathi} {et~al.}(2010){Tripathi}, {Mason}, \&
  {Klimchuk}}]{Tripathi10}
{Tripathi}, D., {Mason}, H.~E., \& {Klimchuk}, J.~A. 2010, \apj, 723, 713

\bibitem[{{Ugarte-Urra} \& {Warren}(2012)}]{Ugarte12}
{Ugarte-Urra}, I. \& {Warren}, H.~P. 2012, \apj, 761, 21

\bibitem[{{Vasyliunas}(1968)}]{Vasyliunas68}
{Vasyliunas}, V.~M. 1968, in Astrophysics and Space Science Library, Vol.~10,
  Physics of the Magnetosphere, ed. {R.~D.~L.~Carovillano \& J.~F.~McClay}, 622

\bibitem[{{Veck} {et~al.}(1984){Veck}, {Strong}, {Jordan}, {Simnett},
  {Cargill}, \& {Priest}}]{Veck84}
{Veck}, N.~J., {Strong}, K.~T., {Jordan}, C., {et~al.} 1984, \mnras, 210, 443

\bibitem[{{Viall} \& {Klimchuk}(2011{\natexlab{a}})}]{Viall11}
{Viall}, N.~M. \& {Klimchuk}, J.~A. 2011{\natexlab{a}}, \apj, 738, 24

\bibitem[{{Viall} \& {Klimchuk}(2011{\natexlab{b}})}]{Viall12}
{Viall}, N.~M. \& {Klimchuk}, J.~A. 2011{\natexlab{b}}, \apj, 738, 24

\bibitem[{{Viall} \& {Klimchuk}(2015)}]{Viall15}
{Viall}, N.~M. \& {Klimchuk}, J.~A. 2015, \apj, 799, 58

\bibitem[{{Vissers} {et~al.}(2015){Vissers}, {Rouppe van der Voort}, {Rutten},
  {Carlsson}, \& {De Pontieu}}]{Vissers15}
{Vissers}, G.~J.~M., {Rouppe van der Voort}, L.~H.~M., {Rutten}, R.~J.,
  {Carlsson}, M., \& {De Pontieu}, B. 2015, ArXiv e-prints

\bibitem[{{Vocks} {et~al.}(2008){Vocks}, {Mann}, \& {Rausche}}]{Vocks08}
{Vocks}, C., {Mann}, G., \& {Rausche}, G. 2008, \aap, 480, 527

\bibitem[{{Warren} {et~al.}(2011){Warren}, {Brooks}, \&
  {Winebarger}}]{Warren11}
{Warren}, H.~P., {Brooks}, D.~H., \& {Winebarger}, A.~R. 2011, \apj, 734, 90

\bibitem[{{Warren} {et~al.}(2014){Warren}, {Ugarte-Urra}, \&
  {Landi}}]{Warren14}
{Warren}, H.~P., {Ugarte-Urra}, I., \& {Landi}, E. 2014, \apjs, 213, 11

\bibitem[{{Warren} \& {Winebarger}(2003)}]{Warren03}
{Warren}, H.~P. \& {Winebarger}, A.~R. 2003, \apjl, 596, L113

\bibitem[{{Warren} {et~al.}(2012){Warren}, {Winebarger}, \&
  {Brooks}}]{Warren12}
{Warren}, H.~P., {Winebarger}, A.~R., \& {Brooks}, D.~H. 2012, \apj, 759, 141

\bibitem[{{Winebarger} {et~al.}(2013){Winebarger}, {Walsh}, {Moore}, {De
  Pontieu}, {Hansteen}, {Cirtain}, {Golub}, {Kobayashi}, {Korreck}, {DeForest},
  {Weber}, {Title}, \& {Kuzin}}]{Winebarger13}
{Winebarger}, A.~R., {Walsh}, R.~W., {Moore}, R., {et~al.} 2013, \apj, 771, 21

\bibitem[{{Winebarger} {et~al.}(2003){Winebarger}, {Warren}, \&
  {Seaton}}]{Winebarger03}
{Winebarger}, A.~R., {Warren}, H.~P., \& {Seaton}, D.~B. 2003, \apj, 593, 1164

\bibitem[{{Young} {et~al.}(2012){Young}, {O'Dwyer}, \& {Mason}}]{Young12}
{Young}, P.~R., {O'Dwyer}, B., \& {Mason}, H.~E. 2012, \apj, 744, 14

\bibitem[{{Young} {et~al.}(2009){Young}, {Watanabe}, {Hara}, \&
  {Mariska}}]{Young09}
{Young}, P.~R., {Watanabe}, T., {Hara}, H., \& {Mariska}, J.~T. 2009, \aap,
  495, 587

\end{thebibliography}

%-------------------------------------------------------------------
\begin{appendix}

%__________________________________________________________________
\section{Influence of the excitation atomic data on the resulting period-averaged spectra}
\label{Appendix:A}

Recently, the CHIANTI database underwent a significant revision containing new excitation cross-sections for many of the coronal Fe ions in the new version 8 \citep{DelZanna15b}. Since we use here the atomic data corresponding to the CHIANTI version 7.1, for which the excitation rates are available in the KAPPA database, we now estimate the influence of these new atomic data on the resulting period-averaged spectrum.

The period-averaged intensity $\left<I_{i,jk}\right>$ of a particular spectral line $\lambda_{jk}$ is given by 
\begin{equation}
 \left<I_{i,jk}\right> = \frac{1}{P} \int_0^{P/2} I^{\kappa=2}_{i,jk}(t) dt +\frac{1}{P} \int_{P/2}^P I^{\mathrm{Mxw}}_{i,jk}(t) dt\,,
 \label{Eq:avg_int}
\end{equation}
where $t$ is time and $I^{\kappa=2}_{i,jk}$ and $I^{\mathrm{Mxw}}_{i,jk}$ are the corresponding time-dependent intensities during the respective half-periods $P/2$ where the distribution is a $\kappa=2$ or a Maxwellian, respectively. Using Eqs. (5)--(6), we get
\begin{eqnarray}
 \nonumber \left<I_{i,jk}\right> &=& \frac{1}{P} \frac{hc}{\lambda_{jk}} \frac{A_{jk}}{n_\mathrm{e}} A_\mathrm{Fe} n_\mathrm{H} n_\mathrm{e} \left(\frac{n_j}{n_i}\right)^{\kappa=2} \int_0^{P/2} Y_i(t) dt + \,\\
              && + \frac{1}{P} \frac{hc}{\lambda_{jk}} \frac{A_{jk}}{n_\mathrm{e}} A_\mathrm{Fe} n_\mathrm{H} n_\mathrm{e} \left(\frac{n_j}{n_i}\right)^{\mathrm{Mxw}} \int_{P/2}^P Y_i(t) dt \,.
              \label{Eq:avg_int_2}
\end{eqnarray}
The expression $n_j/n_i$ represents the excited \textit{fraction} of an ion $+i$ with an electron on the upper level $j$. These expressions can be brought before the respective integrals since the excitation timescales are much shorter than the ionization and recombination timescales \citep{Phillips08}, i.e., the level population adjusts nearly instantaneously to reflect the new distributions of electrons, which is either a Maxwellian or a $\kappa$-distribution within the respective half-periods (see Sect. \ref{Sect:2.2}). Therefore, the $n_j/n_i$ expression is nearly constant throughout each half-period.

Using this result, the ratio of period-averaged intensities calculated using the excitation cross-sections from CHIANTI v8 to CHIANTI v7.1 are then obtained as 
\begin{eqnarray}
 \nonumber \frac{\left<I_{i,jk}\right>_\mathrm{v8}}{\left<I_{i,jk}\right>_{\mathrm{v7.1}}} =
 \frac{ {\left( \frac{n_j}{n_i}\right)^{\kappa=2}_{\mathrm{v8}} \int_0^{P/2} Y_i(t) dt} + {\left( \frac{n_j}{n_i}\right)^{\mathrm{Mxw}}_{\mathrm{v8}} \int_{P/2}^P Y_i(t) dt} }
      { {\left( \frac{n_j}{n_i}\right)^{\kappa=2}_{\mathrm{v7.1}} \int_0^{P/2} Y_i(t) dt} + {\left( \frac{n_j}  {n_i}\right)^{\mathrm{Mxw}}_{\mathrm{v7.1}} \int_{P/2}^P Y_i(t) dt} } = \\
 = \frac{ \left( \frac{n_j}{n_i}\right)^{\mathrm{Mxw}}_{\mathrm{v8}} \left( \frac{\left( \frac{n_j}{n_i}\right)^{\kappa=2}_{\mathrm{v8}}} {\left( \frac{n_j}{n_i}\right)^{\mathrm{Mxw}}_{\mathrm{v8}}} \int_0^{P/2} Y_i(t) dt + \int_{P/2}^P Y_i(t) dt   \right) }
        {\left( \frac{n_j}{n_i}\right)^{\mathrm{Mxw}}_{\mathrm{v7.1}} \left( \frac{\left( \frac{n_j}{n_i}\right)^{\kappa=2}_{\mathrm{v7.1}}} {\left( \frac{n_j}{n_i}\right)^{\mathrm{Mxw}}_{\mathrm{v7.1}}} \int_0^{P/2} Y_i(t) dt + \int_{P/2}^P Y_i(t) dt   \right)}\,.
        \label{Eq:int_ratio}
\end{eqnarray}
Assuming now that the ratio of excitation fractions $(n_j/n_i)^{\kappa=2} /(n_j/n_i)^{\mathrm{Mxw}}$ does not strongly depend on the atomic datasets used (CHIANTI v8 or CHIANTI v7.1), we see that
\begin{equation}
 \frac{\left<I_{i,jk}\right>_\mathrm{v8}}{\left<I_{i,jk}\right>_{\mathrm{v7.1}}} \approx \frac{\left( \frac{n_j}{n_i}\right)^{\mathrm{Mxw}}_{\mathrm{v8}}} {\left( \frac{n_j}{n_i}\right)^{\mathrm{Mxw}}_{\mathrm{v7.1}}}\,.
 \label{Eq:int_ratio_approx}
\end{equation}
This assumption is justified if the behavior of the respective collision strengths with incident energy $E$ does not change much between the two atomic datasets. That this is the case can be seen, for example, from  the fact that the density-sensitive line ratios do not change substantially for different $\kappa$ or different atomic datasets \citep[see][]{Dudik14b}, with the exception of \ion{Fe}{XII} \citep[see also][]{DelZanna12}.

Finally we note that the EM-loci curves shown in Figs. \ref{Fig:EMs_1} and \ref{Fig:EMs_5} will not be affected by the atomic datasets, since the change in period-averaged intensities in the numerator and contribution functions in the denominator cancel out under the above mentioned assumption.

\end{appendix}

\end{document}